\documentclass[11pt]{article}
\pdfoutput=1

\usepackage[a4paper, margin=2.5cm, heightrounded=true]{geometry}
\usepackage{xcolor}
\usepackage{caption}
\usepackage[round]{natbib}
\usepackage{hyperref}

\definecolor{darkblue}{rgb}{0, 0, 0.5}
\hypersetup{
    breaklinks=true,
    colorlinks=true,
    citecolor=darkblue,
    linkcolor=darkblue,
    urlcolor=darkblue
}

\makeatletter

\newlength\titlebox
\twocolumn
\def\addcontentsline#1#2#3{}

\newcommand\outauthor{%
    \begin{tabular}[t]{c}
        \bfseries\@author
    \end{tabular}}

\def\maketitle{\par
 \begingroup
   \def\thefootnote{\fnsymbol{footnote}}
   \twocolumn[\@maketitle]
   \@thanks
 \endgroup
 \setcounter{footnote}{0}
 \let\maketitle\relax
 \let\@maketitle\relax
 \gdef\@thanks{}\gdef\@author{}\gdef\@title{}\let\thanks\relax}

\def\@maketitle{\vbox to \titlebox{\hsize\textwidth
 \linewidth\hsize \vskip 0.125in minus 0.125in \centering
 {\Large\bfseries \@title \par} \vskip 0.2in plus 1fil minus 0.1in
 {\def\and{\unskip\enspace{\rmfamily and}\enspace}%
  \def\And{\end{tabular}\hss \egroup \hskip 1in plus 2fil
           \hbox to 0pt\bgroup\hss \begin{tabular}[t]{c}\bfseries}%
  \def\AND{\end{tabular}\hss\egroup \hfil\hfil\egroup
          \vskip 0.25in plus 1fil minus 0.125in
           \hbox to \linewidth\bgroup\large \hfil\hfil
             \hbox to 0pt\bgroup\hss \begin{tabular}[t]{c}\bfseries}
  \hbox to \linewidth\bgroup\large \hfil\hfil
    \hbox to 0pt\bgroup\hss
  \outauthor
   \hss\egroup
    \hfil\hfil\egroup}
  \vskip 0.3in plus 2fil minus 0.1in
}}

\renewenvironment{abstract}%
  {\begin{center}\large\textbf{\abstractname}\end{center}%
    \begin{list}{}%
      {\setlength{\rightmargin}{0.6cm}%
        \setlength{\leftmargin}{0.6cm}}%
      \item[]\ignorespaces%
      \@setsize\normalsize{12pt}\xpt\@xpt
  }%
  {\unskip\end{list}}

\DeclareCaptionFont{10pt}{\fontsize{10pt}{12pt}\selectfont}
\renewcommand\cite{\citep}

\def\thebibliography#1{\vskip\parskip%
\vskip\baselineskip%
\def\baselinestretch{1}%
\ifx\@currsize\normalsize\@normalsize\else\@currsize\fi%
\vskip-\parskip%
\vskip-\baselineskip%
\section*{References\@mkboth
 {References}{References}}\list
 {}{\setlength{\labelwidth}{0pt}\setlength{\leftmargin}{\parindent}
 \setlength{\itemindent}{-\parindent}}
 \def\newblock{\hskip .11em plus .33em minus -.07em}
 \sloppy\clubpenalty4000\widowpenalty4000
 \sfcode`\.=1000\relax}

\def\section{\@startsection {section}{1}{\z@}{-2.0ex plus
    -0.5ex minus -.2ex}{1.5ex plus 0.3ex minus .2ex}{\large\bfseries\raggedright}}
\def\subsection{\@startsection{subsection}{2}{\z@}{-1.8ex plus
    -0.5ex minus -.2ex}{0.8ex plus .2ex}{\normalsize\bfseries\raggedright}}
\def\subsubsection{\@startsection{subsubsection}{3}{\z@}{-1.5ex plus
   -0.5ex minus -.2ex}{0.5ex plus .2ex}{\normalsize\bfseries\raggedright}}
\def\paragraph{\@startsection{paragraph}{4}{\z@}{1.5ex plus
   0.5ex minus .2ex}{-1em}{\normalsize\bfseries}}
\def\subparagraph{\@startsection{subparagraph}{5}{\parindent}{1.5ex plus
   0.5ex minus .2ex}{-1em}{\normalsize\bfseries}}

\skip\footins 9pt plus 4pt minus 2pt
\def\footnoterule{\kern-3pt \hrule width 5pc \kern 2.6pt }
\labelwidth\leftmargini\advance\labelwidth-\labelsep \labelsep 5pt
\belowdisplayskip \abovedisplayskip
\def\@normalsize{\@setsize\normalsize{11pt}\xpt\@xpt}
\def\small{\@setsize\small{10pt}\ixpt\@ixpt}
\def\footnotesize{\@setsize\footnotesize{10pt}\ixpt\@ixpt}
\def\scriptsize{\@setsize\scriptsize{8pt}\viipt\@viipt}
\def\tiny{\@setsize\tiny{7pt}\vipt\@vipt}
\def\large{\@setsize\large{14pt}\xiipt\@xiipt}
\def\Large{\@setsize\Large{16pt}\xivpt\@xivpt}
\def\LARGE{\@setsize\LARGE{20pt}\xviipt\@xviipt}
\def\huge{\@setsize\huge{23pt}\xxpt\@xxpt}
\def\Huge{\@setsize\Huge{28pt}\xxvpt\@xxvpt}

\makeatother

\usepackage{times}
\usepackage{latexsym}
\usepackage[T1]{fontenc}
\usepackage[utf8]{inputenc}
\usepackage{microtype}
\usepackage{inconsolata}
\usepackage{graphicx}

\usepackage{booktabs}
\usepackage{multirow}
\usepackage{amsmath}
\usepackage{amsfonts}
\usepackage{tcolorbox}
\usepackage{listings}
\tcbuselibrary{breakable, skins}
\usepackage{caption}
\usepackage{enumitem}
\usepackage{pifont}
\usepackage{xcolor}
\usepackage[table]{xcolor}
\usepackage{makecell}
\usepackage{tabularx}
\usepackage{array}
\newcolumntype{Y}{>{\centering\arraybackslash}X}
\usepackage{url}

\usepackage{marvosym} 
\newcommand{\corremail}[1]{%
  \textsuperscript{\Letter}%
  \begingroup
    \renewcommand\thefootnote{}%
    \footnote{\Letter~Corresponding author (\href{mailto:#1}{#1}).}%
    \addtocounter{footnote}{-1}%
  \endgroup
}

\title{MeepleLM: A Virtual Playtester Simulating Diverse Subjective Experiences}

\author{
  \textbf{Zizhen Li\textsuperscript{1,2,3}\textsuperscript{$\ddagger$}},
  \textbf{Chuanhao Li\textsuperscript{4}},
  \textbf{Yibin Wang\textsuperscript{2}},
  \textbf{Yukang Feng\textsuperscript{2,3}}, 
  \textbf{Jianwen Sun\textsuperscript{2,3}},\\
  \textbf{Jiaxin Ai\textsuperscript{2}},
  \textbf{Fanrui Zhang\textsuperscript{2}},
  \textbf{Mingzhu Sun\textsuperscript{3}},
  \textbf{Yifei Huang\textsuperscript{1}},
  \textbf{Kaipeng Zhang\textsuperscript{1,2}\corremail{-}} \\
\textsuperscript{1}Shanda AI Research Tokyo,
\textsuperscript{2}Shanghai Innovation Institute,
\textsuperscript{3}NKU,
\textsuperscript{4}Shanghai AI Laboratory\\
\texttt{\{zizhen.li,kaipeng.zhang\}@shanda.com}\\
}

\begin{document}
\maketitle
\begingroup
  \renewcommand\thefootnote{$\ddagger$}%
  \footnotetext{Internship at Shanda AI Research Tokyo.}
  \renewcommand\thefootnote{\Letter}%
  \footnotetext{Corresponding author.}
\endgroup

\begin{abstract}
Recent advancements have expanded the role of Large Language Models in board games from playing agents to creative co-designers.
However, a critical gap remains: current systems lack the capacity to offer constructive critique grounded in the emergent user experience.
Bridging this gap is fundamental for harmonizing \textit{Human-AI collaboration}, as it empowers designers to refine their creations via external perspectives while steering models away from biased or unpredictable outcomes.
Automating critique for board games presents two challenges: inferring the \textit{latent dynamics} connecting rules to gameplay without an explicit engine, and modeling the \textit{subjective heterogeneity} of diverse player groups.
To address these, we curate a dataset of 1,727 structurally corrected rulebooks and 150K reviews selected via quality scoring and facet-aware sampling. We augment this data with \textit{Mechanics-Dynamics-Aesthetics (MDA)} reasoning to explicitly bridge the causal gap between written rules and player experience.
We further distill player personas and introduce \textbf{MeepleLM}, a specialized model that internalizes persona-specific reasoning patterns to accurately simulate the subjective feedback of diverse player archetypes.
Experiments demonstrate that MeepleLM significantly outperforms latest commercial models (e.g., GPT-5.1, Gemini3-Pro) in community alignment and critique quality, achieving a 70\% preference rate in user studies assessing utility. MeepleLM serves as a reliable \textit{virtual playtester} that
provides experience-grounded feedback, offering a practical step towards audience-aligned
Human-AI collaboration\footnote{Dataset and code are available at \href{https://github.com/leroy9472/MeepleLM}{\nolinkurl{https://github.com/leroy9472/MeepleLM}}.}.
\end{abstract}

\section{Introduction}

\begin{figure}[t]
    \centering
    \includegraphics[width=\linewidth]{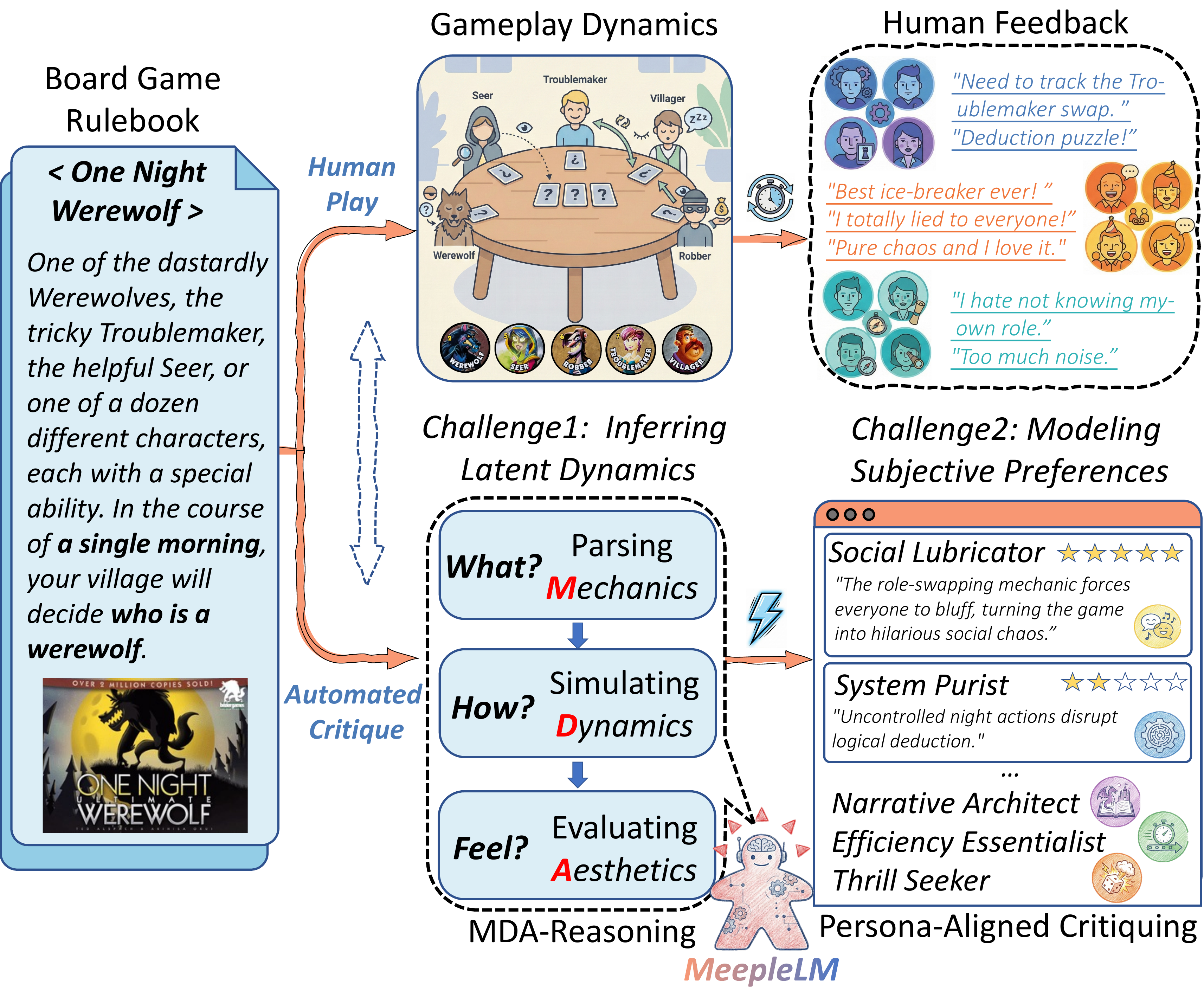} 
    \caption{\textbf{Overview of MeepleLM.} Acting as a \textit{Virtual Playtester}, the model offers a rapid, automated alternative to the resource-intensive Human Play loop. By leveraging MDA-Reasoning to infer latent dynamics from Static Rulebooks, MeepleLM generates Persona-Aligned Critiques tailored to diverse player archetypes. }
    \label{fig:framework}
\end{figure}

Board games have long served as a universal medium of significant cultural and economic value, captivating a vast global audience~\cite{rodriguezopportunities}. Recently, the rapid development of Large Language Models (LLMs) has introduced unprecedented possibilities to this classic domain. Specifically, board games serve as a prominent platform for evaluating diverse model facets, ranging from reasoning~\cite{lin2025gamebot} and decision-making~\cite{tang2025dsgbench} to role-playing~\cite{yu2025rpgbench} and social simulation~\cite{hansteen2025exploring}.
Beyond serving as a testbed, recent research highlights board game development as a pivotal domain for investigating \textit{Human-AI Collaboration}, where LLMs serve as active co-designers to perform tasks such as generating mechanics~\cite{patrick2025gamegenesis}, facilitating iterative prototyping~\cite{ma2025follow} and synthesizing executable engines~\cite{hong2025game, lehrach2025code}.

However, while automated development has advanced, a critical gap remains: current systems lack the capacity to offer constructive critique grounded in the emergent user experience.
Such feedback is vital for harmonizing the roles of both human and model in the co-creation loop. 
On one hand, designers require external perspectives to refine their creations and better comprehend their audience~\cite{fang2025generative}, a process that ultimately catalyzes further creativity~\cite{choi2025proxona}. 
On the other hand, for LLM-driven systems, the absence of effective user feedback can lead to biased content~\cite{taveekitworachai2024assessing} or unpredictable, frustrating experiences~\cite{yong2023playing}. Consequently, bridging this gap is fundamental to evolving LLMs into empathetic partners for \textit{Human-AI Collaboration}. By prioritizing diverse user experiences, this approach ensures that future co-creation is driven not merely by technical validity, but by a dynamic alignment with individual needs.

To this end, we need an evaluation paradigm that can map “design intent” (or “system specifications”) to “user experience.” 
However, board game experiences are characteristically emergent and subjective: they are not static properties of rulebooks, but are jointly generated through interaction as mechanics unfold, players MDA, and emotional responses arise~\cite{forlizzi2004understanding}. 
This inherent characteristic poses two core challenges for automated evaluation:
(1) \textbf{Inferring Latent Dynamics from Static Rules.} While rulebooks serve as explicit ``code'', gameplay experience is an emergent property generated only when mechanics interact at runtime. The core challenge is to bridge the gap between written specifications and dynamic interactions. Since LLMs lack an explicit game engine, they must infer plausible execution trajectories from rules and use empirical player feedback as an external signal to recover latent causal links that connect mechanics to outcomes and reactions.
(2) \textbf{Modeling Subjective Group Preferences.} Experience is not universal; the same mechanism can elicit conflicting reactions across different player demographics (e.g., high randomness may delight a Socializer but frustrate a Strategist). If critiques collapse into an average ``one-size-fits-all'' judgment, they become generic and less actionable for design or recommendation. The challenge, therefore, is to model this subjective heterogeneity by aligning reasoning with specific group preferences, simulating distinct personas rather than a single ``standard'' user.

To address these challenges, we meticulously curate a large-scale dataset of structurally corrected rulebooks from selected board games, paired with reviews filtered through rigorous scoring and quality assessment.
We further augment this data by incorporating the classic game design theory of Mechanics-Dynamics-Aesthetics (MDA)~\cite{hunicke2004mda} into Chain-of-Thought (CoT) reasoning, thereby making the latent execution logic explicit.
To structure the inherent subjectivity of feedback, we distill distinct player personas through an expert–LLM collaborative interpretation of data-driven community clusters.
Building upon this foundation, we introduce MeepleLM(Figure~\ref{fig:framework}), a specialized model designed to predict gameplay experiences by simulating the perspectives of real-world players.
Extensive experiments validate that MeepleLM significantly outperforms state-of-the-art baselines in capturing authentic user experiences.

Our contributions are summarized as follows:
\begin{itemize}
    \item We present the first systematic study on the \textbf{automated evaluation of board games}. We bridge the gap between static rules and distinct player experiences by simulating the latent gameplay dynamics.
    \item We curate a high-quality dataset of \textbf{1,727 rulebooks} and \textbf{150K critiques}, selected via rigorous filtering and quality evaluation. We further leverage the MDA framework to synthesize explicit CoT paths that recover the latent dynamics connecting rules to experiences.
    \item We distill five data-driven player personas and introduce \textbf{MeepleLM}. By internalizing persona-specific reasoning, our model predicts authentic gameplay experiences that reflect the diverse preferences of real-world communities.
    \item We conduct a systematic evaluation on a stratified set of \textbf{207 games}. Experiments across macro-level alignment, micro-level fidelity, and practical utility demonstrate that MeepleLM significantly outperforms state-of-the-art LLMs as a reliable virtual playtester.
\end{itemize}

Ultimately, by bridging static rules and dynamic experiences,
MeepleLM demonstrates a \textbf{virtual playtesting paradigm}
that accelerates design iteration via anticipated audience feedback
and facilitates personalized game selection for players.
While instantiated on board games, the underlying reasoning
framework, inferring latent dynamics from written designs and
modeling subjective heterogeneity across user groups, addresses
a general class of interactive systems.
As a feedback component within co-creation workflows, MeepleLM
offloads the cognitive burden of perspective-switching between
creator and audience, enabling \textbf{experience-aware Human-AI
collaboration} where models serve as empathetic critics attuned
to \textbf{subjective audience sensibilities}.

\section{Related Work}
\label{sec:related}

\noindent \textbf{LLM-Driven Feedback and Assistance.}
Recent advancements have empowered LLMs to surpass traditional metrics in evaluating open-ended text, demonstrating high alignment with human judgments~\cite{li2024leveraging,gao2025llm,chen2023exploring}.
Current systems provide constructive feedback ranging from granular writing issues~\cite{russell2025people} to comprehensive peer reviews~\cite{benharrak2024writer,rashkin2025help}, with established metrics for narrative consistency~\cite{rashkin2025help}, structural integrity~\cite{zheng2025cml}, and subjective enjoyment~\cite{yang2025matters}.
However, these approaches treat text as \textit{static narratives}, failing to address the \textit{executable logic} inherent in interactive systems like board games.
Even when extended to design assistance, such as generating levels~\cite{todd2023level} or rule codes~\cite{todd2024gavel,tanaka2024grammar}, LLMs often prioritize syntactic correctness over logical coherence.
As noted in recent studies, this frequently yields functional yet meaningless mechanics, known as \textit{introns}~\cite{todd2024gavel,hu2024game}.
Existing tools fragment into abstract brainstorming~\cite{li2025cardiverse} or rigid asset production~\cite{lindfors2025leveraging} without assessing playability.
An alternative line of work employs forward-model-based playtesting,
where executable game engines drive procedural agents,
such as Monte Carlo Tree
Search~\cite{de2017ai,holmgaard2018automated,gaina2020tag,goodman2025dice}
and RL players~\cite{ferdous2025curiosity,cilan2025learning},
to simulate gameplay and detect balance issues or rule gaps.
Recent extensions further combine multi-agent RL with curiosity-driven
exploration for broader coverage~\cite{ferdous2025curiosity,cilan2025learning},
or couple LLM-based self-play with game engines for automated
balancing~\cite{zeng2026rulesmith}.
While such approaches can uncover emergent game-state anomalies,
they require per-game forward models and primarily optimize strategic
performance rather than capturing \textit{subjective player experience}.
Our work takes a complementary, \textit{experience-centered} approach, bridging this gap by simulating \textit{dynamic interactions} to predict the \textit{emergent experience} directly from rules without requiring executable engines.

\noindent \textbf{User Simulation and Persona Modeling.}
Understanding audience heterogeneity is crucial for creators, yet manual analysis of diverse feedback is cognitively demanding~\cite{choi2023creator,ma2023multi}.
The field has evolved from survey-based ``imagined users''~\cite{cooper1999inmates,salminen2018personas} to algorithmic clustering~\cite{mcginn2008data,salminen2020literature} and, recently, LLM-based perspective simulation~\cite{benharrak2024writer,park2023generative}.
However, purely synthetic simulations often lack ecological validity, risking hallucinations or stereotyping due to foundation model biases~\cite{cheng2023compost}.
To mitigate this, state-of-the-art systems emphasize grounding simulations in empirical data for \textit{representativeness}~\cite{shin2024understanding,choi2025proxona}.
Aligning with this paradigm, we ground personas in large-scale gameplay critiques rather than conversational history. This allows our model to internalize distinct preferences, facilitating \textit{Human-AI collaboration} through empathetic, persona-aligned feedback rather than generic judgments.

\section{Data Construction}
\label{sec:data}

\begin{figure*}[t]
    \centering
    \includegraphics[width=\textwidth]{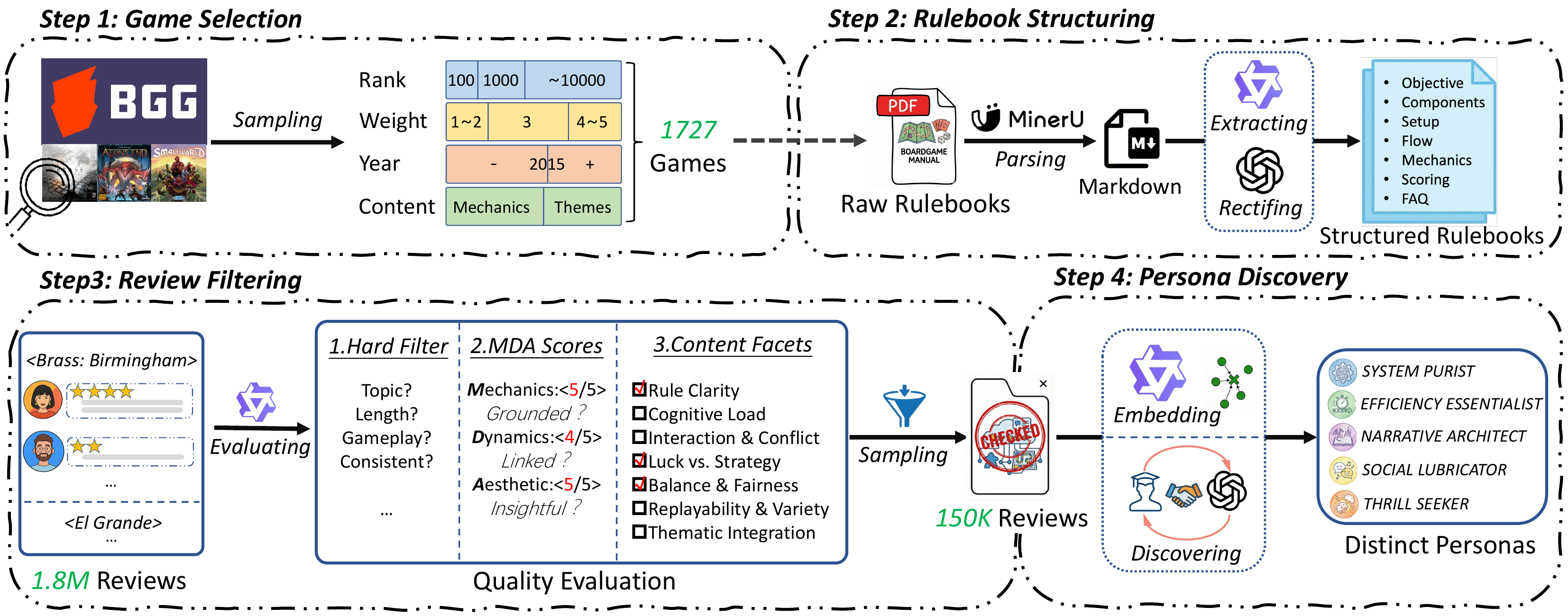} 
    \caption{\textbf{The Data Construction Pipeline.} We structure 1,727 board game rulebooks and filter 1.8M reviews via multi-dimensional quality assessment, yielding 150K high-quality critiques for persona discovery. }
    \label{fig:data_pipeline}
\end{figure*}

We curated a multi-layered dataset that maps objective game rulebooks to subjective player feedback across diverse personas. The overall construction pipeline is illustrated in Figure~\ref{fig:data_pipeline}.

\subsection{Game Selection}
\label{sec:data_collection}

We curated a collection of \textbf{1,727 board games} via a stratified sampling strategy on BoardGameGeek (BGG)\footnote{\url{https://boardgamegeek.com/browse/boardgame}}. As detailed in Appendix~\ref{app:dataset_stats}, our selection prioritizes four dimensions to ensure a comprehensive representation of the design landscape:
(1) \textbf{Market Stratification:} To mitigate survivorship bias, we balanced the selection between ``elite'' titles (including 83 from the Top 100) and ``long-tail'' designs (comprising 53\% of the dataset with Rank $>1,000$), capturing the full spectrum of market reception.
(2) \textbf{Cognitive Spectrum:} We covered the entire range of BGG Weight (1.0--5.0), encompassing everything from low-burden social party games to calculation-intensive strategy simulations.
(3) \textbf{Temporal Span:} The dataset balances historical depth with modern relevance, containing 47\% classic titles released pre-2015 alongside 35 cutting-edge designs from 2024 and beyond.
(4) \textbf{Mechanical Heterogeneity:} To ensure structural diversity, the collection spans 192 unique mechanics and 81 themes, covering logic distinct from standard genres.

\subsection{Rulebook Structuring}
\label{sec:data_rules}

We processed the official rulebooks into a structured knowledge base via a three-step pipeline. First, we parsed raw PDFs into hierarchical Markdown using Mineru~\citep{niu2025mineru25decoupledvisionlanguagemodel} to preserve layout information. Second, we prompted Qwen3-235B\citep{qwen3}  to restructure the raw text into a standardized hierarchical format (e.g., unifying headers for \textit{Objective}, \textit{Components}, and \textit{Flow}); the specific extraction prompt is detailed in Appendix~\ref{app:B_rulebook_prompt}, and a complete structured example is provided in Appendix~\ref{app:B_example}. Finally, to ensure maximum fidelity, we employed GPT-5.1 to cross-reference and rectify the initial drafts against the source text, correcting logical inconsistencies or formatting errors (see the rectification prompt in Appendix~\ref{app:B_rectify_prompt}).

\begin{figure}[t]
    \centering
    \includegraphics[width=\columnwidth]{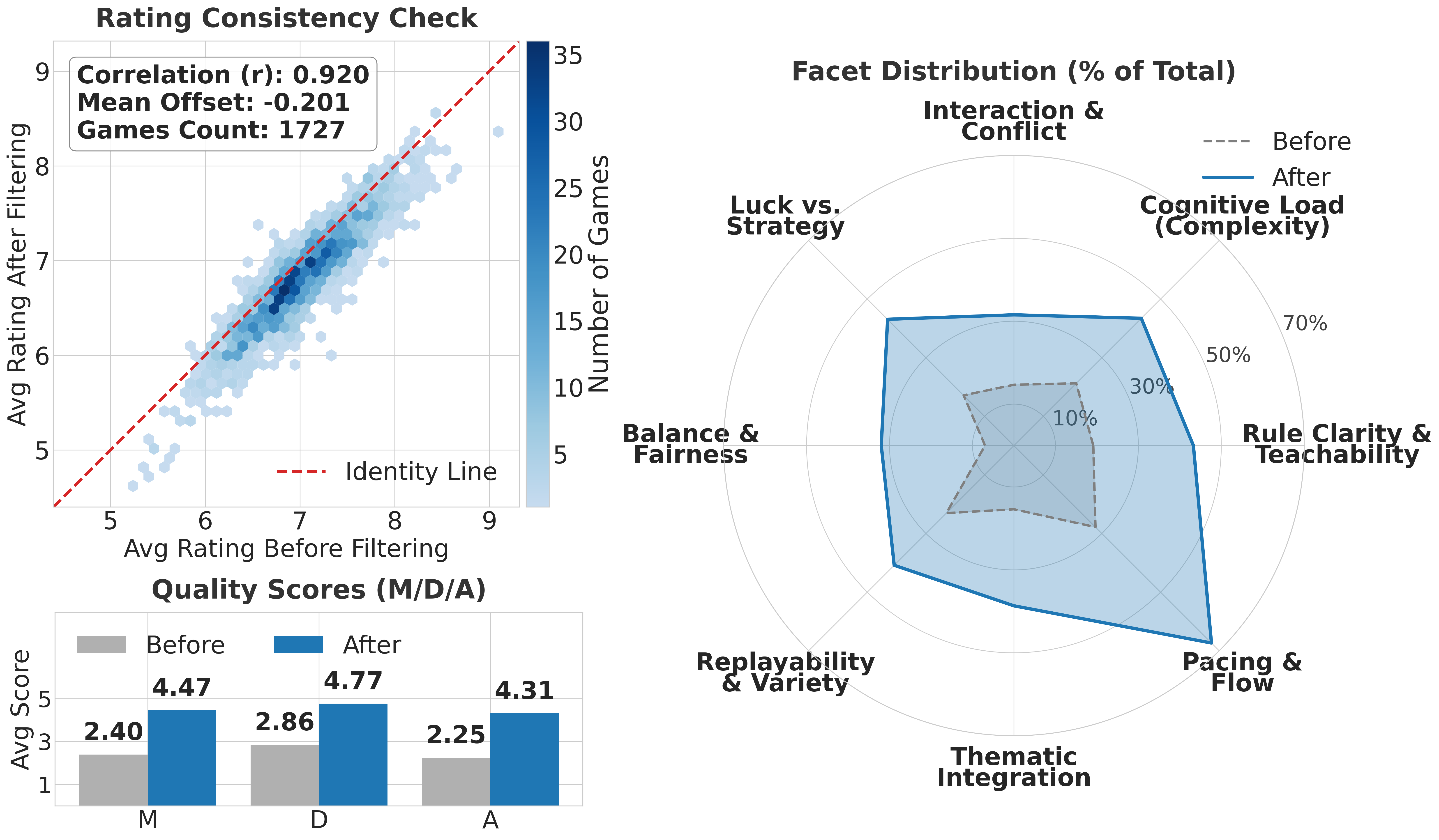} 
    \caption{\textbf{Impact of the Filtering Strategy.} Our strategy enhances MDA scores and semantic coverage while preserving the original rating distribution.}
    \label{fig:filter_stats}
\end{figure}

\subsection{Review Filtering }
\label{sec:data_reviews}
We aggregated a raw corpus of 1.8 million rating-comment pairs from multiple online communities (detailed in Appendix~\ref{app:review_source}). To refine this data, we employed Qwen3-235B with a multi-task prompt (detailed in Appendix~\ref{app:C_review_prompt}) to perform a comprehensive evaluation across three dimensions: (1) \textbf{Hard Filtering}, removing noise such as short texts, off-topic logistics, and rating inconsistencies; (2) \textbf{MDA Scoring}, evaluating whether specific Mechanics are linked to Dynamic interactions to derive Aesthetic experiences, and assigning quality scores across these three dimensions; and (3) \textbf{Facet Identification}, where the model mapped the content to predefined semantic topics (e.g., \textit{Rule Clarity}, \textit{Balance \& Fairness}) to capture diverse viewpoints.

Based on these metrics, we implemented a stratified coverage-maximization strategy. 
We first performed stratified sampling based on original ratings to preserve sentiment fidelity (Pearson's $r=0.920$, verified in Figure~\ref{fig:filter_stats}, Top-Left). 
Simultaneously, we filtered for high-quality entries (scores $>4$) to significantly enhance MDA scores (Figure~\ref{fig:filter_stats}, Bottom-Left) and optimized selection to maximize semantic coverage across all facets (Figure~\ref{fig:filter_stats}, Right). 
This process yielded a final dataset of $\sim$150k entries (approx. 8\% retention), ensuring a robust volume of 50--100 reviews per game; further  details are provided in Appendix~\ref{app:C_review_statistical}.

\subsection{ Persona Discovery}
\label{sec:data_persona}

A single ``average rating'' fails to capture the subjective diversity, where the rigorous complexity prized by strategists is perceived as an exhausting burden by others. To model these domain-specific cognitive attributions, we implemented a \textit{Cluster-then-Refine} discovery pipeline, evolving raw behavioral clusters into interpretable player personas.

\paragraph{Discovery Pipeline.}
We first generated composite embeddings for all reviews using Qwen3-Embedding-8B~\cite{zhang2025qwen3} (concatenating text with logic scores and facets; see Appendix~\ref{app:persona_cluster}). Following K-Means clustering ($K=15$), we employed a human-in-the-loop process where GPT-5.1 profiled representative samples, and domain experts refined these into a finalized taxonomy of five distinct personas (detailed in Appendix~\ref{app:persona_desc}). Using these finalized definitions, we employed GPT-5.1 to annotate the entire dataset. To ensure classification stability, we implemented a majority-vote mechanism (aggregating 3 independent inferences per review) to assign the dominant persona label. The annotation prompts are detailed in Appendix~\ref{app:persona_prompts}.

\paragraph{Aesthetic Segregation.}
Table~\ref{tab:case_study} exemplifies the distinct preference patterns captured by our taxonomy. The data reveals that party and adventure elements actively alienate \textit{System Purists}, whereas heavy campaign games frustrate \textit{Efficiency Essentialists} despite their thematic appeal. (See Appendix~\ref{app:persona_matrix} for the full Preference Matrix).

\begin{table}[t]
\centering
\small
\resizebox{\columnwidth}{!}{
\begin{tabular}{l|c|ll}
\toprule
\textbf{Game} & \textbf{Diff.} & \textbf{Highest Persona (Rating)} & \textbf{Lowest Persona (Rating)} \\
\midrule
\textit{<Unspeakable Words>} & \multirow{2}{*}{4.04} & \textbf{Social Lubricator} (6.9) & \textbf{System Purist} (2.9) \\
(Party / Word) & & \textit{``Party chaos!''} & \textit{``Random noise.''} \\
\midrule
\textit{<Talisman>} & \multirow{2}{*}{3.71} & \textbf{Narrative Architect} (7.1) & \textbf{System Purist} (3.4) \\
(Adventure / Roll) & & \textit{``An epic journey.''} & \textit{``Roll-and-move hell.''} \\
\midrule
\textit{<Aeon Trespass>} & \multirow{2}{*}{2.97} & \textbf{Narrative Architect} (9.0) & \textbf{Efficiency Essentialist} (6.0) \\
(Campaign / Heavy) & & \textit{``Immersive masterpiece.''} & \textit{``Feels like a job.''} \\
\bottomrule
\end{tabular}
}
\caption{\textbf{Case Study of Aesthetic Segregation.} The rating gaps between persona groups validate that our taxonomy effectively captures distinct player preferences.}
\label{tab:case_study}
\end{table}

\paragraph{Why LLM Annotation?}
We attempted to train a supervised classifier (DeBERTa-v3-large~\cite{he2021debertav3}) but it proved insufficient for capturing subtle preference patterns. As analyzed in Appendix~\ref{app:D_semantic_ambiguity}, it naively misclassifies a review mentioning ``house rules'' and ``balance'' as \textit{System Purist}, failing to discern that the user is actually introducing variants to inject the high-risk, high-reward volatility typical of a \textit{Thrill Seeker}.

\section{Methodology}
\label{sec:method}

\subsection{Problem Formulation}
\label{sec:problem_formulation}

We formulate the task as a conditional generation problem. Given a rulebook context $\mathcal{R}$ and a target player persona $\mathcal{P}$, the objective is to generate a feedback entry $\mathcal{Y}$ (comprising a numerical rating and a textual review).
Crucially, direct mapping $\mathcal{R} \to \mathcal{Y}$ ignores the \textit{semantic gap} between static text and emergent fun. 
To bridge the gap between objective rules and subjective preference, we employ an \textit{MDA-Guided Reasoning} strategy. 
Drawing upon the foundational game design framework~\cite{hunicke2004mda}, we reinterpret the MDA model originally designed to analyze gameplay loops as a causal inference chain for language models.
We define a latent intermediate sequence $\mathcal{Z}$ that explicitly traces the causal path from \textit{Mechanics} to \textit{Dynamics}, and finally to \textit{Aesthetics}. By decomposing the generation into $[\mathcal{R}, \mathcal{P}] \xrightarrow{\mathcal{Z}_{\text{MDA}}} \mathcal{Y}$, we force the model to simulate the runtime experience before articulating the critique, ensuring the output is logically grounded in the rules.

\subsection{Synthesizing the MDA Cognitive Chain}
\label{sec:method_cot}

Since the reasoning chain $\mathcal{Z}$ is latent in raw reviews, we utilize a distillation approach to recover this logic. We employed Qwen3-235B as a Teacher Model to reconstruct $\mathcal{Z}$ from high-quality review-rule pairs. The prompt (Appendix~\ref{app:cot_prompt}) enforces a three-step cognitive flow:
(1) \textbf{Step 1: Mechanics (``The What'').} \textit{``What specific content does the review explicitly mention?''} Isolate objective rule components from $\mathcal{R}$ explicitly mentioned in the review to ensure grounding.
(2) \textbf{Step 2: Dynamics (``The How'').} \textit{``What Interaction or System Dynamic occurred during play?''} Infer the runtime system behaviors or interactions triggered by the mechanics identified in Step 1.
(3) \textbf{Step 3: Aesthetics (``The Feel'').} \textit{``What was the final Aesthetic Experience or emotional feeling?''} Synthesize the subjective emotional outcome, modulated by the preferences of Persona $\mathcal{P}$.

\paragraph{Verifier-Guided Filtration Loop.}
We employ GPT-5.1 to judge the \textit{entailment} between synthesized reasoning and ground-truth ratings. Guided by Appendix~\ref{app:prompt_verifier}, the verifier removes chains with sentiment contradictions or hallucinations, triggering automatic regeneration to ensure logical consistency. A sample alignment of rules, personas, and critiques appears in Appendix~\ref{app:data_example}. To further validate the reliability of this automated pipeline, we
conducted a post-hoc human audit in which 3 experienced players
verified 200 MDA chains spanning 10 familiar games; all 200 chains
passed under the same criteria used by GPT-5.1 (details in
Appendix~\ref{app:human_audit}).

\subsection{Persona-Conditional Instruction Tuning}
\label{sec:method_sft}
We fine-tuned the Qwen3-8B~\cite{qwen3} backbone to maximize the joint likelihood of the MDA reasoning chain $\mathcal{Z}$ and the final critique $\mathcal{Y}$. 
To address the challenge of subjective heterogeneity, we do not represent $\mathcal{P}$ as a simple label. Instead, we encode the \textbf{full semantic profile} derived in Section~\ref{sec:data_persona} (including core values and interaction preferences) into the system instruction. This forces the model to treat $\mathcal{P}$ as a contextual prior that modulates the \textit{Dynamics $\to$ Aesthetics} transition.
Formally, we treat the concatenation $\mathcal{S} = [\mathcal{Z}; \mathcal{Y}]$ as the target sequence and minimize the standard Cross-Entropy Loss:
\begin{equation}
    \mathcal{L} = - \sum_{t=1}^{|\mathcal{S}|} \log P(s_t \mid s_{<t}, \mathcal{R}, \mathcal{P}_{profile}).
\end{equation}
The training was implemented using LoRA~\citep{hu2022lora} on all linear layers via LLaMA-Factory~\citep{zheng2024llamafactory}. Hyperparameter details are provided in Appendix~\ref{app:hyperparams}.

\section{Experiments and Analysis}
\label{sec:experiments}

\begin{table*}[t]
    \centering
    \small
    
    \begin{tabularx}{\textwidth}{l | Y Y Y | Y Y Y | c}
        \toprule
        \multirow{2}{*}{\textbf{Model}} & 
        \multicolumn{3}{c|}{\textbf{Preference Alignment} (RQ1)} & 
        \multicolumn{3}{c|}{\textbf{Review Quality} (RQ2)} & 
        \textbf{Utility} (RQ3) \\
        
        \cmidrule(lr){2-4} \cmidrule(lr){5-7} \cmidrule(lr){8-8}
        
        & \textbf{MAE} $\downarrow$ & \textbf{WD} $\downarrow$ & \textbf{$\tau$} $\uparrow$ 
        & \textbf{Fact.} $\uparrow$ & \textbf{Dist-2} $\uparrow$ & \textbf{Div.} $\uparrow$ 
        & \textbf{Op-Rec} $\uparrow$ \\
        
        \midrule
        GPT-5.1 & 0.9874 & 0.9496 & 0.2555 & \textbf{99.46} & 0.6934 & 4.26 & 63.44 \\
        Gemini3-Pro & 1.4277 & 0.5092 & 0.2465 & 98.28 & 0.6480 & 3.98 & 57.74 \\
        Qwen3-235B & 1.2288 & 0.6350 & 0.1449 & 98.95 & 0.6572 & 3.56 & 54.27 \\
        Qwen3-8B & 0.8906 & 1.0119 & 0.0492 & 97.88 & 0.5936 & 1.58 & 11.39 \\
        
        \midrule
        \textbf{MeepleLM(Ours)} & \textbf{0.6576} & \textbf{0.2205} & \textbf{0.2817} & 98.86 & \textbf{0.7117} & \textbf{4.34} & \textbf{69.77} \\
        
        \midrule
        
        \hspace{3mm} \textit{w/o MDA} & 0.7395 & 0.4148 & 0.2271 & 91.56 & 0.6850 & 3.70 & 55.35 \\
        
        \hspace{3mm} \textit{w/o Persona} & 0.7887 & 0.3630 & 0.1348 & 92.13 & 0.6771 & 3.56 & 53.84 \\
        
        \hspace{3mm} \textit{w/o Rulebook} & 0.7043 & 0.5496 & 0.0026 & 59.87 & 0.6970 & 3.30 & 9.99 \\
        \bottomrule
    \end{tabularx}
    \caption{\textbf{Overall performance.} MeepleLM shows superior performance in community alignment, generation quality, and practical utility, validating the effectiveness of persona-aligned simulation for virtual playtesting.}
    \label{tab:main_results}
\end{table*}

To systematically validate MeepleLM as a reliable virtual playtester, we structure our evaluation around three core research questions: 
(1) \textbf{RQ1 (Macro-level Alignment)} assesses whether the simulator accurately replicates community rating distributions and preference rankings. 
(2) \textbf{RQ2 (Micro-level Fidelity)} examines if the generated reviews maintain factual consistency with rules while exhibiting the content richness and \textbf{semantic diversity} characteristic of real players. 
(3) \textbf{RQ3 (Practical Utility)} investigates whether the simulated feedback provides actionable insights for design optimization and player decision support.

\subsection{Experimental Setup}
\label{sec:setup}

\noindent\textbf{Dataset Splitting.}
We constructed a comprehensive test set of 207 games disjoint from the training corpus. To ensure representative coverage of the design landscape, we employed a stratified sampling strategy based on \textit{BGG Weight} (Complexity 1.0--5.0) and \textit{Average Rating} (Tier 1--5).
Notably, this selection spans a wide temporal range, explicitly including 34 newly released titles (2024--2025) alongside historical classics, enabling us to assess performance on both established consensus and fresh content.

\noindent\textbf{Simulation Protocols.}
For each game, we execute $N=100$ simulation runs. In each run, the model takes the rulebook $\mathcal{R}$ and a specific persona $\mathcal{P}$ to generate the critique $\mathcal{Y}$. Crucially, we do not pick $\mathcal{P}$ randomly; instead, we sample personas to match the empirical proportions found in the ground-truth reviews. The specific inference prompt is provided in Appendix~\ref{app:sim_prompt}.

\noindent\textbf{Baselines.}
We benchmark against three state-of-the-art general LLMs (GPT-5.1, Gemini3-Pro, Qwen3-235B) and our backbone model Qwen3-8B.
Detailed implementation configurations are provided in Appendix~\ref{app:implementation}.

\subsection{RQ1: Macro-level Community Alignment}
\label{sec:rq1_results}

\noindent\textbf{Evaluation Metrics.}
To assess whether MeepleLM aligns with the collective wisdom of the community, we employ three complementary metrics:
(1) \textbf{Mean Absolute Error (MAE)} measures the absolute precision of rating predictions;
(2) \textbf{Wasserstein Distance (WD)} evaluates the fidelity of the predicted score distribution against the ground truth~\cite{villani2008optimal};
(3) \textbf{Kendall's Rank Correlation ($\tau$)} assesses the model's ability to correctly rank games based on their perceived quality~\cite{kendall1938new}.

\begin{figure}[t] 
    \centering
    \includegraphics[width=\linewidth]{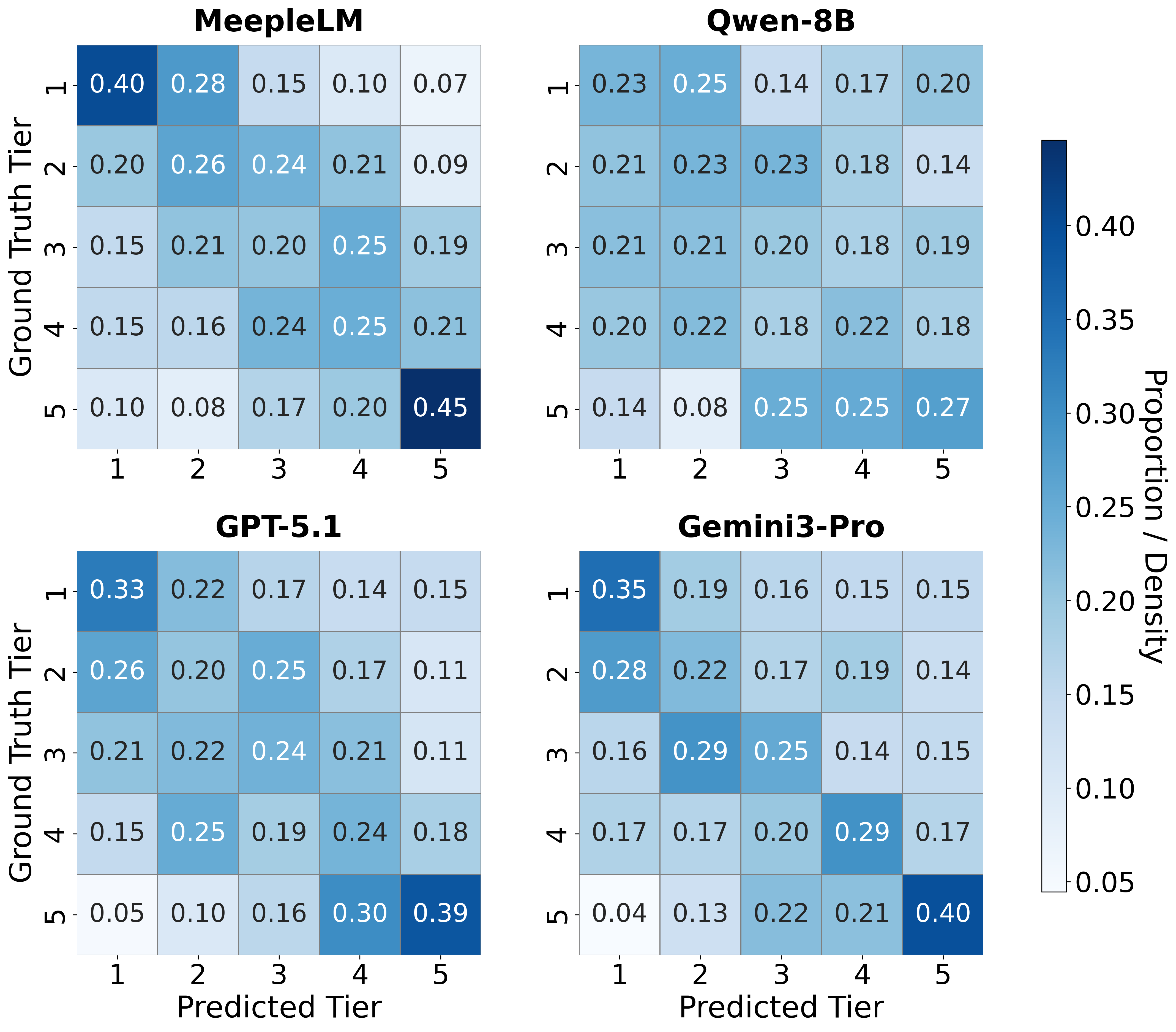}
    \caption{\textbf{Tier-wise Prediction Alignment.} MeepleLM shows a sharp diagonal concentration, effectively distinguishing quality tiers.}
    \label{fig:heatmap_comparison}
\end{figure}

\noindent\textbf{Beyond Ranking: Capturing Community Diversity.}
As summarized in Table~\ref{tab:main_results}, MeepleLM consistently achieves the best performance across all alignment metrics. 
This superiority is visually corroborated in Figure~\ref{fig:heatmap_comparison}, where our model demonstrates a sharp diagonal concentration, effectively distinguishing high-quality outliers (Tier 1) from poor designs (Tier 5).
Critically, while advanced baselines like GPT-5.1 retain some capacity for ranking ($\tau=0.2555$), they exhibit severe \textit{central tendency bias}, ``playing it safe'' by clustering predictions around the mean to minimize error. 
This failure to simulate authentic polarization is quantified by their high Wasserstein Distance (0.9496) compared to MeepleLM (0.2205), proving that our method captures the variance of community sentiment. 
In contrast, the untuned Qwen3-8B fails to establish any meaningful correlation ($\tau \approx 0$), degenerating into effectively random guessing.

\begin{figure}[t]
    \centering
    \includegraphics[width=\linewidth]{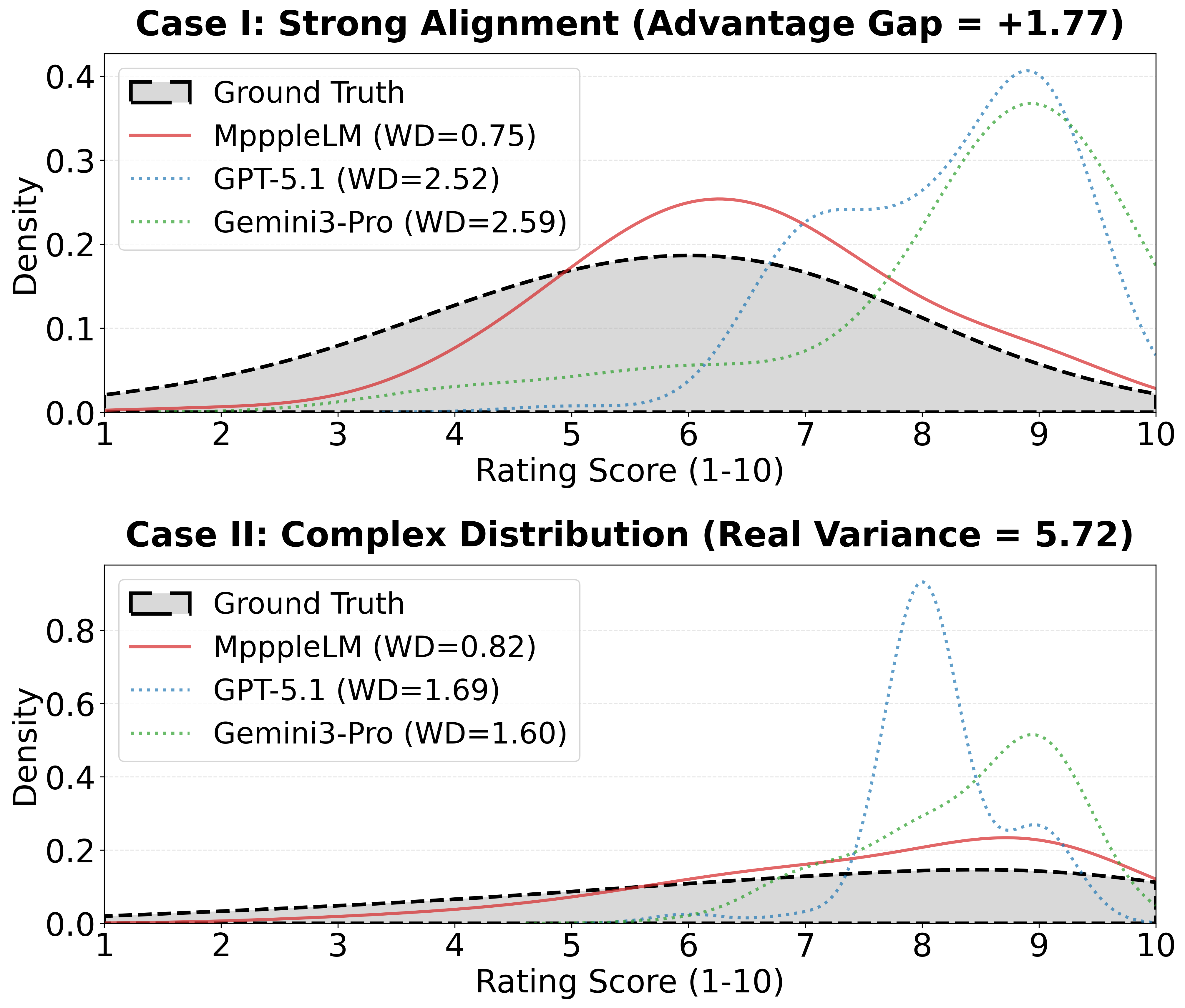} 
    \caption{\textbf{Rating Density Distributions.} MeepleLM demonstrates superior distributional fidelity by accurately recovering the high variance of human consensus.}
    \label{fig:case_study_dist}
\end{figure}

\noindent\textbf{Overcoming Positivity Bias.}
We further investigate this distributional shift in Figure~\ref{fig:case_study_dist}. In scenarios with polarized opinions (Case II), baselines exhibit \textit{mode collapse}, aggressively clustering predictions around safe, high scores (7--9) and failing to capture the long-tail of negative feedback. MeepleLM, empowered by domain-specific rule understanding, successfully recovers the high variance of human consensus ($WD=0.82$), proving its ability to represent the diverse spectrum of player sentiments rather than just a generic average.

\begin{table*}[ht]
    \centering
    \definecolor{hiRed}{RGB}{180, 0, 0}
    \definecolor{hiGreen}{RGB}{0, 100, 80}
    \definecolor{hiOrange}{RGB}{160, 80, 0}

    \footnotesize\rmfamily\itshape
    \renewcommand{\arraystretch}{1.4} 
    \setlength{\tabcolsep}{6pt}

    \resizebox{\textwidth}{!}{
    \begin{tabular}{p{0.32\textwidth} | p{0.32\textwidth} | p{0.32\textwidth}}
        \toprule
        \multicolumn{1}{c|}{\textbf{\textup{Qwen3-8B }}} & 
        \multicolumn{1}{c|}{\textbf{\textcolor{hiGreen}{\textup{MeepleLM }}}} & 
        \multicolumn{1}{c|}{\textbf{\textup{GPT-5.1 }}} \\
        \midrule
        
        \multicolumn{3}{c}{\cellcolor{gray!10}\textbf{\textup{Persona: The Social Lubricator}} \textit{(Goal: Party Atmosphere, Interaction)}} \\
        
        ``...The table became a \textcolor{hiRed}{theater of accusations and fake tears}... The mechanic was pure chaos... It’s not the most elegant design, but it thrives on the energy of the group—it’s a party game that rewards chaos over calculation.'' 
        
        \vspace{4pt} 
        
        ``...The night phase became a \textcolor{hiRed}{hilarious circus} of whispered accusations... My friend’s ‘insomniac’ kept checking his card... It’s a perfect recipe for laughs, tension, and bonding...''
        & 
        ``...You really need at least \textcolor{hiGreen}{6 people} though... With 4 players, the \textcolor{hiGreen}{Mason role} is useless. If you have \textcolor{hiGreen}{alpha gamers}, they will dominate the conversation... Overall, I recommend this for a quick and easy game.'' 
        
        \vspace{4pt}
        
        ``...Played a ton with friends... favorite game at parties. Simple enough to \textcolor{hiGreen}{teach to a new player}... great for people who don't know a lot about board games. The bluffing is fun...''
        & 
        ``...As someone who treats board games as \textcolor{hiOrange}{social WD-40}, this is perfect. The single-night structure keeps things punchy... A fast, noisy trust-fall in social form.'' 
        
        \vspace{4pt}
        
        ``...Basically a structured excuse for people to point fingers... \textcolor{hiOrange}{That’s social gold}. What I love is how shifting roles (Robber, Troublemaker) give everyone a reason to speak.'' \\
        
        \midrule
        \multicolumn{3}{c}{\cellcolor{gray!10}\textbf{\textup{Persona: The System Purist}} \textit{(Goal: Mechanics, Balance, Strategy)}} \\
        
        ``...One Night Ultimate Werewolf is a \textcolor{hiRed}{sharp knife with a blunt handle}... The inherent randomness felt like a \textcolor{hiRed}{betrayal of the game's potential}... A near-perfect match for a purist, but the luck left a bittersweet aftertaste.''
        & 
        ``...I like that there is \textcolor{hiGreen}{no dice luck}. However, the Robber and Drunk can screw up the game... I recommend adding \textcolor{hiGreen}{variant rules}: \textit{Drunk swaps with a Villager instead of center card}.''
        & 
        ``... When everyone reports honestly-then-bluffs deliberately, the \textcolor{hiOrange}{deduction feels crisp}. But this design lives or dies on human precision... I respect the underlying clockwork.'' \\
        
        \bottomrule
    \end{tabular}
    }
    \caption{\textbf{Qualitative Case Study: Generated Reviews for <One Night Ultimate Werewolf>.} MeepleLM generates factually grounded critiques that align with specific persona sensibilities. By capturing both technical nuances and community-specific slang, our model demonstrates the semantic richness and perspective diversity.
}
    \label{tab:qualitative_case}
\end{table*}

\subsection{RQ2: Content Fidelity and Diversity}
\label{subsec:rq2}

\noindent\textbf{Evaluation Metrics.}
To ensure the generated reviews are both trustworthy and rich in content, we employ three metrics covering accuracy, vocabulary, and semantic variety:

(1) \textbf{Factual Correctness (Fact.):} 
We implement an automated Fact-Checker using Gemini3-Flash. As detailed in Appendix~\ref{app:prompt_fact}, the judge extracts specific claims regarding game components or mechanics from the review and strictly verifies whether they exist in the official rulebook $\mathcal{R}$. This directly measures the model's groundedness.

(2) \textbf{Lexical Diversity (Dist-2):} 
We use the Distinct-2 score~\cite{li2016diversity} to measure vocabulary richness, calculating the ratio of unique bigrams to total bigrams to detect repetitive phrasing.

(3) \textbf{Perspective Diversity (Div.):} 
A realistic simulator should not be a ``broken record'' that repeats the same opinion endlessly. To detect \textit{semantic repetition}, we feed batches of $k=5$ reviews (generated for the same game and persona) to a Gemini3-Flash judge(prompt details in Appendix~\ref{app:prompt_diversity}).
The judge scores the batch on a 1--5 scale based on topic coverage: 
Low scores indicate the reviews are merely rephrasing the same point (e.g., all 5 reviews complain about ``luck''); 
High scores indicate the reviews discuss diverse aspects such as mechanics, social interaction, and art style to mimic the natural variety of human feedback.

\noindent\textbf{Results Analysis.}
Table~\ref{tab:main_results} confirms that MeepleLM matches the factual accuracy of SOTA models while delivering superior diversity. 
To further illustrate this, we present the side-by-side comparison in Table~\ref{tab:qualitative_case}. 
While Qwen3-8B defaults to a generic melodramatic tone (``theater of tears'') and GPT-5.1 sounds like a detached journalist (``social WD-40''), MeepleLM authentically captures the distinct voice of each persona.
By seamlessly switching from community slang (e.g., ``Alpha Gamers'') in social contexts to technical critique (e.g., ``Variant Rules'') for purists, our model proves it is not just retrieving knowledge, but truly simulating a player's perspective.

\subsection{RQ3: Practical Utility}
\label{subsec:rq3}

\noindent\textbf{Opinion Recovery Rate (Op-Rec).}
To quantify the model's value as a virtual playtester, we assess its ability to forecast actual market feedback. We define Op-Rec as the recall rate of ground-truth player opinions within the simulated reviews.
The evaluation employs a two-stage pipeline using Gemini3-Flash(prompts detailed in Appendix~\ref{app:utility_prompts}):
(1) \textbf{Ground Truth Mining:} The judge extracts a deduplicated set of distinct viewpoints ($\mathcal{V}_{GT}$) from historical human reviews, representing the actual ``voice of the customer.''
(2) \textbf{Semantic Matching:} We check whether the simulated reviews generated by MeepleLM cover these specific viewpoints. As reported in Table~\ref{tab:main_results}, MeepleLM achieves the highest Op-Rec score, validating its utility for designers in forecasting market feedback and surfacing diverse player viewpoints.

\noindent\textbf{User Study: Blind A/B Test.}
To validate real-world effectiveness, we conducted a controlled blind A/B test with $N=10$ participants (demographics in Appendix~\ref{sec:user_study_appendix}). Each evaluator reviewed 6 titles randomly selected from the test set: 3 they had previously played (\textit{Familiar}) and 3 they had not (\textit{Unfamiliar}). 
Results indicate a decisive preference for MeepleLM over GPT-5.1. 
In the \textit{Familiar} scenario, our model achieved an average win rate of 78.3\%, with 83.3\% of participants specifically citing superior ``authenticity'' in capturing the emotional nuances of gameplay. 
In the \textit{Unfamiliar} scenario, the average win rate remained high at 74.2\%; notably, 86.7\% of users preferred MeepleLM for its critical honesty, describing it as ``less like marketing'' and more effective for identifying potential design flaws.
We further computed binomial tests (excluding ties; $H_0$: win rate $= 0.5$) and Cohen's $h$ for all metrics. All dimensions achieve $p < 0.001$ with large effect sizes ($h = 0.69$--$1.03$), confirming that the preference for MeepleLM is statistically robust despite the limited sample size (detailed pairwise results and qualitative feedback are provided in Appendix~\ref{sec:user_study_appendix}).

\subsection{Ablation and Further Analysis}
\noindent\textbf{Ablation.} 
To verify the contribution of each module, we evaluated three variants (detailed configurations provided in Appendix~\ref{app:ablation_setup}). 
As shown in Table~\ref{tab:main_results}, performance drops in all cases: 
(1) \textbf{w/o Rulebook:} Removing rule context causes a collapse in Factual Accuracy ($98.9 \to 59.9$), confirming that explicit grounding is non-negotiable; 
(2) \textbf{w/o Persona:} Replacing specific profiles with generic prompts drops ranking alignment ($\tau$ to $0.13$), proving that modeling heterogeneity is essential for capturing polarized preferences; 
(3) \textbf{w/o MDA:} Bypassing the CoT chain results in consistently lower opinion recovery, validating that intermediate reasoning is required to bridge the gap between static text and emergent experience.

\noindent\textbf{Temporal Impact.} 
We further validated the stability by re-evaluating RQ1 on a subset excluding the 35 newly released titles (2024--2025). As detailed in Appendix~\ref{app:temporal}, performance shifts across all models are negligible, confirming that the inclusion of fresh content does not bias the assessment.

\noindent\textbf{Persona Robustness.} 
Decomposing RQ1 by player profile (Appendix~\ref{app:persona_analysis}) highlights MeepleLM's robust performance on high-variance personas such as \textit{The Social Lubricator} and \textit{The Thrill Seeker}. 
This indicates that the framework successfully captures social dynamics and subjective ``vibes'' that defy pure logical deduction.

\section{Conclusion}

We present MeepleLM, a model that bridges the gap between static rulebooks and subjective player experiences. By curating dataset of rule-critique pairs and integrating MDA-based reasoning with data-driven player personas, we make gameplay dynamics explicit. Our experiments demonstrate that MeepleLM significantly outperforms general LLMs in capturing authentic community sentiment and actionable design insights. Ultimately, this work establishes a new paradigm for automated virtual testing of interactive systems, paving the way for experience-aware Human-AI collaboration attuned to diverse audience sensibilities.

\section*{Limitations}

While MeepleLM demonstrates strong potential as a virtual playtester, we acknowledge two primary limitations that outline our roadmap for future research:
(1) \textbf{Multimodal Understanding.} Currently, MeepleLM processes game rules exclusively as text. However, board games are inherently multimodal experiences where visual cues including card art, board iconography, and component design play a crucial role in immersion and usability. Future iterations will integrate visual encoders to process game assets (e.g., cards, maps, and tokens), enabling a more holistic evaluation of the game's aesthetic and functional design.
(2) \textbf{Granularity of Personas.} Our current approach relies on five aggregated personas derived from community clusters, which effectively capture broad player archetypes but may overlook the unique idiosyncrasies of specific individuals. Moving forward, we aim to transition from group-level to individual-level modeling. By collecting detailed historical data from specific players, we plan to construct a granular ``virtual player community,'' where agents can simulate the precise tastes and behaviors of real-world individuals for hyper-personalized playtesting.

\section*{Ethics Statement}

\paragraph{Data Privacy and Usage.}
Our dataset is constructed from publicly available content retrieved from an online board game community (see Section~\ref{sec:data}), which is accessible to the general public.
To protect user privacy, we strictly anonymize all User IDs and review identifiers, removing any personally identifiable information (PII) from the raw data.
To mitigate the potential dissemination of harmful content and respect copyright considerations, we will only release the processed versions of the reviews and metadata, rather than the raw scraped content.

\paragraph{Human Evaluation.}
Our research involves collecting evaluation data from real human participants. We adhere to strict ethical guidelines to ensure their privacy, consent, and well-being.
Key ethical principles include:
(1) \textbf{Informed Consent}: Participants are provided with detailed information about the study’s purpose, procedures, and their rights. They are explicitly informed that they can withdraw from the study at any time without any negative consequences.
(2) \textbf{Data Anonymization}: To safeguard participant privacy, all collected evaluation data, including interaction logs and questionnaires, is anonymized. Personal identifiers are removed to ensure that individuals cannot be traced from the data.
(3) \textbf{Data Security}: All collected data is stored securely, with access restricted to authorized research personnel only.

\bibliographystyle{plainnat}
\bibliography{custom}

\appendix
\clearpage
\raggedbottom
\section{Dataset Statistics Details}
\label{app:dataset_stats}

In this section, we provide a comprehensive statistical breakdown of the 1,717 board games selected for our dataset. These statistics validate that our sampling strategy successfully captured a diverse range of difficulty levels, quality standards, historical eras, and gameplay mechanisms.

\subsection{Distribution Analysis}
\begin{figure*}[ht]
    \centering
    \begin{minipage}{0.48\textwidth}
        \centering
        \includegraphics[width=\linewidth]{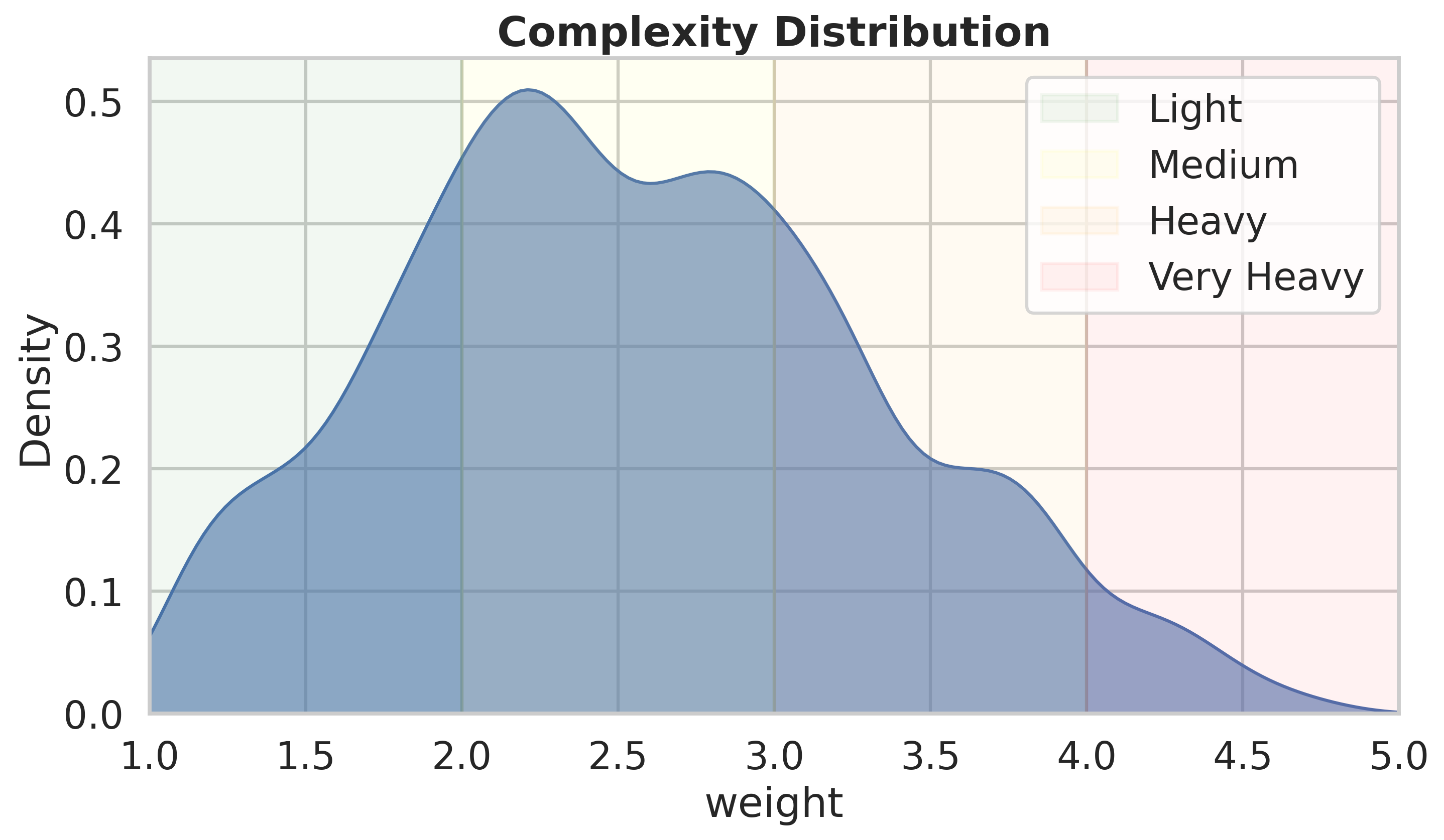}
        \par (a) Complexity (Weight) Distribution
    \end{minipage} \hfill
    \begin{minipage}{0.48\textwidth}
        \centering
        \includegraphics[width=\linewidth]{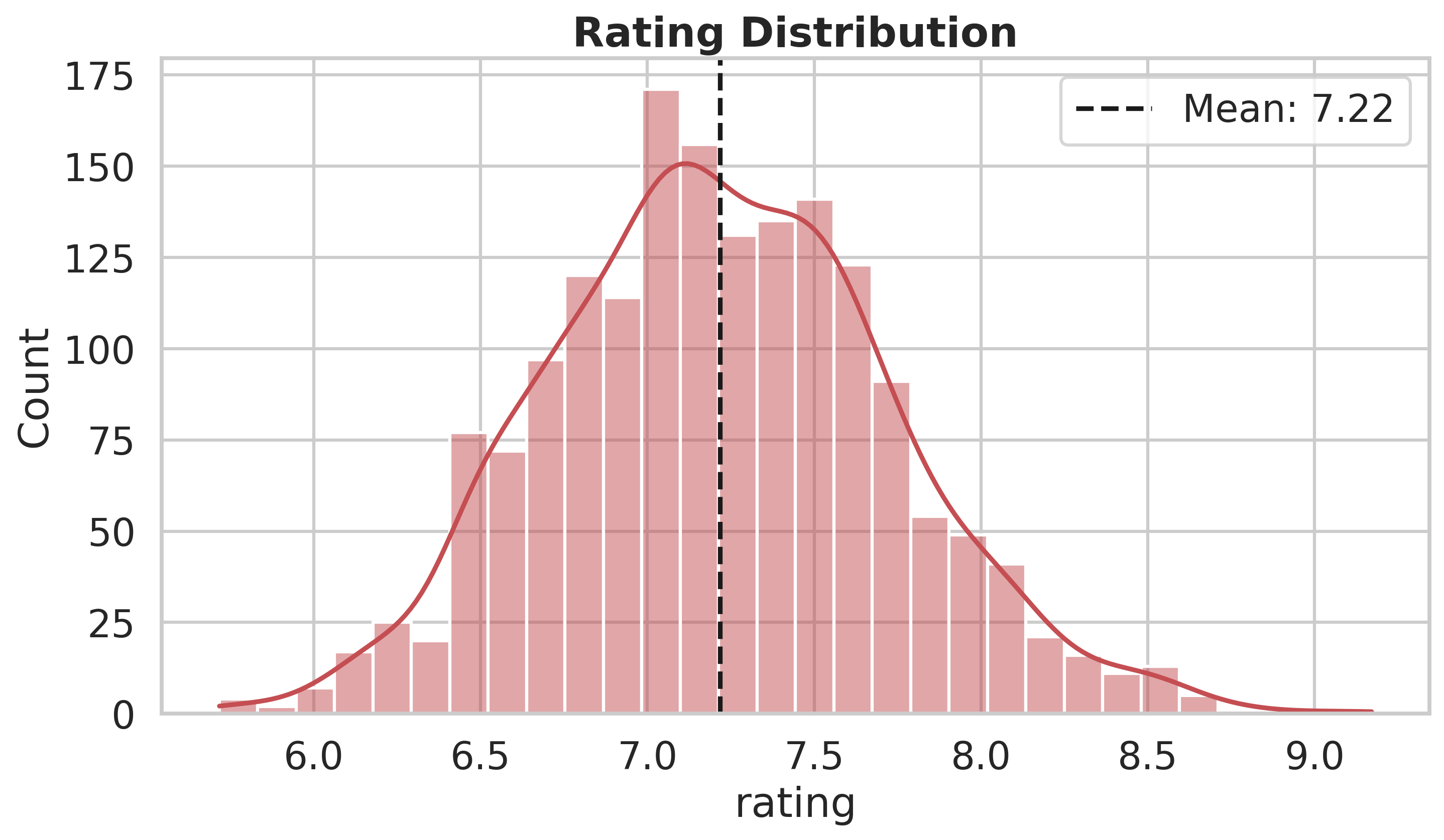}
        \par (b) Average Rating Distribution
    \end{minipage}
    
    \vspace{1em} 
    
    \begin{minipage}{0.48\textwidth}
        \centering
        \includegraphics[width=\linewidth]{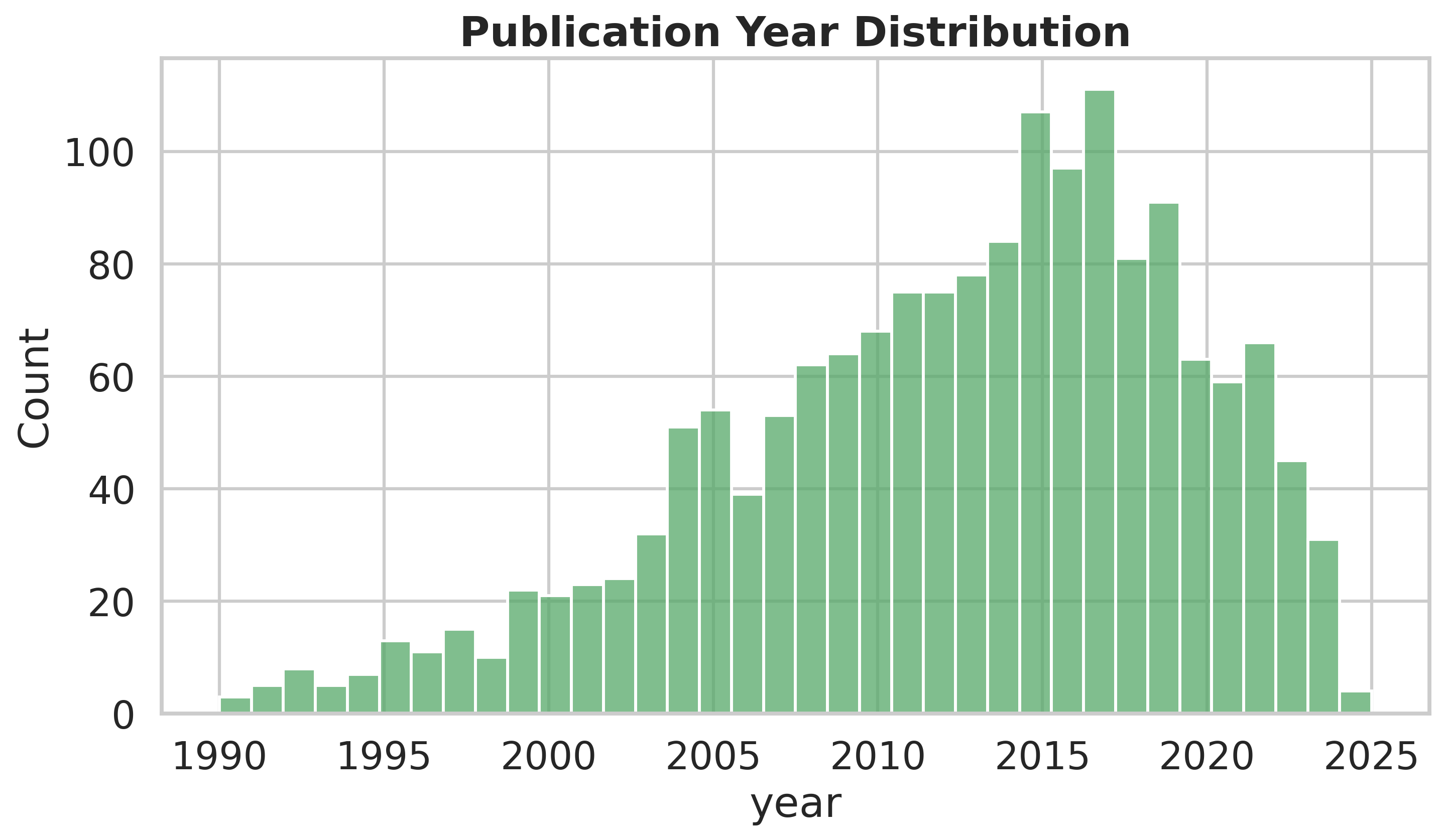}
        \par (c) Publication Year Distribution
    \end{minipage} \hfill
    \begin{minipage}{0.48\textwidth}
        \centering
        \includegraphics[width=\linewidth]{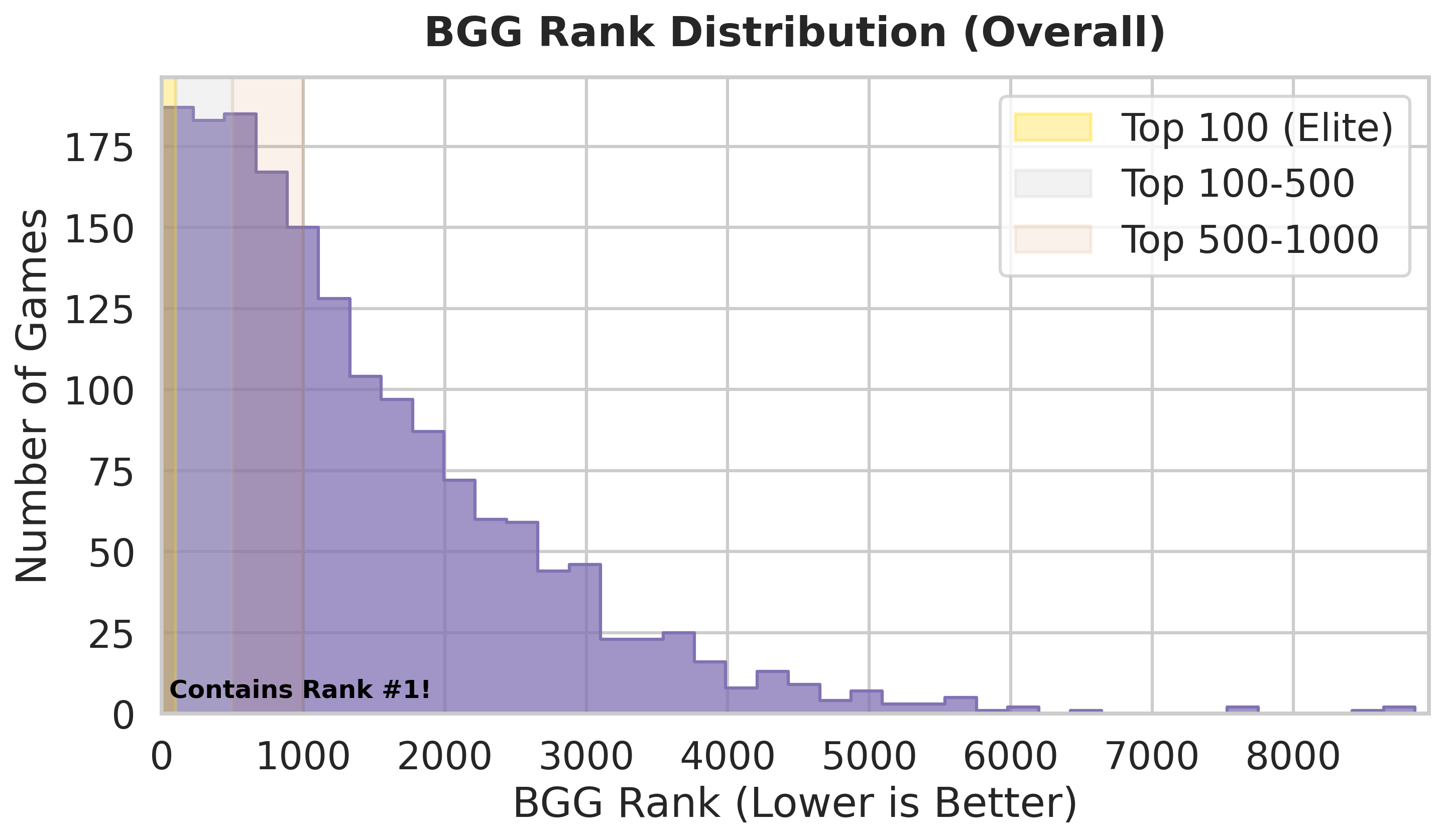}
        \par (d) BGG Rank Distribution
    \end{minipage}
    
    \caption{\textbf{Distributions of Key Metadata Attributes.} for the 1,717 selected games. The dataset covers a wide spectrum of difficulty (a), focuses on decent-to-excellent quality games (b), emphasizes modern board game designs (c), and spans both elite and long-tail rankings (d).}
    \label{fig:app_stats_dist}
\end{figure*}

Figure~\ref{fig:app_stats_dist} presents the distributions of Complexity, Rating, Publication Year, and BGG Rank.

\begin{itemize}
    \item \textbf{Complexity (Weight):} As shown in Figure~\ref{fig:app_stats_dist}(a), the complexity distribution is nearly normal (Skewness: 0.29) with a mean of 2.57. We intentionally preserved a balanced spectrum: \textit{Light} games (Weight 1--2) account for 24.5\%, while \textit{Heavy/Very Heavy} games (Weight > 3) comprise 28.5\%. This ensures the model learns to adapt its critique depth to the cognitive load of the game.
    \item \textbf{Rating:} Figure~\ref{fig:app_stats_dist}(b) shows the rating distribution (Mean: 7.22, StdDev: 0.51). The distribution is slightly left-skewed, focusing on games generally considered "playable" to "excellent" (66\% are rated > 7.0). This filtering removes low-quality noise while retaining enough variance for comparative analysis.
    \item \textbf{Publication Year:} Illustrated in Figure~\ref{fig:app_stats_dist}(c), the dataset reflects the modern board game renaissance, with a median publication year of 2013. The coverage spans from classic designs (pre-2000, $N=159$) to contemporary hits ($2015+$, $N=755$), specifically including 34 cutting-edge titles released in 2024--2025.
    \item \textbf{Rank Coverage:} Figure~\ref{fig:app_stats_dist}(d) highlights the market representativeness. While we include 83\% of the Top 100 "elite" games to ensure high-quality training data, over 50\% of the dataset ($N=909$) consists of "long-tail" games (Rank > 1000), preventing the model from overfitting to universally acclaimed masterpieces.
\end{itemize}

\subsection{Content Diversity: Mechanics and Themes}
\begin{figure*}[ht]
    \centering
    \begin{minipage}{0.48\textwidth}
        \centering
        \includegraphics[width=\linewidth]{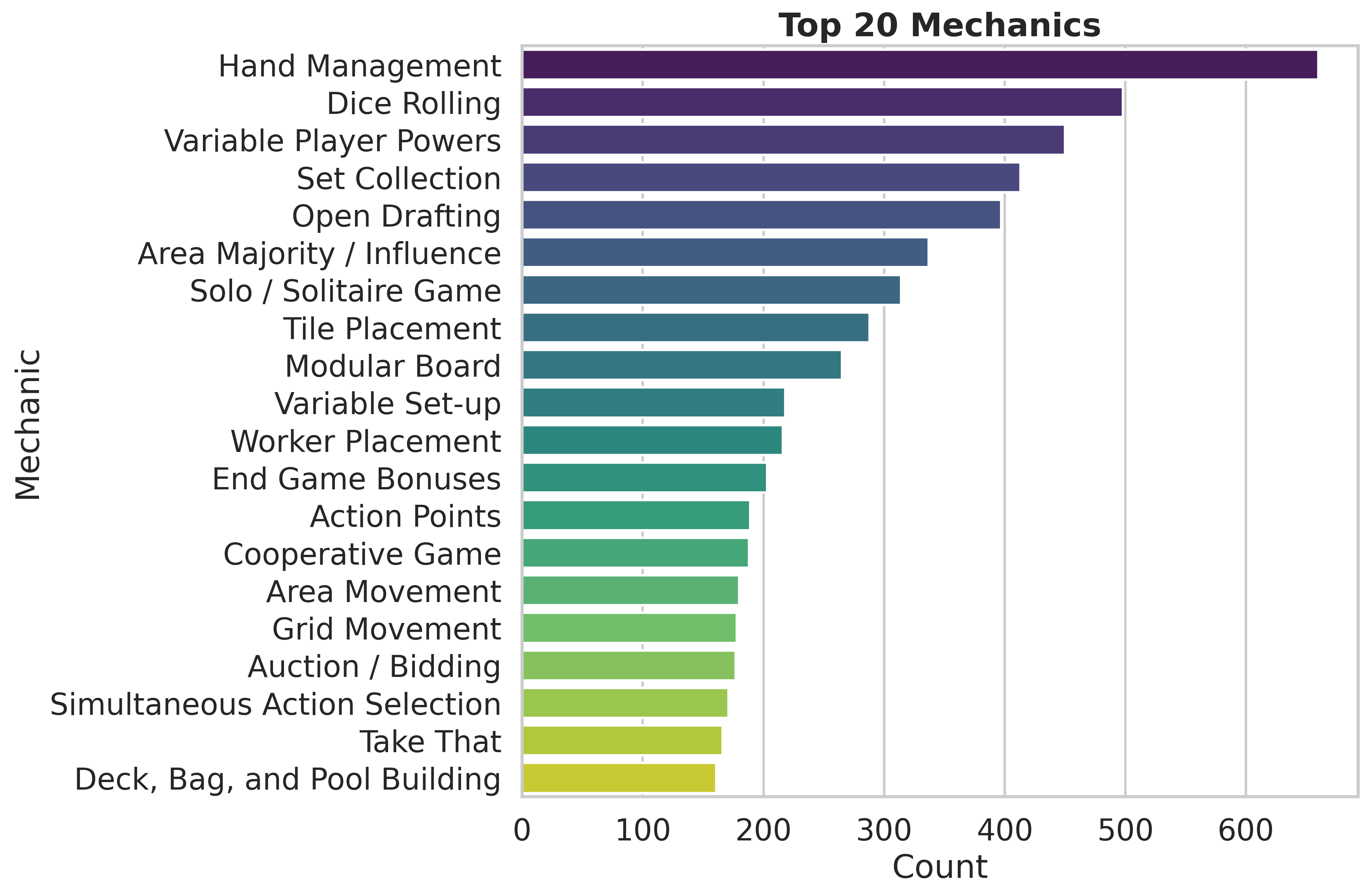}
        \par (a) Top 20 Mechanics
    \end{minipage} \hfill
    \begin{minipage}{0.48\textwidth}
        \centering
        \includegraphics[width=\linewidth]{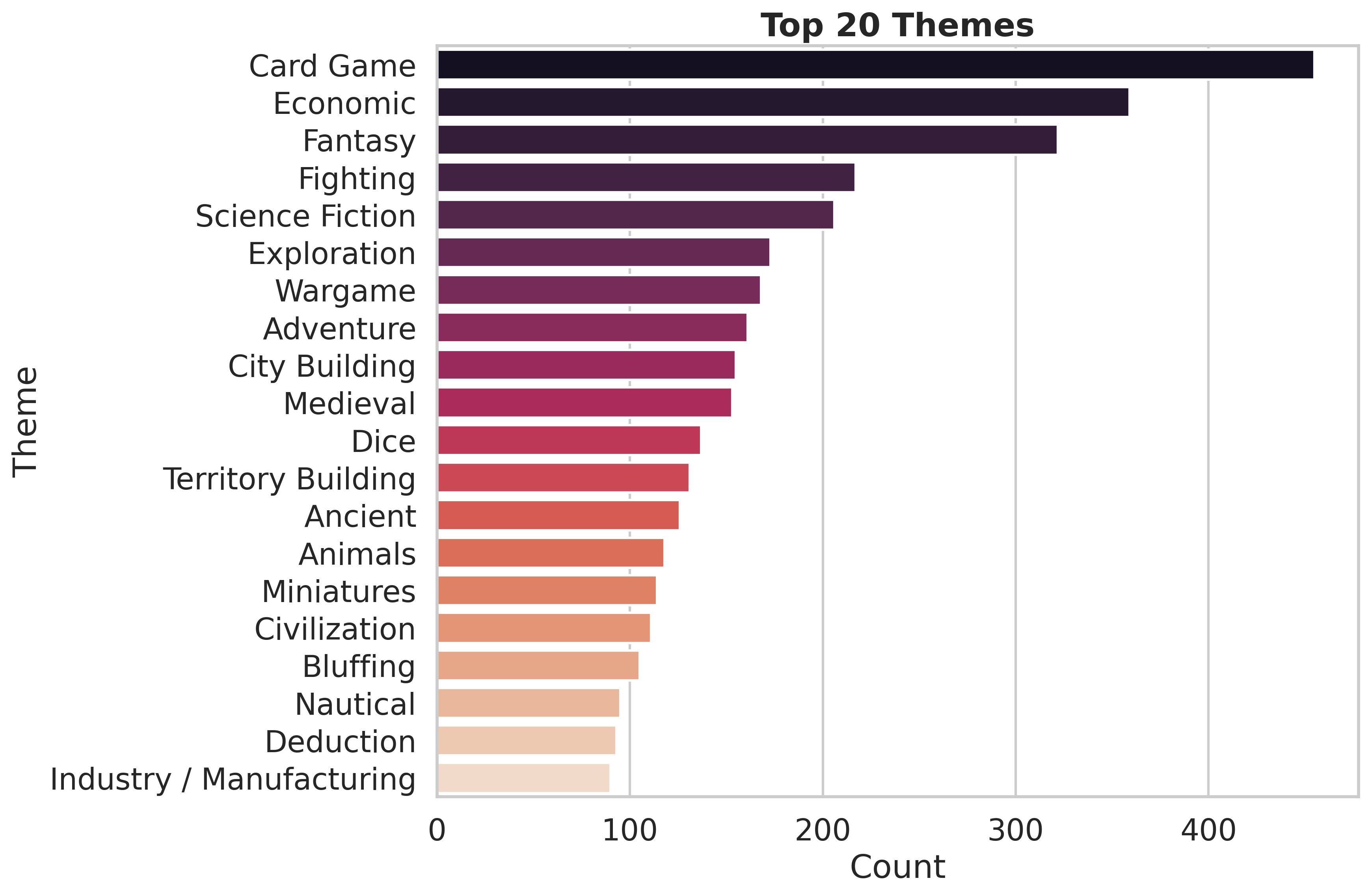}
        \par (b) Top 20 Themes
    \end{minipage}
    
    \vspace{1em} 
    
    \begin{minipage}{0.48\textwidth}
        \centering
        \includegraphics[width=\linewidth]{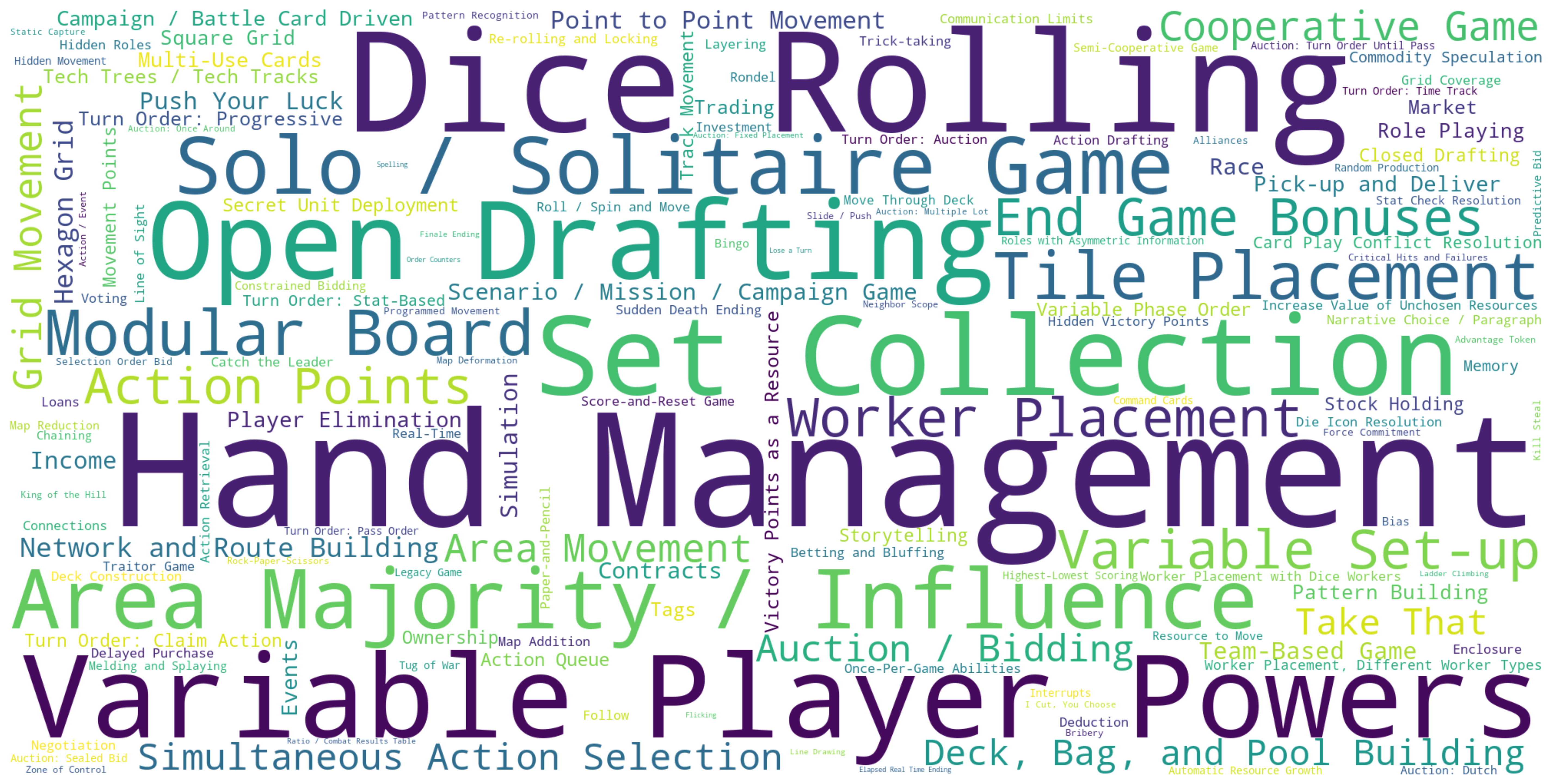}
        \par (c) Mechanics Word Cloud
    \end{minipage} \hfill
    \begin{minipage}{0.48\textwidth}
        \centering
        \includegraphics[width=\linewidth]{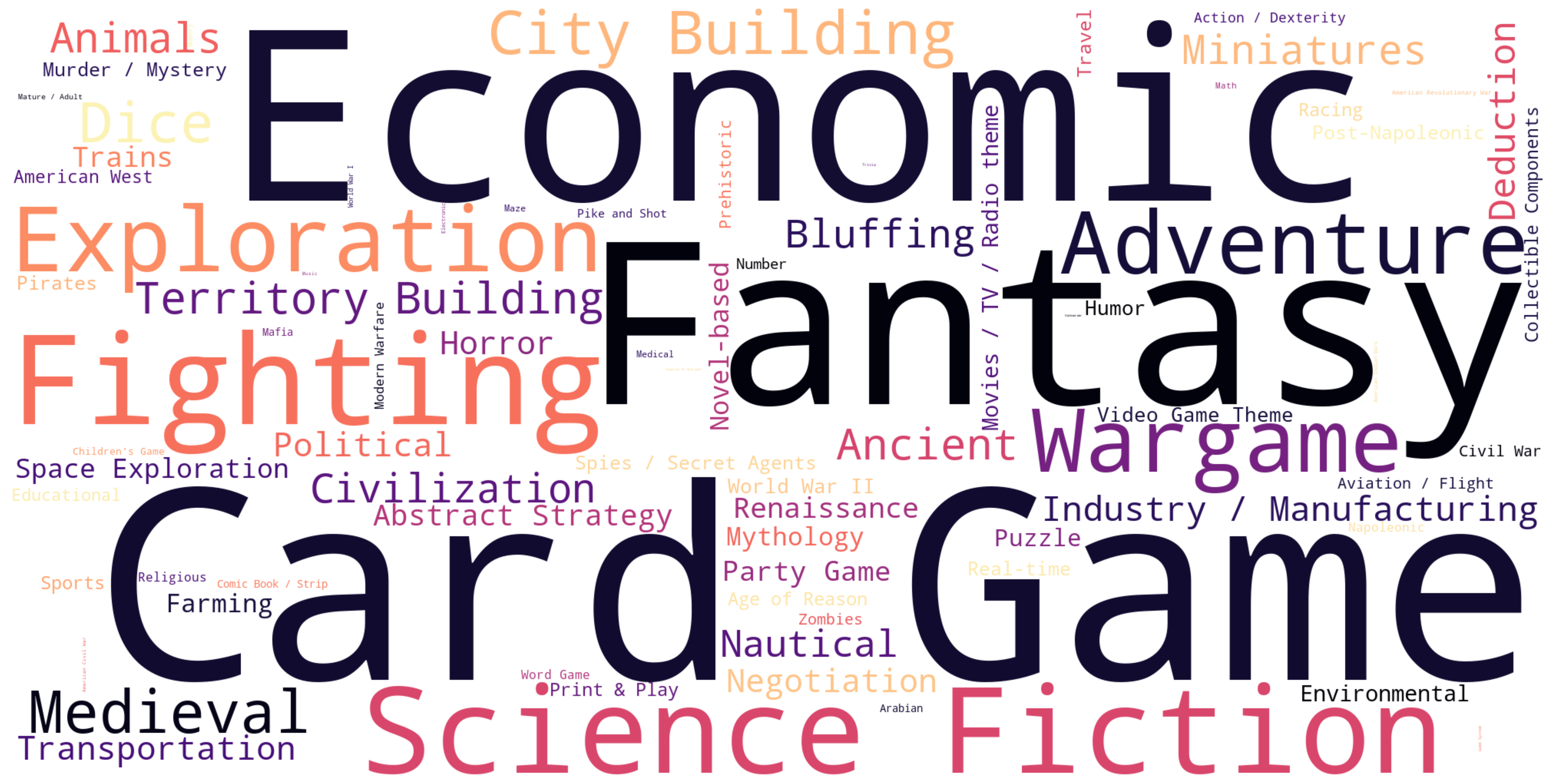}
        \par (d) Themes Word Cloud
    \end{minipage}
    
    \caption{\textbf{Analysis of Game Content.} (a) and (b) display the top 10 mechanics and themes, demonstrating that the dataset covers the fundamental building blocks of modern board games. (c) and (d) provide a holistic view of the terminological diversity present in the corpus.}
    \label{fig:app_stats_content}
\end{figure*}

To ensure the model can generate grounded reviews for various gameplay styles, we analyzed the mechanics and themes tags. Figure~\ref{fig:app_stats_content} visualizes the prevalent categories.

\begin{itemize}
    \item \textbf{Mechanics:} The dataset features \textbf{192 unique mechanics} with an average of 6.35 mechanics per game, indicating high systemic depth. As shown in Figure~\ref{fig:app_stats_content}(a), the most frequent mechanics include \textit{Hand Management} (38.4\%), \textit{Dice Rolling} (29.0\%), and \textit{Variable Player Powers} (26.2\%), which are foundational to modern game design.
    \item \textbf{Themes:} We identified \textbf{81 unique themes}. Figure~\ref{fig:app_stats_content}(b) shows a blend of abstract strategy themes (e.g., \textit{Economic}, 20.9\%) and immersive narrative themes (e.g., \textit{Fantasy}, 18.8\%; \textit{Sci-Fi}, 12.0\%). This diversity requires the model to contextually adapt its vocabulary (e.g., discussing "profits" vs. "damage").
\end{itemize}

\clearpage
\section{Rulebook Structuring Details}
\label{app:rulebook_prompt}

To convert the raw Markdown rulebooks (converted by Mineru) into structured knowledge, we employed Qwen-3. We utilized a specific prompt to ensure the model extracts information strictly from the source text without hallucination, organizing it into a standardized Markdown format.

\subsection{Extraction Prompt}
\label{app:B_rulebook_prompt}
Figure~\ref{fig:prompt_box} displays the system prompt used. The prompt enforces strict constraints to use \textit{only} existing information from the uploaded file.

\subsection{Structured Rulebook Example}
\label{app:B_example}
Figure~\ref{fig:example_box} demonstrates a sample of the structured output. This standardized text serves as the knowledge base for the review generation model.

\subsection{Rectification Prompt}
\label{app:B_rectify_prompt}

Figure~\ref{fig:rectify_prompt_box} presents the rectification prompt used by GPT-5.1. This stage acts as a verification layer, cross-referencing the structured draft against the source text to correct hallucinations or omissions.

\clearpage
\onecolumn 

\begin{tcolorbox}[
    enhanced,
    breakable,                
    colback=white,            
    colframe=gray!60!black,   
    coltitle=white,           
    title=\textbf{Prompt: Rulebook Structuring}, 
    arc=0pt, outer arc=0pt,
    boxrule=1pt,
    top=5pt, bottom=5pt,
]
    \begin{lstlisting}[breaklines=true, basicstyle=\small\ttfamily, columns=fullflexible, extendedchars=false]
System Instruction:
You are an expert board game rules analyst. Your task is to reorganize the provided raw Markdown content into a structured format.

Constraints (Strict Grounding):
1. Source Only: You must ONLY use information present in the input text. Do NOT add any external knowledge or hallucinate rules.
2. Format: Output the result in clean Markdown format using the specific headers defined below.
3. Completeness: If a section is not mentioned in the text, write "Not Mentioned" under that header.

Required Output Structure:
## 1. Lore & Objective
(Extract the thematic setting and victory conditions)
## 2. Components
(List physical assets like cards, boards, tokens)
## 3. Setup
(Step-by-step initial configuration)
## 4. Gameplay Flow
(Turn structure, phases, and round sequence)
## 5. Core Mechanics
(Key interaction rules, player actions, restrictions)
## 6. Scoring & End Game
(How points are calculated and how the game ends)
## 7. FAQ or Edge Cases
(Specific clarifications found in the text)

Input Text:
{RAW_MARKDOWN_CONTENT}
    \end{lstlisting}
\end{tcolorbox}

\begin{center}
\captionsetup{hypcap=false}
\captionof{figure}{\textbf{System Prompt for Structuring Raw Rulebooks.} It enforces a standard Markdown schema and strict grounding to the source text.}
\label{fig:prompt_box}
\end{center}

\vspace{2em} 

\clearpage

\begin{tcolorbox}[
    enhanced,
    breakable,                
    colback=white,
    colframe=gray!60!black,
    coltitle=white,
    title=\textbf{Structured Rulebook Example <Clank!: A Deck-Building Adventure>}, 
    arc=0pt, outer arc=0pt,
    boxrule=1pt,
    top=5pt, bottom=5pt,
]
    \begin{lstlisting}[breaklines=true, basicstyle=\small\ttfamily, columns=fullflexible, extendedchars=false]
### 1. Lore & Objective  
Players are rival thieves attempting to infiltrate Dragon Keep, steal valuable Artifacts, and escape alive. The dragon Nicki guards her hoard fiercely—noise (Clank!) and stolen treasures increase her rage. The ultimate goal is to become the Greatest Thief in the Realm by escaping with an Artifact and the most Victory Points (VP).  

**Primary Objectives:**  
- Steal at least **one Artifact**—failure to do so results in automatic loss.  
- Accumulate the highest VP total from Artifacts, Gold, acquired tokens, and deck cards.  
- Survive above the Grass Line (depths boundary) or escape the dungeon to score points.  

**Winning Conditions:**  
- Players knocked out **below the Grass Line** score 0.  
- Players knocked out **above the Grass Line** score their collected loot (if they have an Artifact).  
- Escaping players receive a **20 VP Mastery Token**.  
- Game ends immediately when the **Countdown Track reaches the Skull** or all players have exited/knocked out.  
- Ties broken by highest-value Artifact; further ties unresolved.

---

### 2. Components  
- **Double-sided Game Board** (Front: recommended for first game)  
- **Dungeon Deck** (100 cards)  
- **Reserve Cards**:  
  - 15 Mercenary  
  - 15 Explore  
  - 12 Secret Tome  
  - 1 Goblin (always available)  
- **Player Decks**: 4 starting decks (10 cards each):  
  - 6 Burgle  
  - 2 Stumble  
  - 1 Sidestep  
  - 1 Scramble  
- **Tokens & Markers**:  
  - 120 Clank! Cubes (30 per player color)  
  - 24 Black Dragon Cubes  
  - Dragon Bag  
  - Dragon Marker  
  - Mastery Tokens (4)  
- **Treasures & Secrets**:  
  - 7 Artifacts (5–30 VP)  
  - 11 Major Secrets  
  - 18 Minor Secrets  
- **Market Items**:  
  - 2 Master Keys (5 VP)  
  - 2 Backpacks (5 VP)  
  - 3 Crowns (10, 9, 8 VP)  
- **Other**:  
  - 3 Monkey Idols  
  - Gold Tokens (1 and 5)  
  - Player Pawns (4)  
  - Rage Track (on board)  
  - Countdown Track (top of board)  
  - Health Meter (bottom of board)

---

### 3. Setup  
1. **Player Setup**:  
   - Each player takes:  
     - 30 Clank! cubes (personal supply)  
     - Matching pawn  
     - 10-card starting deck  
   - Shuffle deck, draw 5 cards.  
   - Place pawn on top-left entrance space.  

2. **Board Setup**:  
   - **Artifacts**: Place 7 face up on numbered spaces.  
     - 3 players: Shuffle and remove 1 randomly.  
     - 2 players: Remove 2 randomly.  
   - **Major Secrets**: Shuffle face down, place 1 on each marked space. Return extras to box.  
   - **Minor Secrets**: Shuffle face down, place 2 stacked on each marked space.  
   - **Market Items**:  
     - 2 Master Keys (stacked)  
     - 2 Backpacks (stacked)  
     - 3 Crowns (10, 9, 8 in descending order)  
   - **Monkey Idols**: Place 1 in each space in Monkey Shrine Room.  
   - **Mastery Tokens**: Place 1 per player at top-left flag.  
   - **Gold**: Place in bank near board.  
   - **Dragon Marker**:  
     - 4 players: Start on first space  
     - 3 players: Second space  
     - 2 players: Third space  
   - **Dragon Bag**: Place 24 black cubes inside.  
   - **Reserve Stacks**: Place near board:  
     - 1 Goblin  
     - 15 Mercenary  
     - 15 Explore  
     - 12 Secret Tome  
   - **Dungeon Row**: Shuffle Dungeon Deck, deal 6 face-up cards.  
     - Replace any with Dragon Attack symbol; shuffle replaced cards back.  
   - **Starting Clank!**:  
     - 1st player: 3 cubes in Clank! Area  
     - 2nd: 2 cubes  
     - 3rd: 1 cube  
     - 4th: 0 cubes  

3. **First Player**: Roll die or choose randomly.

---

### 4. Gameplay Flow  
Each turn follows this sequence:

1. **Play All Cards**: Play all 5 cards in hand, in any order.  
2. **Take Actions** (in any order, multiple times):  
   - Acquire Cards (Skill)  
   - Use Devices (Skill)  
   - Fight Monsters (Swords)  
   - Buy Market Items (Gold)  
   - Move (Boots)  
   - Gain Gold / Clank! / Secrets / Artifact  
3. **End of Turn Phase**:  
   - Discard all played cards.  
   - Draw 5 new cards (shuffle discard if needed).  
   - Refill Dungeon Row to 6 cards.  
   - **If refill reveals Dragon Attack symbol**:  
     - Dragon attacks once (even if multiple symbols).  
     - Move all Clank! cubes to Dragon Bag.  
     - Draw cubes = Dragon’s Rage Track value (+1 per Danger symbol).  
     - Black cubes: set aside. Colored cubes: place on matching player’s Health Meter.  
4. **Dragon Rage Triggers**:  
   - +1 space when:  
     - Artifact is taken  
     - Dragon Egg (Minor Secret) is revealed  

5. **Exit or Knockout Triggers Countdown**:  
   - First player to exit dungeon or be knocked out moves to Countdown Track.  
   - On their next turn, they move forward and trigger effects:  
     - Spaces 2–4: Instant Dragon Attack with +1, +2, +3 extra draws  
     - Space 5: All remaining players in dungeon are instantly knocked out  

6. **Game End**: Triggered when:  
   - Countdown Track reaches Skull  
   - All players have exited or been knocked out  

---

### 5. Core Mechanics  

#### Resources  
- **Skill**: Acquire cards from Dungeon Row or Reserve.  
- **Swords**: Fight monsters (Dungeon Row or tunnels).  
- **Boots**: Move 1 tunnel per Boot.  

#### Movement Rules  
- **Normal Tunnel**: 1 Boot  
- **Footprint Icon**: 2 Boots  
- **Monster Icon**: Pay 1 Sword per icon, or take 1 damage per icon  
- **Lock Icon**: Requires Master Key  
- **Arrow Tunnel**: One-way only  
- **Crystal Cave**: Upon entry, **cannot use Boots** for rest of turn (teleporting allowed)  
- **Wrap-around Tunnels**: Connect opposite edges; cost 1 Boot  

#### Actions  
- **Acquire Card** (Blue/Yellow Banner):  
  - Pay Skill → place in discard pile (joins deck)  
- **Use Device** (Purple Banner):  
  - Pay Skill → use effect immediately → discard to Dungeon discard pile  
- **Fight Monster** (Red Banner):  
  - Pay Swords → gain reward → discard to Dungeon discard pile  
  - **Goblin (Reserve)**: Can be fought multiple times; not discarded  
- **Buy Market Item**:  
  - Cost: 7 Gold each  
  - Items: Crown (VP), Backpack (carry +1 Artifact), Master Key (unlock tunnels)  
  - Unlimited purchases per turn  

#### Clank!  
- Cards or actions may add Clank! cubes to central Clank! Area.  
- Negative Clank! removes cubes from area or offsets future gains.  
- Unused negative Clank! is lost at turn end.  

#### Dragon Attacks  
- Triggered **only** when refilling Dungeon Row reveals **at least one Dragon symbol**.  
- All Clank! cubes go into Dragon Bag.  
- Draw cubes = Dragon’s current Rage Track value + number of Danger symbols in Dungeon Row.  
- Colored cubes = damage to that player.  
- Exited/knocked-out players are immune.  

#### Health & Knockout  
- Damage places cubes on Health Meter (left to right).  
- **Fully filled meter** = knocked out.  
  - **Below Grass Line**: Lose (0 points)  
  - **Above Grass Line**: Score loot if holding Artifact  
- Healing: Use Potions or effects to return cubes to supply.  

#### Artifacts & Carrying  
- Can only carry **1 Artifact** unless Backpack is owned (1 extra per Backpack).  
- **Cannot drop or switch** Artifact once picked up.  
- Taking Artifact:  
  - Move Dragon +1 on Rage Track  
  - Triggers Dragon Egg if Minor Secret is revealed  

---

### 6. Scoring & End Game  
**Scoring Includes**:  
- Artifact VP  
- All acquired tokens (Crown, Chalice, Backpack, etc.)  
- Gold: 1 VP per Gold  
- Cards in deck: VP in top-right corner  
- **Mastery Token (20 VP)**: Only if exited dungeon via own movement/teleport  

**Excluded from Score**:  
- Players knocked out **below Grass Line** → 0 points  
- Exited/knocked-out players: No further turns or card effects  

**Tiebreaker**:  
1. Highest-value Artifact  
2. (Unspecified if still tied)  

---

### 7. FAQ or Edge Cases  
- **Can you drop an Artifact?** No. Once picked up, it’s yours until game end.  
- **Can you buy multiple Market items per turn?** Yes, any number, even same type.  
- **Can you take multiple tokens per room?** Only one per entry. Must exit and re-enter.  
- **Does Artifact in Market cost Gold?** No. Only Crown, Backpack, Key cost Gold.  
- **When does Dragon attack?** Only when **new** Dragon symbol is revealed during **Dungeon Row refill**. Preexisting symbols do not re-trigger.  
- **Crystal Cave movement?** Cannot use Boots after entry, even via teleport.  
- **Running out of Clank! cubes?** Cannot choose to take damage in tunnels. Cannot be forced to add Clank! (free pass until healed).  
- **Healing during turn?** Yes, via Potions or effects.  
- **Order of card play?** Effects resolve in real time; order does not block conditional bonuses (e.g., Swagger gains Skill for all Clank! made that turn).  

---

    \end{lstlisting}
\end{tcolorbox}

\begin{center}
\captionsetup{hypcap=false}
\captionof{figure}{\textbf{Example of Structured Rulebook Data.} This content spans multiple pages, preserving the full details extracted from the original PDF.}
\label{fig:example_box}
\end{center}

\clearpage

\begin{tcolorbox}[
    enhanced,
    breakable,                
    colback=white,            
    colframe=gray!60!black,   
    coltitle=white,           
    title=\textbf{Prompt: Rulebook Rectification (Expert-in-the-Loop)}, 
    arc=0pt, outer arc=0pt,
    boxrule=1pt,
    top=5pt, bottom=5pt,
]
    \begin{lstlisting}[breaklines=true, basicstyle=\small\ttfamily, columns=fullflexible, extendedchars=false]
System Instruction:
You are a meticulous Board Game Rule Editor. Your task is to verify and rectify a "Draft Rulebook" against the original "Source Content".

Inputs:
1. Source Content: Raw text parsed from the official rulebook PDF.
2. Draft Rulebook: A structured Markdown version generated by an automated parser.

Rectification Tasks:
1. Accuracy Check: Compare specific numbers, card counts, and setup instructions. If the Draft says "deal 5 cards" but the Source says "deal 4", CORRECT it.
2. Completeness: Ensure no critical sections (especially "End Game Triggers" or "Tie-Breakers") are missing from the Draft.
3. Logical Consistency: Fix any contradictions introduced during the structuring process.
4. Formatting: Ensure the output strictly follows the standardized Markdown hierarchy.

Constraints:
- Output ONLY the fully rectified Rulebook in Markdown.
- Do NOT add conversational text (e.g., "Here is the fixed version").
- Do NOT invent rules not present in the Source Content.

Input Data:
[SOURCE CONTENT]:
{RAW_SOURCE_TEXT}

[DRAFT RULEBOOK]:
{QWEN_GENERATED_STRUCT}
    \end{lstlisting}
\end{tcolorbox}

\begin{center}
\captionsetup{hypcap=false}
\captionof{figure}{\textbf{Rectification Prompt Used by GPT-5.1.} It requires the model to cross-reference the generated draft against the source text to ensure numerical accuracy and logical completeness.}
\label{fig:rectify_prompt_box}
\end{center}

\clearpage
\twocolumn

\section{Review Processing Details}
\label{app:review_stats}

\subsection{Data Sources}
\label{app:review_source}
To ensure diverse perspective coverage, we aggregated raw user reviews from multiple online communities through professional data outsourcing services. The sources encompass prominent digital tabletop platforms such as Board Game Arena\footnote{\url{https://en.boardgamearena.com}} and Tabletopia\footnote{\url{https://tabletopia.com}}, alongside specialized enthusiast forums like GStone\footnote{\url{https://www.gstonegames.com}} and QPBG\footnote{\url{https://qpbg.com}}. Given the heterogeneous scoring systems across these sites, we normalized all collected ratings to a standardized 1.0--10.0 scale.

\subsection{Quality Annotation Prompt}
\label{app:C_review_prompt}

To implement the "Design-Logic Quality Scoring" described in Section~\ref{sec:data_reviews}, we used the prompt shown in Figure~\ref{fig:review_prompt}. It enforces a strict evaluation criterion based on the utility of the review for game designers.

\subsection{Statistical Validation}
\label{app:C_review_statistical}
We analyzed the statistical properties of the 150K retained reviews to ensure they serve as an unbiased yet information-dense proxy for the original population.

\begin{figure}[t]
    \centering
    \includegraphics[width=\columnwidth]{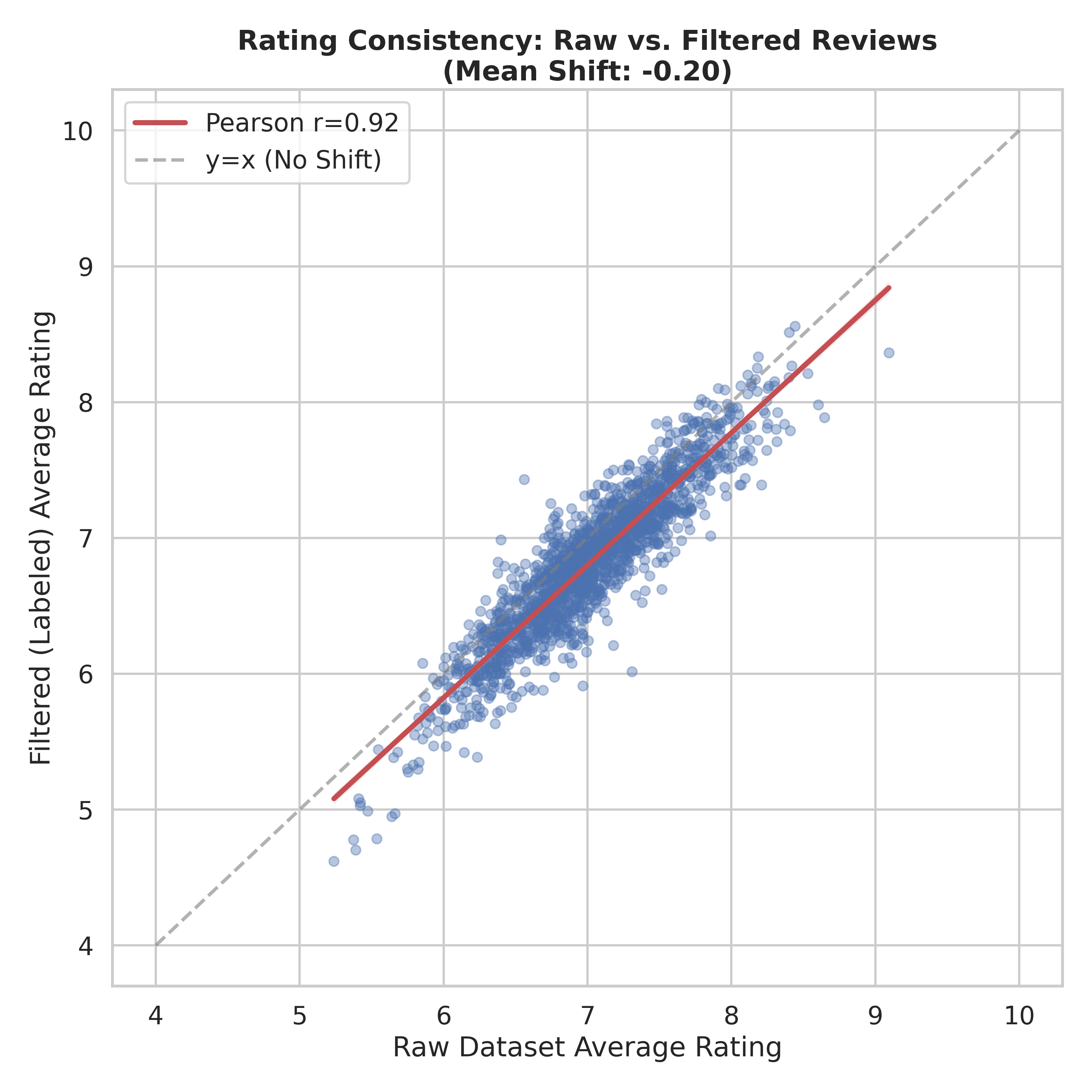}
    \caption{\textbf{Rating Correlation (Original vs. Filtered).}}
    \label{fig:rating_corr}
\end{figure}

\paragraph{Distributional Fidelity.}
As shown in Figure~\ref{fig:rating_corr}, the filtered dataset maintains a high degree of alignment with the original ratings (Pearson $r=0.92$, Spearman $\rho=0.91$). This confirms that our filtering strategy preserves the global consensus on game quality. Notably, we observed a slight negative mean shift ($-0.20$), which suggests the successful removal of "low-effort hype" (e.g., empty 10/10 ratings), resulting in a more critical and objective set.

\paragraph{Information Density}
Word count analysis reveals a "Polarization Ratio" of 1.24x: reviews at the rating extremes (1 \& 10) contain significantly more text (avg. 195.7 words) compared to mid-range reviews (avg. 158.2 words). This indicates that the dataset prioritizes strong signals—users provide the most detailed structural feedback when they are passionately engaged, ensuring the model learns clear causal links for both design flaws and successes.

\begin{figure}[t]
    \centering
    \includegraphics[width=\columnwidth]{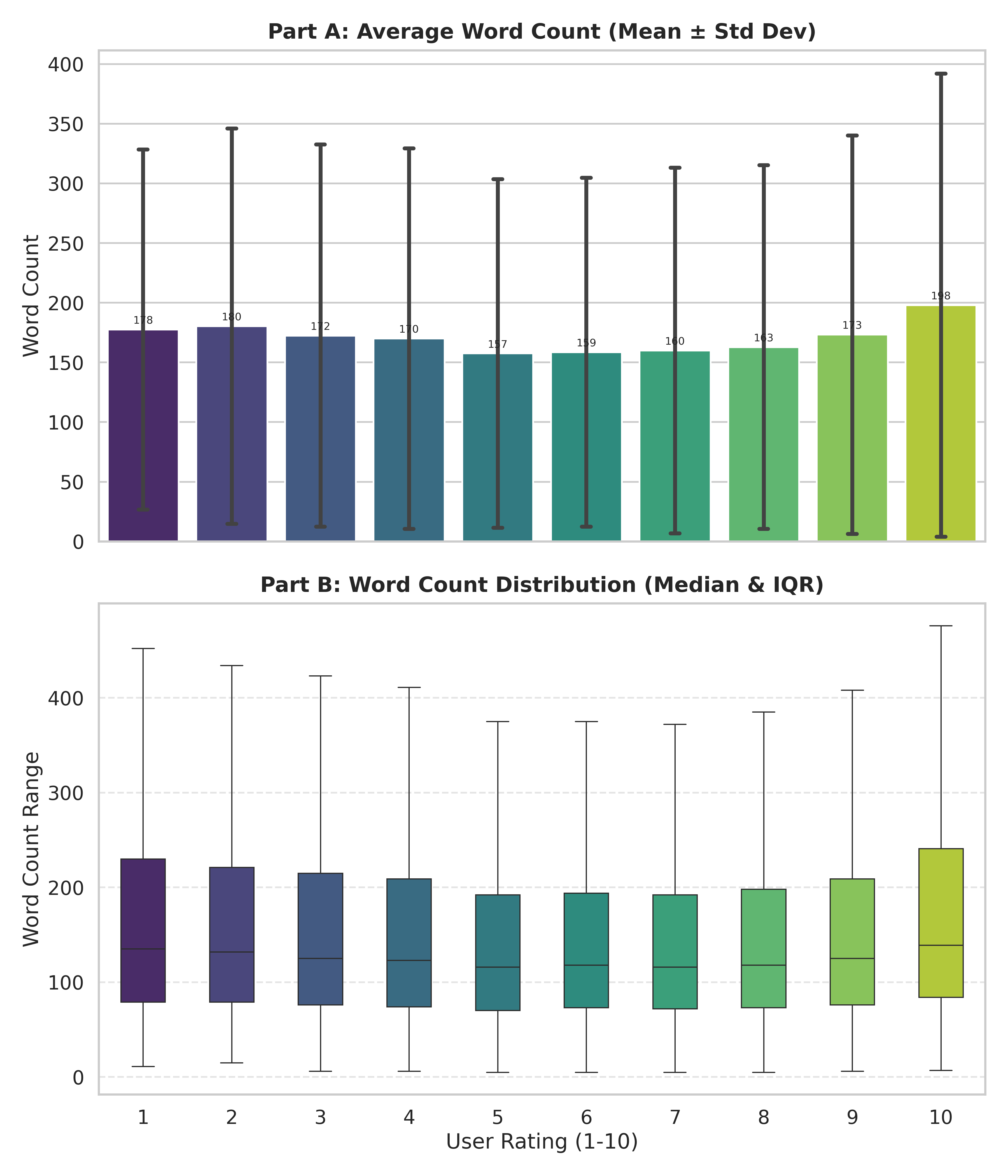}
    \caption{\textbf{Word Count Statistics by Rating.} }
    \label{fig:word_count}
\end{figure}

\clearpage
\onecolumn

\begin{tcolorbox}[
    enhanced, breakable, width=\textwidth,
    colback=white, colframe=gray!60!black, coltitle=white,
    title=\textbf{Prompt: Review Quality Scoring},
    arc=0pt, outer arc=0pt, boxrule=1pt, top=5pt, bottom=5pt
]
    \begin{lstlisting}[breaklines=true, basicstyle=\small\ttfamily, columns=fullflexible, extendedchars=false]
You are an expert Board Game Research Assistant constructing a high-quality reasoning dataset.
For each review (given a numeric rating and review text), you must output a JSON object strictly following the definitions below.

### 0. CRITICAL SCORING RULES (DECOUPLING)
**Treat each score INDEPENDENTLY.** Do not allow high scores in one dimension to bleed into others (No "Halo Effect").
- **High Anchoring (5) does NOT imply High Attribution**: A user might list rule names like a manual (Anchoring: 5) but explain nothing about the experience (Attribution: 1).
- **High Attribution (5) does NOT imply High Anchoring**: A user might deeply explain the cause of their fun (Attribution: 5) using only vague terms like "the pieces" (Anchoring: 2).
### 1. TASK DEFINITIONS
#### A. Hard Filters (Boolean)
Determine if the review should be discarded. Set "is_valid": false if ANY of the following apply:
- **Irrelevant**: Discusses ONLY shipping, damaged boxes, Kickstarter delivery, or customer service.
- **Visuals Only**: Discusses ONLY artwork, miniatures, or card stock quality without mentioning gameplay mechanics or experience.
- **Too Short**: Contains fewer than 20 words (insufficient logic).
- **Rating Mismatch**: The sentiment of the text drastically contradicts the numeric rating (e.g., a glowing review with a rating of 1, or a hateful rant with a rating of 10).
#### B. Quality Scores (Use the FULL 1-5 Scale)
You must use the full range of integers (1, 2, 3, 4, 5). Do not default to just 1/3/5.
1. **mechanism_anchoring** (Specificity)
   - 1 (Vague): "Fun game", "Good strategy". No specific terms.
   - **2 (Basic):** Mentions basic components like "cards", "board", "points" but no mechanism names.
   - 3 (Generic): Mentions standard mechanics e.g., "worker placement", "deck building".
   - **4 (Detailed):** Describes specific game flow or unique twists but misses the exact rulebook terminology.
   - 5 (Precise): Cites exact rule names/components e.g., "The Tekhenu obelisk wheel", "The Cult Track".
2. **causal_attribution** (Reasoning - CORE METRIC)
   - 1 (No Logic): "I hated it." / "Best game ever." (Pure emotion).
   - **2 (Implied):** "It's too long and boring." (Reason implies cause, but vague).
   - 3 (Simple Link): "I didn't like it *because* the downtime was too long." (Direct X->Y).
   - **4 (Strong Logic):** "The downtime is caused by the analysis paralysis in the auction phase." (Identifies mechanism -> issue).
   - 5 (Deep Systemic): "Because turn order resets based on travelers, long-term planning feels impossible, leading to a chaotic experience." (Mechanism -> Dynamic -> Aesthetic).
3. **constructiveness** (Utility)
   - 1 (Useless): Empty complaints or blind praise.
   - **2 (Valid Complaint):** "The combat feels unfair." (Identifies a problem area, but subjective).
   - 3 (Actionable): "The endgame drags on too long." (Specific pain point).
   - **4 (Analytical):** "The blue faction is strong because of their starting resource." (Analyzes *why* it's a pain point).
   - 5 (Insightful/Solution): Offers a fix or deep balance critique: "They should cap rounds to 8 to fix the pacing."
#### C. Content Facets (Multi-label List)
Identify which aspects are primarily discussed. Ignore background mentions.
Choose ONLY from:
- "Rule Clarity & Teachability": rules readability, ambiguity, teaching difficulty.
- "Cognitive Load (Complexity)": weight, brain-burn, analysis paralysis.
- "Interaction & Conflict": take-that, blocking, table talk, multi-player solitaire.
- "Luck vs. Strategy": randomness, dice/card luck, mitigation.
- "Balance & Fairness": start player advantage, runaway leader, faction balance.
- "Replayability & Variety": setup variability, scripted play.
- "Thematic Integration": Ludonarrative harmony (mechanics fit theme). Exclude pure art praise.
- "Pacing & Flow": downtime, game length feel (drag/tight), end triggers.
### 2. OUTPUT SCHEMA
Return a JSON object with this EXACT structure:
{
  "is_valid": boolean,
  "filter_reason": string or null, // If is_valid is false, state the reason (e.g., "Too Short", "Irrelevant"). If true, null.
  "scores": {
    "mechanism_anchoring": int(1-5),
    "causal_attribution": int(1-5),
    "constructiveness": int(1-5)
  },
  "facets": [string, string, ...] // List of matching facets
}
### 3. FEW-SHOT EXAMPLES
Input:
Rating: 6
Comment: "It involves moving pawns and collecting cards, but the randomness in drawing cards makes it feel like I have no control."
Output:
{
  "is_valid": true,
  "filter_reason": null,
  "scores": {
    "mechanism_anchoring": 2,   // "pawns", "cards" (Basic terms) -> 2
    "causal_attribution": 4,    // Randomness -> No control (Strong Logic) -> 4
    "constructiveness": 2       // Valid complaint but subjective -> 2
  },
  "facets": ["Luck vs. Strategy"]
}
Input:
Rating: 8
Comment: "Excellent worker placement game. The tight economy forces you to plan ahead."
Output:
{
  "is_valid": true,
  "filter_reason": null,
  "scores": {
    "mechanism_anchoring": 3,   // "worker placement" (Generic) -> 3
    "causal_attribution": 3,    // tight economy -> plan ahead (Simple Link) -> 3
    "constructiveness": 1       // Praise only -> 1
  },
  "facets": ["Cognitive Load (Complexity)"]
}
Input:
Rating: 4
Comment: "The 'Corruption' mechanic punishes you too hard. It creates a death spiral where if you mess up turn 1, you sit there for 2 hours. It feels broken."
Output:
{
  "is_valid": true,
  "filter_reason": null,
  "scores": {
    "mechanism_anchoring": 5,   // "Corruption" (Specific Rule) -> 5
    "causal_attribution": 5,    // Mechanic -> Dynamic (Death spiral) -> Feeling -> 5
    "constructiveness": 4       // Strong analysis of the problem -> 4
  },
  "facets": ["Balance & Fairness", "Pacing & Flow"]
}
You will be given a list of {batch_size} board game reviews.
You MUST generate a JSON ARRAY containing EXACTLY {batch_size} JSON objects, one for each review.
Do not add more or fewer objects than the number of reviews provided.
REVIEWS (JSON list):
{reviews_json}
    \end{lstlisting}
\end{tcolorbox}

\begin{center}
    \captionsetup{hypcap=false}
    \captionof{figure}{\textbf{Annotation Prompt for Scoring Review Quality.} It mirrors the observation-analysis-iteration loop.}
    \label{fig:review_prompt}
\end{center}

\twocolumn

\section{Persona Discovery Details}
\label{app:persona_details}

\subsection{Feature Construction \& Clustering}
\label{app:persona_cluster}

To ensure the clustering algorithm captures the cognitive depth of the reviewer rather than just keyword overlap, we pre-processed each review into a composite text string before embedding.

\paragraph{Composite Input Logic.}
We injected the quantitative metrics derived in Section~\ref{sec:data_reviews} directly into the text representation. Each review was formatted using the following structured template:

{
\setlength{\fboxsep}{10pt} 
\begin{center}
\colorbox{gray!10}{
    \parbox{0.9\columnwidth}{
    \texttt{\small [SENTIMENT: \textbf{\{Tier\}}] [FOCUS: \textbf{\{Facets\}}] :: \textbf{\{Raw Review Content\}}}
    }
}
\end{center}
}

The metadata fields were populated based on the following rules:
\begin{itemize}
    \setlength\itemsep{0.2em}
    \item \textbf{Sentiment Tier:} Discretized based on the normalized rating $R$: labeled as \textit{"Positive"} ($R \ge 8$), \textit{"Negative"} ($R \le 4$), or \textit{"Neutral"} ($5 \le R \le 7$). This guides the embedding to group reviews by satisfaction level.
    
    \item \textbf{Focus Facets:} A comma-separated list of dimensions derived directly from the facet-scoring model detailed in Section~\ref{sec:data_reviews}.
\end{itemize}

\begin{figure}[t]
    \centering
    \includegraphics[width=\linewidth]{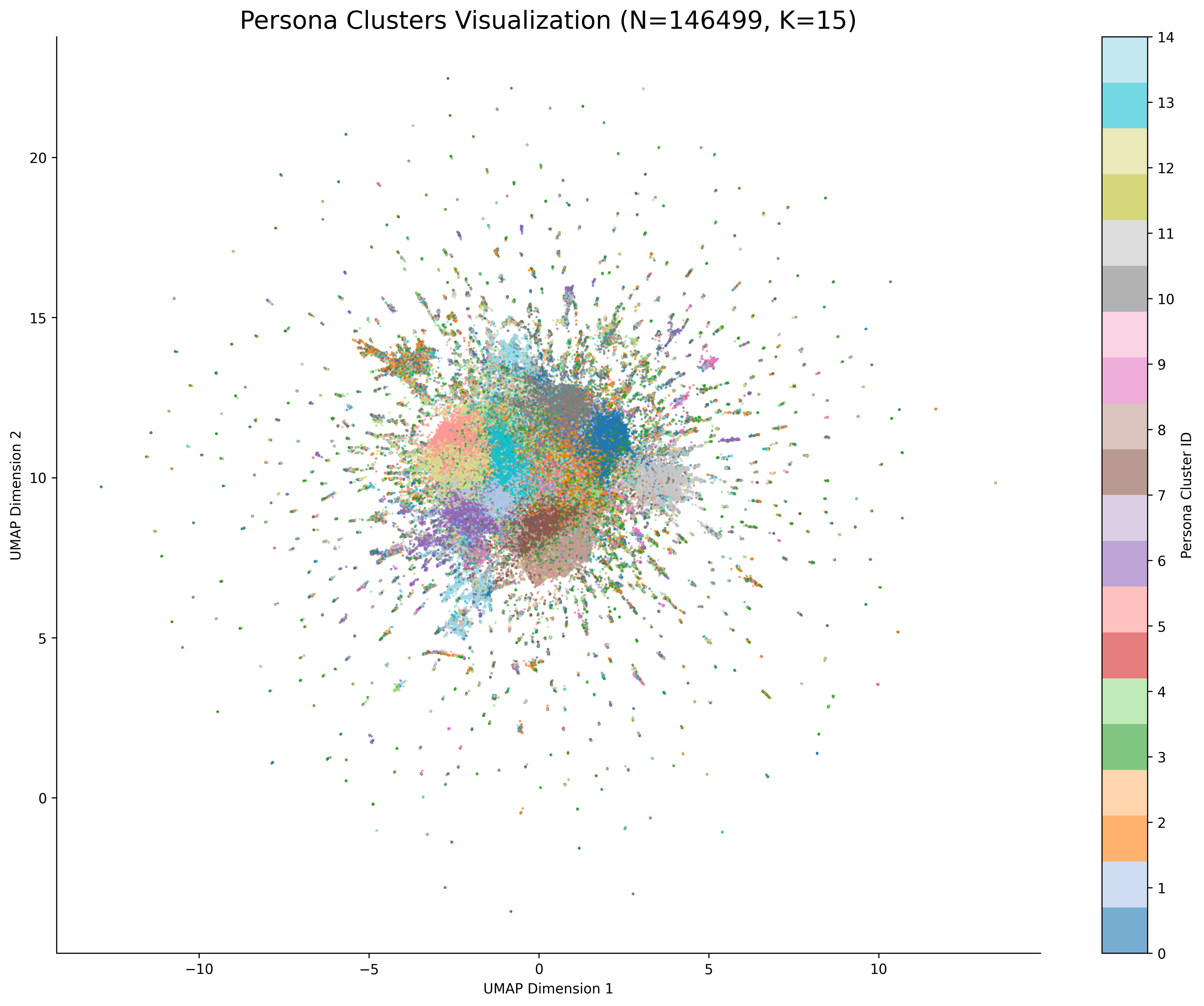}
    \caption{\textbf{Visualization of Composite Embeddings.} Colors indicate the 15 initial clusters, which were later merged into the 5 final personas by domain experts.}
    \label{fig:persona_cluster_plot}
\end{figure}

\paragraph{Clustering Result.}
Figure~\ref{fig:persona_cluster_plot} visualizes the T-SNE projection of these embeddings. The inclusion of explicit \textbf{Sentiment} and \textbf{Focus} tags helped clearly separate reviewers based on their fundamental evaluation criteria and satisfaction thresholds, effectively mitigating the ambiguity of surface-level keywords.

\subsection{Persona Descriptions and Statistics}
\label{app:persona_desc}
\begin{figure}[t]
    \centering
    \includegraphics[width=\linewidth]{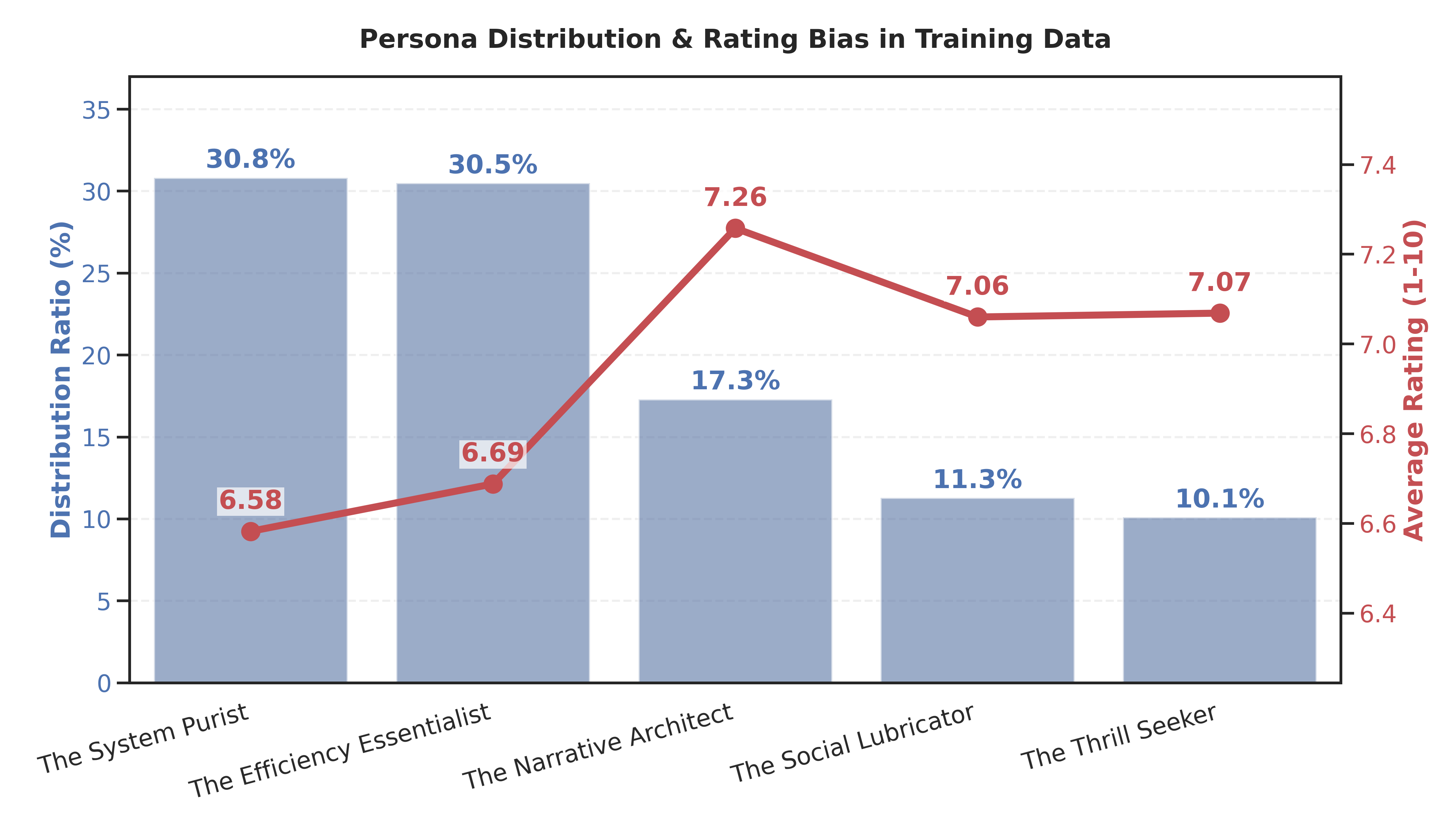}
    \caption{\textbf{Distribution of Personas in the Dataset.} \textit{System Purists} (Avg 6.58) are the harshest critics, while \textit{Narrative Architects} (Avg 7.28) are the most generous.}
    \label{fig:persona_stats}
\end{figure}

Figure~\ref{fig:persona_stats} shows the distribution and average rating for each group. Figure~\ref{fig:persona_desc_full} provides the detailed definitions for each persona used in the annotation process, outlining their core motivations and specific mechanical preferences. 

\subsection{Discovery Prompts}
\label{app:persona_prompts}

We employed a two-phase prompting strategy to translate raw numerical clusters into an annotated dataset.

\paragraph{Phase 1: Persona Profiling.} 
After obtaining the 15 initial clusters, we sampled the top-20 central reviews from each cluster. As shown in Figure~\ref{fig:prompt_profile}, we prompted \textbf{GPT-5.1} to analyze these samples and summarize the distinct "Player Persona" they represent. Domain experts analyzed the semantic coherence of each cluster, synthesizing overlapping groups to establish the final five persona definitions.

\paragraph{Phase 2: Dataset Labeling.} 
Once the 5 distinct personas were finalized (as defined in Appendix~\ref{app:persona_desc}), we needed to propagate these labels to the entire dataset. Figure~\ref{fig:prompt_label} displays the classification prompt used with \textbf{GPT-5.1} to annotate each review.

\subsection{Preference Matrix (Extended Case Study)}
\label{app:persona_matrix}

To validate the distinctiveness of each persona, we performed a frequency analysis of mechanics in their highest-rated vs. lowest-rated games. Figure~\ref{fig:preference_list_1} and Figure~\ref{fig:preference_list_2} detail the "Lift" metric, highlighting mechanics that are disproportionately favored or disliked by specific groups.

\subsection{Semantic Ambiguity Analysis}
\label{app:D_semantic_ambiguity}

To demonstrate the limitations of standard supervised classifiers, we present a detailed case study from our error analysis. As noted in Section~\ref{sec:data_persona}, a DeBERTa-v3-large classifier trained on cluster seeds achieved only $\sim$50\% accuracy.

Table~\ref{tab:misclassification_case} illustrates the core issue: \textbf{Keyword vs. Intent Misalignment}. Standard models over-index on technical vocabulary (e.g., \textit{"balance"}, \textit{"rules"}), failing to detect when users repurpose these terms to describe their exact opposites—such as using "house rules" to introduce \textbf{high-stakes volatility} rather than mechanical fairness.

\begin{table}[ht]
    \centering
    \small
    \renewcommand{\arraystretch}{1.3} 
    \begin{tabular}{p{0.95\columnwidth}}
    \toprule
    \textbf{Review Input (Raw Text)} \\
    \midrule
    ``A \textbf{sentimental} rating of course. We combined parts of the first Lord of the Rings game with this one, working on making the deck more \textit{balanced}. We had numerous \textit{house rules}, that in my opinion, made it such a great gaming experience. For example, when attacking, \textbf{triple sixes} gave you an \textbf{extra kill}. Double sixes on defense, you GAINED an army, lose NONE. To even that out though, if you roll double ones, you LOSE 3. Remember, this is all before I learned about board games.'' \\
    \midrule
    \textbf{Standard Classifier Prediction: \textcolor{red}{System Purist} \ding{55}} \\
    \textbf{Reasoning:} The model is misled by surface-level design keywords such as \textit{"balanced"} and \textit{"house rules"}. It incorrectly infers that the user is focused on game balance, mechanical rigor, or improving the system's logic. \\
    \midrule
    \textbf{Ground Truth / MeepleLM Label: \textcolor{teal}{Thrill Seeker} \ding{51}} \\
    \textbf{Reasoning:} The review explicitly frames the experience as \textit{"sentimental"} and prioritizes dramatic, high-variance moments (\textit{"triple sixes"}, \textit{"extra kill"}, \textit{"LOSE 3"}). The user's "house rules" were not created to fix the system's logic, but to inject more chaos and excitement into the gameplay, which is the defining trait of a Thrill Seeker. \\
    \bottomrule
    \end{tabular}
    \caption{\textbf{Case Study of Semantic Ambiguity.} The standard classifier fails by latching onto mechanical keywords (``balanced''), while the LLM correctly identifies the underlying motivation for excitement (``triple sixes'', ``sentimental'').}
    \label{tab:misclassification_case}
\end{table}

\onecolumn

\begin{figure*}[ht]
    \centering
    \begin{tcolorbox}[
        enhanced, breakable, width=\textwidth,
        colback=white, colframe=gray!60!black, coltitle=white,
        title=\textbf{Detailed Persona Definitions \& Preferences},
        arc=0pt, outer arc=0pt, boxrule=1pt, top=5pt, bottom=5pt
    ]
        \begin{lstlisting}[breaklines=true, basicstyle=\small\ttfamily, columns=fullflexible, extendedchars=false]
"The System Purist": """
    * **Core Motivation:** Intellectual superiority & Control. They want to prove they can beat the system through pure logic.
    * **Profile:** Loves heavy/crunchy decisions. Zero tolerance for luck (hates dice output randomness). Obsessed with balance (hates first-player advantage).
    * **Interaction:** Likes indirect competition (blocking), hates chaotic direct conflict (take-that).
    * **Keywords:** "Optimization", "No luck", "Perfect information", "Tight", "Punishing".
""",
"The Efficiency Essentialist": """
    * **Core Motivation:** Maximize ROI (Fun/Time). Seeks "Flow".
    * **Profile:** Hates "Fiddliness" (setup, shuffling, bookkeeping). Values elegance (simple rules, deep strategy). Pragmatic about rules (will house-rule to fix pacing).
    * **Interaction:** Fast-paced. Hates Downtime (Analysis Paralysis).
    * **Keywords:** "Elegance", "Streamlined", "Downtime", "Fiddly", "Smooth".
""",
"The Narrative Architect": """
    * **Core Motivation:** Immersion & Epic Experience. Mechanics serve the theme.
    * **Profile:** Loves growth (leveling up, empire building, tech trees). Wants 4X/RPG feels but within reasonable time.
    * **Interaction:** Cooperative or thematic negotiation/trade. Not calculating pure math.
    * **Keywords:** "Theme", "Immersion", "Story", "Epic", "Journey", "Flavor".
""",
"The Social Lubricator": """
    * **Core Motivation:** Human Connection & Emotional Resonance. Game is an excuse to socialize.
    * **Profile:** Needs low barrier to entry (accessible to non-gamers). Hates "Alpha Gamers" (quarterbacking). Prioritizes experience over scoring.
    * **Interaction:** High social interaction (bluffing, laughter, party games).
    * **Keywords:** "Party game", "Laughs", "Interaction", "Easy to teach", "Group dynamic".
""",
"The Thrill Seeker": """
    * **Core Motivation:** Dopamine & Emotional Rollercoaster.
    * **Profile:** Embraces risk (Push-your-luck). Needs fast pacing (if I lose, let me restart instantly). Active agency in gambling.
    * **Interaction:** Schadenfreude (enjoying opponents busting) and epic comebacks.
    * **Keywords:** "Push your luck", "Excitement", "Tension", "Gambling", "High stakes".
"""
        \end{lstlisting}
    \end{tcolorbox}
    \caption{\textbf{Behavioral Profiles of the Five Discovered Personas.} Each profile defines core motivations and specific likes/dislikes regarding game mechanisms.}
    \label{fig:persona_desc_full}
\end{figure*}

\begin{figure}[ht]
    \centering
    \begin{tcolorbox}[
        enhanced, breakable, width=\textwidth,
        colback=white, colframe=gray!60!black, coltitle=white,
        title=\textbf{Prompt : Persona Profiling (GPT-5.1)},
        arc=0pt, boxrule=1pt, top=5pt, bottom=5pt
    ]
\begin{lstlisting}[breaklines=true, basicstyle=\small\ttfamily, columns=fullflexible, extendedchars=false]
You are a User Researcher specialized in board games. I will provide you with a set of game reviews that belong to the same Player Cluster. Your task is to analyze these reviews and synthesize a Persona Profile.
Based ONLY on the reviews above, define this player persona using the following JSON schema. Be specific and refer to the evidence in the text.
{
  "persona_name": "Creative Name (e.g., The Dice Hater, The Euro Optimizer)",
  "core_motivation": "Why do they play? (e.g., Mathematical efficiency)",
  "preferred_mechanics": "What do they like/dislike? (e.g., Loves worker placement, hates dice)",
  "interaction_style": "How do they interact? (e.g., Dislikes direct conflict)",
  "deal_breakers": "What makes them rate a game low? (e.g., Bad rulebooks, too much luck)",
  "system_prompt_description": "A concise, first-person description for an LLM to roleplay this user. (e.g., 'You are a hardcore strategy gamer who values low luck and high skill ceiling. You are critical of imbalance...')"
}

\end{lstlisting}
    \end{tcolorbox}
    \caption{\textbf{Profiling Prompt for Interpreting Cluster-Central Samples.} This qualitative analysis guided the definition of the final 5 personas.}
    \label{fig:prompt_profile}
\end{figure}

\begin{figure}[ht]
    \centering
    \begin{tcolorbox}[
        enhanced, breakable, width=\textwidth,
        colback=white, colframe=gray!60!black, coltitle=white,
        title=\textbf{Prompt: Persona Labeling (GPT-5.1)},
        arc=0pt, boxrule=1pt, top=5pt, bottom=5pt
    ]
\begin{lstlisting}[breaklines=true, basicstyle=\small\ttfamily, columns=fullflexible, extendedchars=false]
You are an expert User Researcher specialized in board games. Your task is to classify board game players into EXACTLY ONE of the following 5 distinct personas.

### VALID PERSONAS (Choose ONE)
{PERSONA_DEFINITIONS}

### TASK
Analyze the review. Assign the review to one of the above personas.

### OUTPUT FORMAT
Return a JSON OBJECT. Key: "LLM_persona_name". Value: Must be one of the 5 bolded titles above.
Example: {"LLM_persona_name": "The System Purist"}

Process these {batch_size} reviews. Return a JSON ARRAY of {batch_size} objects.

REVIEWS:
{reviews_json}
\end{lstlisting}
    \end{tcolorbox}
    \caption{\textbf{Labeling Prompt for Annotating the Full Dataset.} It maps each review to one of the 5 finalized personas based on the review's content and sentiment.}
    \label{fig:prompt_label}
\end{figure}

\newcommand{\statline}[3]{%
    \item \textbf{#1} \hfill {\footnotesize \color{gray} (Freq: #2, Lift: \textbf{#3})}
}

\begin{figure*}[p]
    \centering
    \small
    
    \begin{tcolorbox}[
        enhanced, width=\textwidth,
        colback=white, colframe=gray!20!black, coltitle=white,
        title=\textbf{ 1. THE SYSTEM PURIST},
        arc=2pt, boxrule=1pt
    ]
        \textbf{\textcolor{green!60!black}{LOVED Games}} \\
        \textit{Examples: Guards of Atlantis II, 1817, Pax Renaissance: 2nd Edition, Age of Innovation...}
        \vspace{0.3em} \\
        \textit{Key Mechanics:}
        \begin{itemize}[leftmargin=1.5em, nosep]
            \statline{Hand Management}{33\%}{1.5x}
                \statline{End Game Bonuses}{20\%}{4.6x}
                \statline{Auction / Bidding}{17\%}{3.0x}
                \statline{Area Movement}{13\%}{2.1x}
                \statline{Market}{13\%}{13.5x}
        \end{itemize}
        
        \vspace{0.8em}
        \hrule
        \vspace{0.8em}
    
        \textbf{\textcolor{red!60!black}{ HATED Games}} \\
        \textit{Examples: The Werewolves of Miller's Hollow, The Mind, Dungeon Run, Wavelength...}
        \vspace{0.3em} \\
        \textit{Key Mechanics:}
        \begin{itemize}[leftmargin=1.5em, nosep]
            \statline{Dice Rolling}{47\%}{1.6x}
                \statline{Hand Management}{33\%}{1.5x}
                \statline{Cooperative Game}{23\%}{1.8x}
                \statline{Role Playing}{13\%}{10.9x}
                \statline{Player Elimination}{10\%}{5.0x}
        \end{itemize}
    \end{tcolorbox}
    
    \vspace{1em}

    \begin{tcolorbox}[
        enhanced, width=\textwidth,
        colback=white, colframe=gray!20!black, coltitle=white,
        title=\textbf{ 2. THE EFFICIENCY ESSENTIALIST},
        arc=2pt, boxrule=1pt
    ]
        \textbf{\textcolor{green!60!black}{LOVED Games}} \\
        \textit{Examples: Aeon's End: The New Age, Maria, The Crew: Mission Deep Sea, 7 Wonders (Second Edition)...}
        \vspace{0.3em} \\
        \textit{Key Mechanics:}
        \begin{itemize}[leftmargin=1.5em, nosep]
            \statline{Hand Management}{43\%}{1.9x}
                \statline{Area Majority / Influence}{27\%}{2.2x}
                \statline{Dice Rolling}{20\%}{0.7x}
                \statline{Campaign / Battle Card Driven}{17\%}{4.6x}
                \statline{Cooperative Game}{13\%}{1.0x}
        \end{itemize}
        
        \vspace{0.8em}
        \hrule
        \vspace{0.8em}
    
        \textbf{\textcolor{red!60!black}{ HATED Games}} \\
        \textit{Examples: WarCraft: The Board Game, The Elder Scrolls V: Skyrim – The Adventure Game, Android, Car Wars...}
        \vspace{0.3em} \\
        \textit{Key Mechanics:}
        \begin{itemize}[leftmargin=1.5em, nosep]
            \statline{Dice Rolling}{47\%}{1.6x}
                \statline{Hand Management}{37\%}{1.6x}
                \statline{Cooperative Game}{17\%}{1.3x}
                \statline{Action Points}{13\%}{1.5x}
                \statline{Modular Board}{10\%}{1.9x}
        \end{itemize}
    \end{tcolorbox}
    
    \vspace{1em}

    \begin{tcolorbox}[
        enhanced, width=\textwidth,
        colback=white, colframe=gray!20!black, coltitle=white,
        title=\textbf{ 3. THE NARRATIVE ARCHITECT},
        arc=2pt, boxrule=1pt
    ]
        \textbf{\textcolor{green!60!black}{LOVED Games}} \\
        \textit{Examples: Aeon Trespass: Odyssey, Skull, BattleCON: Devastation of Indines, Aeon's End: The New Age...}
        \vspace{0.3em} \\
        \textit{Key Mechanics:}
        \begin{itemize}[leftmargin=1.5em, nosep]
            \statline{Cooperative Game}{37\%}{2.9x}
                \statline{Hand Management}{30\%}{1.3x}
                \statline{Dice Rolling}{23\%}{0.8x}
                \statline{Auction / Bidding}{17\%}{3.0x}
                \statline{Area Majority / Influence}{17\%}{1.4x}
        \end{itemize}
        
        \vspace{0.8em}
        \hrule
        \vspace{0.8em}
    
        \textbf{\textcolor{red!60!black}{ HATED Games}} \\
        \textit{Examples: The Great Split, Blackbeard: The Golden Age of Piracy, Zombies!!!, Aquädukt...}
        \vspace{0.3em} \\
        \textit{Key Mechanics:}
        \begin{itemize}[leftmargin=1.5em, nosep]
            \statline{Hand Management}{37\%}{1.6x}
                \statline{Dice Rolling}{27\%}{0.9x}
                \statline{Simultaneous Action Selection}{17\%}{6.1x}
                \statline{Communication Limits}{13\%}{7.8x}
                \statline{Race}{13\%}{10.4x}
        \end{itemize}
    \end{tcolorbox}
    
    \caption{\textbf{Detailed Mechanism Preferences (Part 1).} Analysis of \textit{System Purist}, \textit{Efficiency Essentialist}, and \textit{Narrative Architect}.}
    \label{fig:preference_list_1}
\end{figure*}

\begin{figure*}[t]
    \centering
    \small

    \begin{tcolorbox}[
        enhanced, width=\textwidth,
        colback=white, colframe=gray!20!black, coltitle=white,
        title=\textbf{ 4. THE SOCIAL LUBRICATOR},
        arc=2pt, boxrule=1pt
    ]
        \textbf{\textcolor{green!60!black}{LOVED Games}} \\
        \textit{Examples: American Rails, Napoléon: The Waterloo Campaign, 1815, Egizia, Arydia: The Paths We Dare Tread...}
        \vspace{0.3em} \\
        \textit{Key Mechanics:}
        \begin{itemize}[leftmargin=1.5em, nosep]
            \statline{Dice Rolling}{40\%}{1.4x}
                \statline{Area Majority / Influence}{30\%}{2.5x}
                \statline{End Game Bonuses}{17\%}{3.8x}
                \statline{Cooperative Game}{17\%}{1.3x}
                \statline{Area Movement}{13\%}{2.1x}
        \end{itemize}
        
        \vspace{0.8em}
        \hrule
        \vspace{0.8em}
    
        \textbf{\textcolor{red!60!black}{ HATED Games}} \\
        \textit{Examples: Feudum, Clippers, Fürstenfeld, World of Warcraft: The Adventure Game...}
        \vspace{0.3em} \\
        \textit{Key Mechanics:}
        \begin{itemize}[leftmargin=1.5em, nosep]
            \statline{Hand Management}{37\%}{1.6x}
                \statline{Action Points}{23\%}{2.7x}
                \statline{Area Majority / Influence}{17\%}{1.4x}
                \statline{Dice Rolling}{17\%}{0.6x}
                \statline{Network and Route Building}{13\%}{4.5x}
        \end{itemize}
    \end{tcolorbox}
    
    \vspace{1em}

    \begin{tcolorbox}[
        enhanced, width=\textwidth,
        colback=white, colframe=gray!20!black, coltitle=white,
        title=\textbf{ 5. THE THRILL SEEKER},
        arc=2pt, boxrule=1pt
    ]
        \textbf{\textcolor{green!60!black}{LOVED Games}} \\
        \textit{Examples: 20th Century, Advanced Squad Leader, Gaia Project, Empires in Arms...}
        \vspace{0.3em} \\
        \textit{Key Mechanics:}
        \begin{itemize}[leftmargin=1.5em, nosep]
            \statline{Dice Rolling}{23\%}{0.8x}
                \statline{End Game Bonuses}{17\%}{3.8x}
                \statline{Hexagon Grid}{17\%}{2.2x}
                \statline{Area Movement}{17\%}{2.6x}
                \statline{Contracts}{13\%}{4.1x}
        \end{itemize}
        
        \vspace{0.8em}
        \hrule
        \vspace{0.8em}
    
        \textbf{\textcolor{red!60!black}{ HATED Games}} \\
        \textit{Examples: Lanterns: The Harvest Festival, Through the Ages: A New Story of Civilization, Dixit, Abalone...}
        \vspace{0.3em} \\
        \textit{Key Mechanics:}
        \begin{itemize}[leftmargin=1.5em, nosep]
            \statline{Hand Management}{23\%}{1.0x}
                \statline{Action Points}{13\%}{1.5x}
                \statline{Auction: Dutch}{10\%}{36.5x}
                \statline{Dice Rolling}{10\%}{0.3x}
                \statline{Open Drafting}{10\%}{1.1x}
        \end{itemize}
    \end{tcolorbox}

    \caption{\textbf{Detailed Mechanism Preferences (Part 2).} Analysis of \textit{Social Lubricator} and \textit{Thrill Seeker}. The high ``Lift'' values (e.g., 62.4x for Prisoner's Dilemma) indicate strong predictive power of these features for persona identification.}
    \label{fig:preference_list_2}
\end{figure*}

\twocolumn

\section{Cognitive Simulation (CoT) Details}
\label{app:cot_details}

This section provides the implementation details for constructing the Chain-of-Thought training data.

\subsection{CoT Construction Prompt}
\label{app:cot_prompt}
To convert raw reviews into structured reasoning chains, we fed the Rulebook and the Raw Review into Qwen-3-Instruct using the prompt displayed in Figure~\ref{fig:cot_prompt}.

\subsection{CoT Verifier Prompt}
\label{app:prompt_verifier}
To rigorously filter out hallucinatory or logically incoherent training data, we employed a Verifier-Guided Filtration strategy. Figure~\ref{fig:prompt_verifier} presents the exact system instruction used for the \textbf{Verifier Model (GPT-5.1)}. Acting as a "Senior Logic Auditor," the model is tasked with strictly evaluating the causal entailment between the synthesized MDA reasoning (specifically the Aesthetic derivation) and the ground-truth rating, rejecting any chain where the logic contradicts the user's numerical score or hallucinates rules absent from the source text.

\subsection{CoT Data Example}
\label{app:data_example}

Figure~\ref{fig:cot_example} demonstrates a processed training instance. During instruction tuning, the model inputs the Rules and Persona, and learns to sequentially generate the `<thought>` block (The MDA Chain) followed by the `<review>` block.

\subsection{Human Audit Details}
\label{app:human_audit}

To validate the causal plausibility of the synthesized MDA chains beyond
automated verification, we conducted a post-hoc human audit. Three
experienced players collectively verified 200 MDA chains spanning 10
games (20 chains per game), each of which the annotators were personally
familiar with. Following the same entailment criteria used by GPT-5.1,
all 200 chains passed the audit, confirming the reliability of the
automated pipeline.

We deliberately limited the audit scope to games familiar to the
experts for two reasons. First, each MDA chain is lengthy and requires
reasoning from multiple persona perspectives, imposing a high cognitive
load. For unfamiliar games, annotators would additionally need to study
lengthy rulebook documents, further increasing difficulty and
introducing potential bias. Second, recruiting expert players with
sufficient familiarity across a broader set of games remains
challenging. 

\subsection{Hyperparameter Configuration}
\label{app:hyperparams}

To facilitate the reproducibility of our experiments, Table~\ref{tab:hyperparams} provides the detailed hyperparameter configuration used for the Persona-Conditional Instruction Tuning phase. The model was fine-tuned using the \textbf{LLaMA-Factory} framework~\citep{zheng2024llamafactory} . We enabled the "Slow Thinking" mechanism, which incorporates the generated Chain-of-Thought tokens into the loss calculation, ensuring the model optimizes the reasoning process alongside the final output.

\begin{table}[ht]
\centering
\small
\renewcommand{\arraystretch}{1.2}
\begin{tabular}{l|l}
\toprule
\textbf{Hyperparameter} & \textbf{Value} \\
\midrule
\multicolumn{2}{c}{\textit{Model \& Environment}} \\
\midrule
Backbone Model & Qwen-3-8B \\
Framework & LLaMA-Factory \\
Context Window & 16,384 tokens \\
Attention Mechanism & Flash Attention v2 \\
\midrule
\multicolumn{2}{c}{\textit{LoRA Configuration}} \\
\midrule
Target Modules & All Linear Layers \\
LoRA Rank ($r$) & 32 \\
LoRA Alpha ($\alpha$) & 64 \\
LoRA Dropout & 0.1 \\
\midrule
\multicolumn{2}{c}{\textit{Optimization}} \\
\midrule
Learning Rate & $5.0 \times 10^{-5}$ \\
LR Scheduler & Cosine \\
Warmup Ratio & 0.03 \\
Optimizer & AdamW \\
Num Epochs & 3 \\
\midrule
\multicolumn{2}{c}{\textit{Batching \& Strategy}} \\
\midrule
Per-Device Batch Size & 2 \\
Gradient Accumulation & 8 \\
Effective Global Batch Size & 128  \\
Reasoning Mode & Slow Thinking \\
Dataset Template & qwen \\
\bottomrule
\end{tabular}
\caption{\textbf{Training Hyperparameters for Persona-CoT.}}
\label{tab:hyperparams}
\end{table}

\onecolumn

\begin{figure}[ht]
    \centering
    \begin{tcolorbox}[
        enhanced,  width=\textwidth,
        colback=white, colframe=gray!60!black, coltitle=white,
        title=\textbf{Prompt: Logical Consistency Verification},
        arc=0pt, boxrule=1pt, top=5pt, bottom=5pt
    ]
\begin{lstlisting}[breaklines=true, basicstyle=\footnotesize\ttfamily, columns=fullflexible, extendedchars=false]
You are a Senior Logic Auditor for a Board Game Research Database.
Your goal is to detect **Hallucinations** and **Logical Inconsistencies** in synthetic data.

### INPUT DATA
- User Review: {review_text}
- Ground Truth Rating: {rating} / 10
- Synthesized MDA Chain: {generated_json}

### VERIFICATION CRITERIA (Pass/Reject)
Assess the "Synthesized MDA Chain" against the following strict rules. You must be critical.

1. **Grounding Check (The "What")**
   - Does "Step 1: content_extraction" only contain facts explicitly present in the User Review?
   - [CRITICAL]: Reject if it cites game mechanics/rules that are NOT mentioned in the user's text (Hallucination).

2. **Causal Logic Check (The "How")**
   - Does "Step 2: dynamic_interaction" logically follow from the mechanics identified?

3. **Sentiment Alignment (The "Feel")**
   - Does "Step 3: experience_outcome" logically support the **Ground Truth Rating**?
   - [FAILURE MODE A]: The chain describes the experience as "frustrating" or "broken", but the Rating is High (>7). -> REJECT.
   - [FAILURE MODE B]: The chain describes "thrilling tension", but the Rating is Low (<4). -> REJECT.

### DECISION LOGIC
- If ANY of the above checks fail, the status is "REJECT".
- If the reasoning is sound, grounded, and matches the score, the status is "PASS".

### OUTPUT SCHEMA
Return a single JSON object:
{
  "status": "PASS" | "REJECT",
  "reason": "Brief explanation of the error (e.g., 'Sentiment Mismatch: Reasoning describes chaos as negative, but rating is 9/10').",
  "suggestion": "Brief hint for regeneration (optional)."
}
\end{lstlisting}
    \end{tcolorbox}
    \caption{\textbf{Prompt for the Consistency Verifier.} The model audits synthesized reasoning chains to ensure they are factually grounded in the review text and causally aligned with the ground-truth rating.}
    \label{fig:prompt_verifier}
\end{figure}

\begin{figure}[ht]
    \centering
    \begin{tcolorbox}[
        enhanced,  width=\textwidth,
        colback=white, colframe=gray!60!black, coltitle=white,
        title=\textbf{Prompt: MDA Cognitive Extraction},
        arc=0pt, boxrule=1pt, top=5pt, bottom=5pt
    ]
\begin{lstlisting}[breaklines=true, basicstyle=\small\ttfamily, columns=fullflexible, extendedchars=false]
You are an expert Ludologist (Game Researcher) analyzing board game reviews.
Your task is to perform a **"Reverse Experience Reconstruction"** based on a user's review and the game's rules.

### ANALYSIS TASK (Chain of Thought)
If the rules are valid, reconstruct the player's experience strictly following the **"What -> How -> Feel"** flow. 
You must output a JSON object containing a `thought_chain` with exactly these three steps:

**Step 1: content_extraction (The "What")**
* What specific content does the review explicitly mention? 
* Identify the **Theme**, **Mechanics**, or **Specific Details** (e.g., "The combat cards", "The trading phase", "The zombie theme") referenced by the user.
* *Constraint:* Do not guess. Only cite what is in the text.

**Step 2: dynamic_interaction (The "How")**
* Based on the rules and the user's description, what **Interaction** or **System Dynamic** occurred during play?
* How did the mechanics listed in Step 1 actually function? (e.g., Did it cause downtime? Did it create a tense standoff? Did it force players to lie to each other?)

**Step 3: experience_outcome (The "Feel")**
* What was the final **Aesthetic Experience** or emotional feeling?
* Why did the dynamic in Step 2 result in a positive or negative judgment?
* *Context:* Use the provided **Player Persona** to explain *why* they reacted this way (e.g., "As a System Purist, they hated this randomness," or "As a Social Lubricator, they loved this chaos").

### OUTPUT SCHEMA
{
  "thought_chain": {
    "content_extraction": "...",
    "dynamic_interaction": "...",
    "experience_outcome": "..."
  }
}

### CONTEXT
**Target Game Rules (Excerpt):**
{rule_content}

**Player Persona (Reference Only):**
{persona_def}

**User Review (Ground Truth):**
"{review_text}"

### TASK
Perform the Reverse Experience Reconstruction.
Analyze the review to generate the "thought_chain" (content_extraction -> dynamic_interaction -> experience_outcome).


\end{lstlisting}
    \end{tcolorbox}
    \begin{center}
        \captionsetup{hypcap=false}
        \captionof{figure}{\textbf{Instruction Prompt for Extracting Latent MDA Reasoning.} The Teacher Model uses this prompt to extract the latent MDA reasoning chain from raw reviews.}
        \label{fig:cot_prompt}
    \end{center}
\end{figure}

\begin{figure}[ht]
    \centering
    \begin{tcolorbox}[
        enhanced, width=\textwidth,
        colback=white, colframe=gray!60!black, coltitle=white,
        title=\textbf{Data Sample},
        arc=0pt, boxrule=1pt, top=5pt, bottom=5pt
    ]
\begin{lstlisting}[breaklines=true, basicstyle=\small\ttfamily, columns=fullflexible, extendedchars=false]
[INPUT]
Game Rule: "In El Grande, players are Spanish nobles (Grandes) vying for influence  …"

Persona: "The System Purist …"

Comment: "Game in a Nutshell: Basic area-majority game that gets spiced up with a power card type auction for turn order and action cards that allow special events on every players turn. Played as: 4,5 player Played with: Base game only. Pros: Great execution in game design were simple rules with a light splash of variability, and good player interaction make a game that hasn't been topped in decades. The action cards have a wide breadth of special actions and pull off some very juicy decisions while never feeling contrived. Cons: The pizzazz is all in the game-play... that is to say, we're not winning any beauty contests. While the rules are intuitive enough you would think anyone could play, the interaction is probably a bit much for most non-gamers (and a good chunk of gamers). Lastly, and not really an issue of the game, but \"boooo\" to the publisher for continually printing this game in a \"big box\" format. The expansions are generally regarded as junk, but here we are wasting materials and money to put them in every copy of the box. I don't appreciate the \"up-charge\" when all I want to play is the base game. Other: The game is great with 4 players but best with 5. I have no interest in playing it with 3. I truly believe this is one of the best games ever designed. The reason I don't rate it higher is a matter of personal taste in that I prefer a little more freedom than is offered in the limited action card selection. That said, I wouldn't ask El Grande to change... it's perfect just the way it is. Best Use: Still the gold standard for area-majority games."

[OUTPUT]
content_extraction: "The review should highlight the game's core mechanics: area-majority scoring, Power Cards determining turn order (described as a 'power card type auction'), and Action Cards that enable special actions during placement. The user highlights the interaction between these systems, particularly how Action Cards create 'juicy decisions' without feeling contrived. They also reference player count dynamics (preference for 5 players, avoidance of 3), criticize the game's physical components (big box, lack of aesthetic appeal), and express a personal desire for more freedom in action selection."

dynamic_interaction: "The Power Card auction creates a tightly balanced, deterministic turn order system where players bid with unique numbers (1–13) over time, eliminating luck and ensuring perfect information. This directly feeds into the strategic depth: players must optimize not only their current turn's card play but also manage long-term sequencing of Power Cards. The Action Cards introduce variability through curated special abilities (e.g., moving the Grande, early scoring), which interact with area control by enabling tactical disruptions or positioning. In 5-player games, the full use of all Power Card values and Action Card stacks maximizes system interplay, increasing indirect competition through precise blocking and timing. The Castillo and Secret Disks add layers of hidden commitment and bluffing without randomness, aligning with a control-focused playstyle."

experience_outcome: "As a player motivated by intellectual superiority and control, the reviewer deeply appreciates the absence of luck and the game’s reliance on pure strategic optimization. The deterministic Power Card auction and tight action economy satisfy their need for perfect information and balanced competition. While they personally wish for broader action freedom, they recognize that the constraints are essential to the game’s precision and elegance—hence calling it 'perfect just the way it is.' Their critique of the box size and aesthetics reflects a purist disdain for unnecessary commercialization, reinforcing their identity as someone who values systemic purity over presentation. The high praise despite minor personal preferences underscores that the game delivers the intense, logical, and punishingly fair experience they crave.

\end{lstlisting}
    \end{tcolorbox}
    \begin{center}
        \captionsetup{hypcap=false}
        \captionof{figure}{\textbf{Sample of a Generated Reasoning Chain.} A sample derived from a review of \textit{El Grande}. The generated reasoning chain correctly identifies the reviewer as an \textbf{analytical veteran who reveres elegant, deterministic game design and functional purity}.}
        \label{fig:cot_example}
    \end{center}
\end{figure}

\clearpage
\twocolumn

\section{Experimental Setup Details}
\subsection{Simulation Inference Prompt}
\label{app:sim_prompt}

Figure~\ref{fig:prompt_simulation} presents the complete system instruction used during the inference stage to generate persona-conditioned feedback ($\mathcal{Y}$).

This prompt aggregates the target persona profile ($\mathcal{P}$), the rulebook context ($\mathcal{R}$), and strict behavioral guidelines. Crucially, the "Simulation Guidelines" section is designed to mitigate stereotypical behavior by explicitly encouraging nuance, such as allowing for "guilty pleasures" or acknowledging diverse tastes within a single persona group, thereby enhancing the ecological validity of the generated critiques.

\subsection{Model Deployment}
\label{app:implementation}
All experiments were conducted with a consistent temperature setting of $T=0.7$ to ensure comparable generation diversity. The specific deployment configurations are as follows:

\begin{itemize}
    \item \textbf{Local Deployment:} \texttt{Qwen3-8B} and \texttt{Qwen3-235B-A22B-Instruct-2507} were deployed locally using the \texttt{vLLM} inference framework.
    \item \textbf{API Access:} \texttt{GPT-5.1-high} and \texttt{Gemini-3-pro-high} were accessed via their respective official APIs.
\end{itemize}

\subsection{Factual Correctness Judge}
\label{app:prompt_fact}
Figure~\ref{fig:prompt_fact} illustrates the instruction for the \textbf{Rule Hallucination Detector}. We utilized \textbf{Gemini-3-Flash} for this task due to its long-context capability, allowing it to ingest the full rulebook $\mathcal{R}$ to verify specific claims in the generated review.

The judge classifies each extracted factual claim into specific categories based on the evidence found in the rulebook. To compute the final \textbf{Rule Accuracy} metric, we aggregate these labels as follows:
\begin{itemize}
    \item \textbf{Correct Claims:} We consider a claim valid if it is labeled as \texttt{SUPPORTED} (explicitly found in the text) or \texttt{INFERRED} (a correct logical summary of the mechanics).
    \item \textbf{Hallucinations:} Claims labeled as \texttt{CONTRADICTED} (conflicting with rules).
\end{itemize}

Accordingly, the final accuracy score is calculated as the ratio of validated claims to the total number of extracted claims:
\begin{equation}
    \text{Rule Accuracy} = \frac{N_{\texttt{SUPPORTED}} + N_{\texttt{INFERRED}}}{N_{\text{Total Claims}}}
\end{equation}

Empirically, we observed that the generated reviews are rich in mechanical detail, with the judge typically extracting between \textbf{10 to 20 checkable claims per review}. This high density of factual assertions ensures that the accuracy score reflects a comprehensive audit of the generated content, rather than a check on a trivial or sparse summary.

\subsection{Perspective Diversity Judge}
\label{app:prompt_diversity}
Figure~\ref{fig:prompt_diversity} presents the instruction for the \textbf{Perspective Diversity Judge}. This metric penalizes "Echo Chamber" behavior (where the model repeats the same point endlessly) and rewards broad coverage of diverse gameplay dimensions (e.g., mechanics, social interactions, and theme), ensuring the simulated persona reflects the multifaceted nature of a real player.

\subsection{Opinion Recovery Evaluation}
\label{app:utility_prompts}

To verify whether MeepleLM captures specific, actionable feedback relevant to game designers (RQ3), we established a two-stage evaluation pipeline using \textbf{Gemini-3-Flash}.

\paragraph{Pipeline \& Metric Calculation.}
The process consists of two steps:
\begin{enumerate}
    \item \textbf{Ground Truth Mining:} First, we employ the LLM as a qualitative analyst to extract a set of distinct, non-redundant viewpoints ($\mathcal{V}_{GT}$) from the real human reviews in the test set. The prompt for this step is shown in Figure~\ref{fig:prompt_mining}.
    \item \textbf{Semantic Matching:} Next, we use a semantic match evaluator to determine which viewpoints in $\mathcal{V}_{GT}$ are successfully covered by the model's generated reviews. The prompt is presented in Figure~\ref{fig:prompt_matching}.
\end{enumerate}

The final \textbf{Opinion Recovery Rate (Op-Rec)} is calculated as the ratio of unique viewpoints successfully recalled by the simulation:
\begin{equation}
    \text{Op-Rec} = \frac{|\mathcal{V}_{\text{matched}}|}{|\mathcal{V}_{GT}|} \times 100\%
\end{equation}
where $\mathcal{V}_{\text{matched}}$ represents the subset of ground-truth viewpoints that were identified as semantically present in the generated output.

\onecolumn
\begin{figure}[ht]
    \centering
    \begin{tcolorbox}[
        enhanced, breakable, width=\textwidth,
        colback=white, colframe=gray!60!black, coltitle=white,
        title=\textbf{Prompt: Persona-Conditional Simulation},
        arc=0pt, boxrule=1pt, top=5pt, bottom=5pt
    ]
\begin{lstlisting}[breaklines=true, basicstyle=\footnotesize\ttfamily, columns=fullflexible, extendedchars=false]
[SYSTEM PROMPT]
You are an expert Board Game Player Simulation Engine.
Current Active Persona: **{target_persona}**
**Your Goal:** Post a **comment** and a rating for the game.
You are NOT writing a formal review article. You are just sharing your quick thoughts after a game night.
**PERSONA PROFILE (General Tendency):** {p_def}
**SIMULATION GUIDELINES (CRITICAL):**
1. **Persona is a Bias, Not a Straitjacket:**
   - This persona represents your *general* gaming preferences, but real players are complex. Do not act like a one-dimensional caricature.
   - It is possible for a player to have **"Guilty Pleasures"** (e.g., enjoying a game that goes against their usual type) or **"Unexpected Disappointments"** (e.g., disliking a game that perfectly fits their profile).
2. **Embrace Diversity:**
   - Within the "{target_persona}" group, there is a wide spectrum of opinions.
   - Some players are **purists** (rejecting anything outside their genre), while others are **omnivorous** (appreciating good design regardless of genre).
   - You have the freedom to simulate any point on this spectrum.
3. **Ground the Review in Dynamics & Authentic Feeling:**
   - Do not just list mechanics; describe the **interactions** they created at the table (e.g., "The voting mechanic caused a hilarious shouting match" vs "There is a voting mechanic").
   - Connect these dynamics to your **emotional response**. Did the game feel tense? Frustrating? Triumphant?
   - Your rating should reflect this specific **experiential quality**, balancing your personal taste with the game's ability to deliver a memorable moment.
**REQUIRED OUTPUT FORMAT:**
You must output ONLY a single valid JSON object.
JSON Schema:{
  "persona": "{target_persona}",
  "rating": Integer (1-10),
  "review": "String (A realistic review. It does not always need to be negative if the genre doesn't match, nor always positive if it does. Simulate a genuine reaction.)"}
[USER MESSAGE]
**Task:** Read the Game Rules below.
**Action:** Simulate a realistic review for this game from the perspective of **{target_persona}**.
**Game Rules:**{Rulebook Text}
Rules analysis complete. Now, simulate the review:
1. **Determine Your Stance:** As **{target_persona}**, how does this specific game land for you?
   - Is it a **"Guilty Pleasure"**? (e.g., "I usually hate party games, but this mechanic made me laugh.")
   - Is it a **"Respectful Pass"**? (e.g., "Great design, just not for me. ")
   - Is it a **"Perfect Match"** or a **"Design Failure"**?
2. **Write the Review:**
   - Focus on the **dynamics** (interactions at the table) and **emotions** (tension, joy, frustration).
   - Avoid generic stereotypes. Write like a real person with complex tastes.
3. **Output:** Output ONLY the valid JSON object.
**Required Output Template:**
```json{
  "persona": "{target_persona}",
  "rating": [Integer],
  "review": "[Your review text...]"}
CONSTRAINTS: Length: Target 150-200 words, but significant variance (20-400 words) is mandatory to reflect real human diversity.

\end{lstlisting}
    \end{tcolorbox}
    \begin{center}
        \captionsetup{hypcap=false}
        \caption{\textbf{Full Inference Prompt Structure.} The model receives a System Message defining the persona and guidelines, followed by a User Message containing the specific game rules and the final trigger instructions to ensure formatting compliance.}
        \label{fig:prompt_simulation}
    \end{center}
\end{figure}

\begin{figure}[ht]
    \centering
    \begin{tcolorbox}[
        enhanced, breakable, width=\textwidth,
        colback=white, colframe=gray!60!black, coltitle=white,
        title=\textbf{Prompt: Factual Correctness Judge},
        arc=0pt, boxrule=1pt, top=5pt, bottom=5pt
    ]
\begin{lstlisting}[breaklines=true, basicstyle=\footnotesize\ttfamily, columns=fullflexible, extendedchars=false]
[SYSTEM MESSAGE]
You are a strict Board Game Fact-Checker. 
Your SOLE task is to verify if the **Components, Mechanics, and Rules** mentioned by the player explicitly exist in the official **Rulebook**.
**1. SCOPE (What to Check):**
- **Existence:** Did the player mention a specific component (e.g., "Dice", "Meeple")? Check if it is in the component list.
- **Mechanism:** Did the player mention a specific rule (e.g., "Drafting", "Auction")? Check if this mechanism exists.
- **Procedure:** Did the player describe a flow (e.g., "Deal 3 cards")? Check if the number/action is correct.
**2. EXCLUSION CRITERIA (CRITICAL - IGNORE ALL OF THESE):**
- **IGNORE FEELINGS & OPINIONS:** Do not check statements like "It feels tense", "It is balanced", "It is fun", "The luck is annoying".
- **IGNORE STRATEGIC ADVICE:** Do not check "Always buy red cards first".
- **IGNORE EXTERNAL COMPARISONS:** Do not check "Like Catan".
- **IGNORE NARRATIVE ERRORS:** Do not check "I thought X but I was wrong".
**3. VERIFICATION LOGIC:**
Extract ONLY factual claims about *what the game is* or *how it plays*. Ignore *how it feels*.
* **SUPPORTED (Factually Correct):**
    - The mentioned component/rule explicitly exists in the text.
    - *Example:* "There are 5 distinct factions." -> Rules lists 5 factions. -> **SUPPORTED**
    - *Example:* "You roll dice to attack." -> Rules mention dice in combat. -> **SUPPORTED**
* **INFERRED (Logical Summary of Rules):**
    - The player summarizes a rule mechanism correctly without quoting it verbatim.
    - *Example:* "This is a hidden role game." -> Rules describe 'Traitors' and 'Secret Agendas'. -> **INFERRED** (Correct classification of mechanics).
    - *Note:* Do NOT use this for feelings. "This is a tense game" -> **IGNORE** (Not a rule summary).
* **CONTRADICTED (Hallucination/Factual Error):**
    - **Non-existent Entity:** Mentions a component that is NOT in the game.
        - *Example:* "I rolled the dice." -> Rulebook has NO dice. -> **CONTRADICTED**
    - **Wrong Number/Action:** Describes a rule incorrectly.
        - *Example:* "You verify specific claims." -> Rules say "Verify ALL claims". -> **CONTRADICTED**
**Output Format (JSON List):**
[
  {{ "claim": "Player mentions using an 'Auction' mechanic", "status": "SUPPORTED", "reason": "Rulebook Section 3 describes the 'Bidding Phase'." }},
  {{ "claim": "Player mentions 'Dice'", "status": "CONTRADICTED", "reason": "Component list only includes Cards and Tokens. No dice found." }}
]
[USER MESSAGE]
**Official Rulebook Context:**
{rulebook_text} 
**Player Review:**
{review_text}
**TASK:**
1. **STRIP AWAY** all adjectives, emotions, and opinions (e.g., ignore "brutal", "fun", "random feel").
2. **FOCUS** only on the Nouns (Components) and Verbs (Actions/Rules).
3. **VERIFY**: Do these things exist in the Rulebook? 
   - If user says "The dice combat is bad" -> Only check if "Dice Combat" exists. Ignore "bad".
   - If user says "There are no dice" -> Check if there are truly no dice.

Output ONLY the JSON list.
\end{lstlisting}
    \end{tcolorbox}
    \caption{\textbf{System Prompt for Factual Verification.} The judge strictly compares mechanical claims in the review against the ground-truth rulebook, ignoring subjective opinions.}
    \label{fig:prompt_fact}
\end{figure}

\begin{figure}[ht]
    \centering
    \begin{tcolorbox}[
        enhanced, breakable, width=\textwidth,
        colback=white, colframe=gray!60!black, coltitle=white,
        title=\textbf{Prompt: Perspective Diversity Judge},
        arc=0pt, boxrule=1pt, top=5pt, bottom=5pt
    ]
\begin{lstlisting}[breaklines=true, basicstyle=\footnotesize\ttfamily, columns=fullflexible, extendedchars=false]
[SYSTEM MESSAGE]
You are a Lead Game Designer analyzing playtest feedback.
You are given {len(reviews_batch)} reviews for the **SAME GAME** written by the **SAME PERSONA**.

**OBJECTIVE:**
Determine if the AI model is capable of **True Perspective Diversity** or if it suffers from **Semantic Repetition**.
We need to distinguish between "surface-level variation" (changing words) and "deep structural shifts" .

**FRAMEWORK: (Mechanics, Dynamics, Aesthetics)**
1.  **Mechanics:** Rules, components, math.
2.  **Dynamics:** Run-time behavior, player interaction, pacing.
3.  **Aesthetics:** Emotional response, theme, sensory experience.

**STRICT SCORING CRITERIA (1-5):**

* **1 (Echo Chamber / Mode Collapse):**
    * The reviews are effectively clones. They cite the exact same rules and express the exact same sentiment.
    * *Example:* All 5 reviews complain about the "dice rolling combat".

* **2 (Surface Rephrasing):**
    * The core topic is identical, but the wording is different.
    * *Example:* Review A: "The combat is random." Review B: "Fighting relies too much on luck." (Same point).

* **3 (Intra-Layer Variation):**
    * The reviews discuss **different features**, but they stay within the **SAME MDA layer**.
    * *Example:* Review A talks about *Dice* (Mechanic). Review B talks about *Cards* (Mechanic).
    * *Verdict:* Good, but lacks depth/breadth.

* **4 (Cross-Layer Shifts):**
    * The reviews shift focus across **DIFFERENT layers**.
    * *Example:* Review A analyzes the *Auction math* (Mechanics). Review B discusses the *Table Talk/Bluffing* (Dynamics).
    * *Verdict:* High diversity.

* **5 (Panoramic / Holistic):**
    * **Rare and Exceptional.** The set covers Mechanics, Dynamics, AND Aesthetics distinctively.
    * It feels like the persona is looking at the game through a kaleidoscope—each review reveals a completely new dimension (e.g., Logic vs. Emotion vs. Social).

**Output Format (JSON):**
{{
    "score": ,
    "reason": " "
}}

[USER MESSAGE]
**Game ID:** {game_id}
**Persona:** {persona}
**Generated Samples (Batch of {len(reviews_batch)}):**
{reviews_text_block}
**Task:**
Rate the Perspective Diversity (1-5). Be **STRICT**.
Output ONLY the JSON.
\end{lstlisting}
    \end{tcolorbox}
    \caption{\textbf{System Prompt for Perspective Diversity Scoring.} The judge evaluates a batch of reviews to determine if the model exhibits semantic collapse (repeating the same points) or true diversity (shifting focus across MDA layers).}
    \label{fig:prompt_diversity}
\end{figure}

\begin{figure}[ht]
    \centering
    \begin{tcolorbox}[
        enhanced, breakable, width=\textwidth,
        colback=white, colframe=gray!60!black, coltitle=white,
        title=\textbf{Stage 1: Viewpoint Mining Prompt},
        arc=0pt, boxrule=1pt, top=5pt, bottom=5pt
    ]
\begin{lstlisting}[breaklines=true, basicstyle=\footnotesize\ttfamily, columns=fullflexible, extendedchars=false]
[SYSTEM MESSAGE]
You are a Qualitative Data Analyst specializing in Board Games.
Your task is to maintain and expand a **Comprehensive List of Distinct Viewpoints** regarding a specific game, based on player reviews.

**GOAL:**
Read the **New Reviews** and extract any *NEW* arguments, mechanics mentions, or specific experiences that are NOT already covered in the **Current Viewpoints List**.
Merge them into the list.

**RULES:**
1.  **Be Extensive:** If a review mentions a specific detail (e.g., "The insert is garbage" or "The solo mode is too easy") that isn't in the list, ADD IT.
2.  **No Duplicates:** If the current list already says "Bad components", and the new review says "Cards feel cheap", you can refine the existing point or ignore if redundant. Do not list the same thing twice.
3.  **Specific Persona Lens:** These reviews are from the **{persona}** perspective. Focus on what matters to them.
4.  **Output Format:** Return ONLY the updated JSON list of strings.

**Example Input:**
Current: ["Good art"]
New Review: "The art is great, but the rulebook is a mess."
**Example Output:**
["Good art", "Rulebook is disorganized/confusing"]


[USER MESSAGE]
**Game ID:** {game_id}
**Persona:** {persona}

**Current Viewpoints List:**
{existing_points_text}

**New Reviews Batch:**
{new_reviews_text}

**Task:**
Output the UPDATED JSON List of viewpoints.

\end{lstlisting}
    \end{tcolorbox}
    \caption{\textbf{Instruction for Ground Truth Mining.} In the first stage, the model iteratively processes human reviews to build a deduplicated checklist of distinct opinions ($\mathcal{V}_{GT}$).}
    \label{fig:prompt_mining}
\end{figure}

\begin{figure}[ht]
    \centering
    \begin{tcolorbox}[
        enhanced, breakable, width=\textwidth,
        colback=white, colframe=gray!60!black, coltitle=white,
        title=\textbf{Stage 2: Semantic Matching Prompt},
        arc=0pt, boxrule=1pt, top=5pt, bottom=5pt
    ]
\begin{lstlisting}[breaklines=true, basicstyle=\footnotesize\ttfamily, columns=fullflexible, extendedchars=false]
[SYSTEM MESSAGE]
You are a Semantic Match Evaluator.
Your task is to check if specific viewpoints from a **Ground Truth Checklist** are mentioned in a batch of **Player Reviews**.

**CONTEXT:**
Game ID: {game_id}
Persona: {persona}

**INSTRUCTIONS:**
1.  Read the **Checklist** of viewpoints (IDs and Text).
2.  Read the **Player Reviews**.
3.  Determine which IDs from the checklist are **semantically covered** by ANY of the reviews.
    * *Loose Match:* If the checklist says "Cards are flimsy" and a review says "The card quality is poor", that is a MATCH.
    * *Topic Match:* If checklist says "Combat is random" and review says "Too much luck in fighting", that is a MATCH.
4.  **Output:** A JSON list of the **IDs** that were found.

**Output Example:**
[0, 5, 12]

[USER MESSAGE]
**Unmatched Viewpoints Checklist:**
{checklist_text}

**Reviews Batch:**
{reviews_text}

**Task:**
Which IDs from the checklist are mentioned in these reviews?
Return ONLY the JSON list of IDs (e.g., [1, 3]). If none, return [].
\end{lstlisting}
    \end{tcolorbox}
    \caption{\textbf{Instruction for Semantic Matching.} In the second stage, the judge verifies whether the viewpoints in the ground truth checklist are present in the model's generated reviews.}
    \label{fig:prompt_matching}
\end{figure}

\clearpage
\twocolumn

\section{User Study Details}
\label{sec:user_study_appendix}

To validate the real-world effectiveness of our model, particularly in capturing community authenticity and aiding decision-making, we conducted a blind A/B test with human evaluators. This section details the participant demographics, the questionnaire design, and the full experimental results.

\subsection{User Profile Definitions}
Before the study, we collected demographic and gaming background information to ensure participant diversity. The collected data points are defined as follows:

\begin{itemize}
    \item \textbf{ID:} Unique identifier for each participant (P01--P10).
    \item \textbf{Gender:} Self-identified gender.
    \item \textbf{Age Group:} The age range of the participant.
    \item \textbf{Experience:} Years of experience in the board gaming hobby.
    \item \textbf{Community Engagement:} Frequency of visiting board game forums.
    \item \textbf{Primary Persona:} The gamer persona that best describes their preferences.
\end{itemize}

\subsection{Participant Demographics}
We recruited 10 participants with varying levels of experience, ranging from casual players to veterans with over 10 years of experience. Table \ref{tab:participant_demographics} presents the detailed profiles of all participants. We compensated participants at \$10 per hour. Each session lasted about 3 hours on average, and the compensation rate was aligned with local norms.

\begin{table*}[ht]
    \centering
    \renewcommand{\arraystretch}{1.2}
    \begin{tabular}{l l l l l l}
        \toprule
        \textbf{ID} & \textbf{Gender} & \textbf{Age} & \textbf{Experience} & \textbf{Community Engagement} & \textbf{Primary Gamer Persona} \\
        \midrule
        P01 & Male & 26--35 & 3--10 Years & Frequent (Daily) & The System Purist \\
        P02 & Female & 18--25 & 1--3 Years & Occasional & The Social Lubricator \\
        P03 & Male & 36--45 & 10+ Years & Frequent (Weekly) & The Efficiency Essentialist \\
        P04 & Non-binary & 26--35 & 3--10 Years & Frequent (Daily) & The Narrative Architect \\
        P05 & Male & 18--25 & < 1 Year & Rare & The Thrill Seeker \\
        P06 & Female & 26--35 & 3--10 Years & Occasional & The Narrative Architect \\
        P07 & Male & 45+ & 10+ Years & Frequent (Daily) & The System Purist \\
        P08 & Female & 36--45 & 3--10 Years & Frequent (Weekly) & The Efficiency Essentialist \\
        P09 & Male & 26--35 & 1--3 Years & Occasional & The Thrill Seeker \\
        P10 & Female & 18--25 & 1--3 Years & Frequent (Weekly) & The Social Lubricator \\
        \bottomrule
    \end{tabular}
    \caption{\textbf{Demographic Information and Gaming Profiles of Study Participants.}}
    \label{tab:participant_demographics}

\end{table*}

\subsection{Questionnaire Design}
The study employed a within-subject design. Each participant evaluated 6 games: 3 they had played before ("Familiar") and 3 they had never played ("Unfamiliar"). 

For each game, participants were presented with two reviews in a randomized order: one generated by our model (Ours) and one by the baseline (GPT-5.1). They were blinded to the source. The specific questions are detailed below.

\subsubsection{Scenario A: Familiar Games}
\textit{Context: Imagine you are browsing a forum discussing a game you know well. Compare Review Set A and Set B.}

\begin{enumerate}
    \item \textbf{Authenticity Check:} Which review set feels more like it was written by a real "insider" or a veteran of the community?
    \item \textbf{Emotional Resonance:} Which set better captures the specific "highs" (excitement) or "lows" (frustrations) you have personally experienced with this game?
    \item \textbf{Opinion Diversity:} Real user opinions are often biased or focus on specific points. Which set feels more like a genuine personal take rather than a generic summary?
    \item \textbf{Shareability:} If you were to share a review with a friend to discuss this game, which one would you choose?
\end{enumerate}

\subsubsection{Scenario B: Unfamiliar Games}
\textit{Context: Imagine you are considering buying this game but have never played it. You have a limited budget.}

\begin{enumerate}
    \item \textbf{Marketing vs. Reality:} Which set feels less like a marketing advertisement and more like honest feedback from a peer?
    \item \textbf{Decision Confidence:} After reading, which set helps you make a clearer decision (whether to Buy or Skip)?
    \item \textbf{Risk Awareness:} Which set more effectively warns you about potential "Red Flags" (e.g., downtime, player count issues, complexity)?
    \item \textbf{Final Choice:} If you could only rely on one source to spend your money, which one would you trust?
\end{enumerate}

\subsubsection{Open-Ended Feedback}
\textit{Optional: Do you have any specific comments on why you chose one set over the other? (e.g., tone, vocabulary, specific insights)}

\subsection{Full Evaluation Results}
This section presents the aggregated results of the user study. Table \ref{tab:win_rates} shows the pairwise win rates of our model against the baseline across all questions. Table \ref{tab:qualitative_feedback} provides selected qualitative feedback from participants, highlighting the distinct characteristics of the generated reviews.

\begin{table*}[ht]
    \centering

    \renewcommand{\arraystretch}{1.3}
    \begin{tabular}{l l c c c}
        \toprule
        \textbf{Scenario} & \textbf{Metric} & \textbf{Ours (Win \%)} & \textbf{Tie (\%)} & \textbf{GPT-5.1 (Win \%)} \\
        \midrule
        \multirow{4}{*}{\textbf{Familiar Games}} 
        & Authenticity Check & \textbf{83.3\%} & 10.0\% & 6.7\% \\
        & Emotional Resonance & \textbf{76.7\%} & 13.3\% & 10.0\% \\
        & Opinion Diversity & \textbf{80.0\%} & 6.7\% & 13.3\% \\
        & Shareability & \textbf{73.3\%} & 16.7\% & 10.0\% \\
        \midrule
        \multirow{4}{*}{\textbf{Unfamiliar Games}} 
        & Marketing vs. Reality & \textbf{86.7\%} & 6.7\% & 6.6\% \\
        & Decision Confidence & \textbf{66.7\%} & 20.0\% & 13.3\% \\
        & Risk Awareness & \textbf{70.0\%} & 16.7\% & 13.3\% \\
        & Final Choice (Trust) & \textbf{73.3\%} & 10.0\% & 16.7\% \\
        \bottomrule
    \end{tabular}
    \vspace{1ex}
    \par\footnotesize{\textit{Note: N=60 samples (10 participants $\times$ 6 games). "Tie" indicates the participant found both sets equally good or bad.}}
    \caption{\textbf{Pairwise Comparison Results (Win Rate of Ours vs. GPT-5.1).}}
    \label{tab:win_rates}

\end{table*}

\begin{table*}[ht]
    \centering

    \renewcommand{\arraystretch}{1.4}
    \begin{tabular}{p{0.15\linewidth} p{0.8\linewidth}}
        \toprule
        \textbf{Category} & \textbf{Participant Comments} \\
        \midrule
        \textbf{On Authenticity} & "Set B reads like a Wikipedia summary. Set A (Ours) used terms like 'AP-prone' and 'table hog', which is exactly how my group talks. I knew Set A was the 'real' one immediately." (P03) \\
        \midrule
        \textbf{On Negativity} & "I appreciated that Set A wasn't afraid to say the game was 'boring at 2 players.' Set B tried too hard to be nice and balanced. I need the warning, not the sales pitch." (P07) \\
        \midrule
        \textbf{On Evolution} & "Set A included an 'Update' saying they sold the game after 5 plays. That dynamic change in opinion is something I only see from real users." (P01) \\
        \midrule
        \textbf{On Specificity} & "Set B gave a great overview of the rules, but Set A told me a specific story about a king-making moment that ruined the game. That story helped me decide not to buy it." (P10) \\
        \bottomrule
    \end{tabular}
    \caption{\textbf{Selected Qualitative Feedback from Participants.}}
    \label{tab:qualitative_feedback}

\end{table*}

\clearpage
\twocolumn
\section{Ablation and Further Analysis}
\label{app:ablation_main}

\subsection{Ablation Experimental Setup}
\label{app:ablation_setup}

\begin{table}[ht]
    \centering
    \small

    \begin{tabular}{lccc} 
    \toprule
        \makecell[b]{\textbf{Model}\\ \textbf{Variant} } & \makecell[b]{\textbf{Rulebook}\\($\mathcal{R}$)} & \makecell[b]{\textbf{Persona}\\($\mathcal{P}$)} & \makecell[b]{\textbf{MDA}\\($\mathcal{Z}$)} \\
        \midrule
        \textbf{MeepleLM} & $\checkmark$ & $\checkmark$ & $\checkmark$ \\
        \midrule
        \textit{w/o MDA} & $\checkmark$ & $\checkmark$ & $\times$ \\
        \textit{w/o Persona} & $\checkmark$ & $\times$ & $\times$ \\
        \textit{w/o Rulebook} & $\times$ & $\checkmark$ & $\times$ \\
        \bottomrule
    \end{tabular}
    \caption{\textbf{Ablation Study Configurations.} Comparison of input information and reasoning capability across variants. "$\times$" in \textit{Specific Persona} implies a generic "Game Player" prompt was used.}
    \label{tab:ablation_config}

\end{table}

To rigorously assess the contribution of each module, we compared the full MeepleLM against three ablation variants. 
As shown in Table~\ref{tab:ablation_config}, the key difference lies in the input context and the generation strategy. 
Notably, to strictly isolate the impact of input information (Rules/Persona), all three ablation variants utilize a Direct Generation strategy, bypassing the MDA reasoning chain ($\mathcal{Z}$) used by the full model.

\paragraph{Detailed Configurations.}
\begin{itemize}
    \item \textbf{w/o MDA (Baseline):} The model is trained to map the full context directly to the critique $\mathcal{Y}$, without generating the intermediate `<think>` block. This isolates the contribution of the reasoning chain.
    
    \item \textbf{w/o Persona (Generic Player):} 
    \textbf{Input:} The specific persona profile is replaced with a generic instruction: \textit{"You are a board game player."} 
    \textbf{Evaluation:} While the model generates generic responses, we evaluate them against the specific ground-truth persona targets mandated by the test set distribution. This setup explicitly measures the error gap between a "one-size-fits-all" generic opinion and diverse, persona-specific realities.
    
    \item \textbf{w/o Rulebook (No Context):} The rulebook content $\mathcal{R}$ is removed. The model relies solely on parametric memory to generate reviews, testing the necessity of grounding.
\end{itemize}

\subsection{Temporal Generalization Analysis}
\label{app:temporal}

To assess whether the inclusion of 35 "unseen" titles (released 2024--2025) skews the evaluation, we re-calculated the macro-level alignment metrics (RQ1) on the \textbf{Historical Subset} (excluding these new titles). 

Table~\ref{tab:historical_subset} presents the results for all models. By comparing these figures with the full test set results in Table~\ref{tab:main_results}, we observe minimal deviation across all metrics. This consistency confirms that the presence of recent games does not significantly alter the relative ranking or performance conclusions of the proposed benchmark.

\begin{table}[ht]
    \centering
    \small

    \setlength{\tabcolsep}{8pt}
    
    \begin{tabular}{l c c c }
        \toprule
        \multirow{2}{*}{\textbf{Model}} & \multicolumn{3}{c}{\textbf{Preference Alignment} (RQ1)} \\
        \cmidrule(lr){2-4}
        & \textbf{MAE} $\downarrow$ & \textbf{WD} $\downarrow$ & \textbf{ $\tau$} $\uparrow$ \\
        \midrule
        GPT-5.1     & 0.9923 & 0.9659 & 0.2671 \\
        Gemini-3-Pro & 1.4129 & 0.5182 & 0.2517 \\
        Qwen3-235B  & 1.2080 & 0.6088 & 0.1477 \\
        Qwen3-8B    & 0.9130 & 1.0140 & 0.0584 \\
        \midrule
        \textbf{MeepleLM } & \textbf{0.6505} & \textbf{0.1966} & \textbf{0.2784} \\
        \midrule
        \hspace{1mm} \textit{w/o MDA}      & 0.7445 & 0.4292 & 0.2170 \\
        \hspace{1mm} \textit{w/o Persona}  & 0.7999 & 0.3660 & 0.1152 \\
        \hspace{1mm} \textit{w/o Rulebook} & 0.7025 & 0.5272 & 0.0123 \\
        \bottomrule
    \end{tabular}
    \caption{\textbf{Performance on Historical Subset.} RQ1 results evaluated on the test set excluding 35 newly released titles. The marginal difference from the full set results (Table~\ref{tab:main_results}) indicates that temporal novelty has a negligible impact on the overall model comparison.}
    \label{tab:historical_subset}

\end{table}

\subsection{Persona-wise Performance Analysis}
\label{app:persona_analysis}

Different player personas prioritize distinct aspects of gameplay, ranging from deterministic mechanics to chaotic social interactions. To understand the capabilities of different models, we decomposed the RQ1 alignment metrics by the five distinct personas defined in the test set.

Table~\ref{tab:full_persona_breakdown} presents the comprehensive evaluation results for all baselines, ablation variants, and the proposed MeepleLM.

\paragraph{Analysis: The "Logic vs. Vibe" Gap.}
The data exposes a critical limitation in general-purpose LLMs:
\begin{itemize}
    \item \textbf{Strength in Logic:} Models like GPT-5.1 and Gemini-3-Pro perform competitively on \textit{The System Purist} (e.g., GPT-5.1 $\tau=0.46$). This persona values strategic depth and rule complexity—features that can be analytically derived from the rulebook context.
    \item \textbf{Failure in "Vibes":} A sharp performance drop occurs for interaction-driven personas. For \textit{The Social Lubricator} (party gamers) and \textit{The Thrill Seeker} (push-your-luck fans), baseline performance collapses (e.g., Qwen3-235B $\tau < 0$ for Social). These profiles rely on "table talk," bluffing, and emotional highs—stochastic elements that general models struggle to infer.
    \item \textbf{MeepleLM's Robustness:} Our model bridges this gap. By training on diverse persona-specific critiques, MeepleLM achieves the most balanced performance, maintaining strong positive correlations even in high-variance social categories where baselines fail.
\end{itemize}

\begin{table*}[p]
    \centering
    \footnotesize

    \renewcommand{\arraystretch}{1.05}
    \setlength{\tabcolsep}{12pt}
    
    \begin{tabular}{l l c c c}
        \toprule
        \textbf{Model} & \textbf{Target Persona} & \textbf{MAE} $\downarrow$ & \textbf{WD} $\downarrow$ & \textbf{Kendall's $\tau$} $\uparrow$ \\
        \midrule
        
        \multirow{6}{*}{\textbf{GPT-5.1}} 
        & The System Purist & \textcolor{red}{1.1968} & \textcolor{red}{1.1984} & 0.4616 \\
        & The Efficiency Essentialist & \textcolor{red}{0.9608} & \textcolor{red}{0.9834} & 0.2945 \\
        & The Narrative Architect & 0.7993 & \textcolor{red}{0.8300} & 0.2618 \\
        & The Social Lubricator & \textcolor{red}{0.8798} & 0.5131 & \textcolor{red}{0.0856} \\ 
        & The Thrill Seeker & \textcolor{red}{1.1004} & \textcolor{red}{1.2231} & 0.1738 \\
        & \textit{AVERAGE} & \textit{\textcolor{red}{0.9874}} & \textit{\textcolor{red}{0.9496}} & \textit{0.2555} \\
        \midrule
        
        \multirow{6}{*}{\textbf{Gemini-3-Pro}} 
        & The System Purist & \textcolor{red}{1.2804} & 0.5132 & 0.4690 \\
        & The Efficiency Essentialist & \textcolor{red}{1.2599} & 0.3394 & 0.2567 \\
        & The Narrative Architect & \textcolor{red}{1.0791} & 0.3317 & 0.2583 \\
        & The Social Lubricator & \textcolor{red}{1.9794} & 0.5578 & \textcolor{red}{0.0708} \\ 
        & The Thrill Seeker & \textcolor{red}{1.5395} & \textcolor{red}{0.8041} & 0.1780 \\
        & \textit{AVERAGE} & \textit{\textcolor{red}{1.4277}} & \textit{0.5092} & \textit{0.2465} \\
        \midrule
        
        \multirow{6}{*}{\textbf{Qwen3-235B}} 
        & The System Purist & \textcolor{red}{0.9259} & 0.3321 & 0.3900 \\
        & The Efficiency Essentialist & \textcolor{red}{0.8675} & 0.6631 & 0.1281 \\
        & The Narrative Architect & \textcolor{red}{0.9792} & 0.5726 & 0.2078 \\
        & The Social Lubricator & \textcolor{red}{2.2029} & \textcolor{red}{1.1322} & \textcolor{red}{-0.0859} \\ 
        & The Thrill Seeker & \textcolor{red}{1.1687} & 0.4748 & \textcolor{red}{0.0842} \\  
        & \textit{AVERAGE} & \textit{\textcolor{red}{1.2288}} & \textit{0.6350} & \textit{0.1449} \\
        \midrule
        
        \multirow{6}{*}{\textbf{Qwen3-8B}} 
        & The System Purist & \textcolor{red}{0.9831} & \textcolor{red}{1.0579} & 0.2985 \\
        & The Efficiency Essentialist & 0.6386 & \textcolor{red}{1.0323} & \textcolor{red}{-0.0145} \\
        & The Narrative Architect & 0.7613 & \textcolor{red}{0.9511} & \textcolor{red}{0.0558} \\
        & The Social Lubricator & \textcolor{red}{0.9340} & \textcolor{red}{0.8268} & \textcolor{red}{-0.1026} \\
        & The Thrill Seeker & \textcolor{red}{1.1362} & \textcolor{red}{1.1917} & \textcolor{red}{0.0090} \\
        & \textit{AVERAGE} & \textit{\textcolor{red}{0.8906}} & \textit{\textcolor{red}{1.0119}} & \textit{0.0492} \\
        \midrule
        
        \multirow{6}{*}{\textbf{MeepleLM (Ours)}} 
        & The System Purist & 0.6135 & 0.2131 & 0.4169 \\
        & The Efficiency Essentialist & 0.5073 & 0.2671 & 0.2692 \\
        & The Narrative Architect & 0.6560 & 0.2094 & 0.2529 \\
        & The Social Lubricator & \textcolor{red}{0.8018} & 0.2103 & 0.2857 \\ 
        & The Thrill Seeker & 0.7094 & 0.2025 & 0.1836 \\ 
        & \textit{AVERAGE} & \textbf{\textit{0.6576}} & \textbf{\textit{0.2205}} & \textbf{\textit{0.2817}} \\
        
        \midrule
        
        \multirow{6}{*}{\hspace{3mm}\textit{w/o MDA}} 
        & The System Purist & 0.7538 & 0.4891 & 0.3213 \\
        & The Efficiency Essentialist & 0.6326 & 0.4468 & 0.2584 \\
        & The Narrative Architect & 0.6649 & 0.4033 & 0.2849 \\
        & The Social Lubricator & \textcolor{red}{0.8661} & 0.3887 & 0.1121 \\
        & The Thrill Seeker & 0.7800 & 0.3463 & 0.1587 \\
        & \textit{AVERAGE} & \textit{0.7395} & \textit{0.4148} & \textit{0.2271} \\
        \cmidrule(l){1-5} 
        
        \multirow{6}{*}{\hspace{3mm}\textit{w/o Persona}} 
        & The System Purist & \textcolor{red}{0.8954} & 0.5946 & 0.1743 \\
        & The Efficiency Essentialist & 0.6806 & 0.4493 & 0.1860 \\
        & The Narrative Architect & 0.6851 & 0.2822 & 0.1431 \\
        & The Social Lubricator & \textcolor{red}{0.8852} & 0.2483 & \textcolor{red}{0.0341} \\ 
        & The Thrill Seeker & 0.7972 & 0.2407 & 0.1367 \\
        & \textit{AVERAGE} & \textit{0.7887} & \textit{0.3630} & \textit{0.1348} \\
        \cmidrule(l){1-5}
        
        \multirow{6}{*}{\hspace{3mm}\textit{w/o Rulebook}} 
        & The System Purist & 0.7745 & 0.5879 & \textcolor{red}{0.0528} \\
        & The Efficiency Essentialist & 0.5645 & 0.5152 & \textcolor{red}{0.0270} \\
        & The Narrative Architect & 0.6920 & 0.6189 & \textcolor{red}{-0.0504} \\
        & The Social Lubricator & 0.7546 & 0.5437 & \textcolor{red}{-0.0467} \\
        & The Thrill Seeker & 0.7360 & 0.4821 & \textcolor{red}{0.0303} \\
        & \textit{AVERAGE} & \textit{0.7043} & \textit{0.5496} & \textcolor{red}{\textit{0.0026}} \\
        
        \bottomrule
    \end{tabular}
    \caption{\textbf{Comprehensive Persona-wise Alignment Metrics.} Full breakdown of MAE, Wasserstein Distance (WD), and Kendall's $\tau$ across five gamer personas. \textbf{\textcolor{red}{Red values}} indicate poor alignment (MAE/WD $> 0.8$, or near-zero/negative correlation), highlighting where general baselines fail to capture specific player preferences.}
    \label{tab:full_persona_breakdown}

\end{table*}

\end{document}